\documentclass{aa}  

\usepackage{graphicx}
\usepackage{txfonts,textcomp}
\usepackage[colorlinks]{hyperref}

\def\aap{A\&A}
\def\apj{ApJ}
\def\apjs{ApJS}

\def \hi {\ion{H}{i}}
\def\h2{H$_2$}

\def\kms{km\,s$^{-1}$}

\def\deg{\hbox{$^\circ$}}
\def\arcmin{\hbox{$^\prime$}}

\def\fdg{\hbox{$.\!\!^\circ$}}
\def\farcm{\hbox{$.\mkern-4mu^\prime$}}

\defcitealias{Kalberla2021}{Paper~I}
\usepackage[normalem]{ulem}

\begin{document}

\title{Aspect ratios of far-infrared and \hi\ filaments in the diffuse
 interstellar medium at high Galactic latitudes}

\subtitle{}

   \author{P.\ M.\ W.\ Kalberla \inst{1}, \and U.\ Haud \inst{2} }

\institute{Argelander-Institut f\"ur Astronomie,
           Auf dem H\"ugel 71, 53121 Bonn, Germany \\
           \email{pkalberla@astro.uni-bonn.de}
           \and
           Tartu Observatory, University of Tartu,
           61602 T\~oravere, Tartumaa, Estonia }

   \authorrunning{P.\,M.\,W. Kalberla \& U.\ Haud } 

   \titlerunning{Aspect ratios and filamentarities of FIR/\hi\ filaments}

   \date{Received 13 October 2022 / Accepted 17 March 2023}

% \abstract{}{}{}{}{} 
% 5 {} token are mandatory
  \abstract 
{Dusty magnetized structures observable in the far-infrared (FIR) at
  high Galactic latitudes are ubiquitous and found to be closely
  related to \hi\ filaments with coherent velocity structures.  }
% aims heading (mandatory) \hi\
{Considering dimensionless morphological characteristics based on
  Minkowski functionals, we determine the
  distribution of filamentarities $ {\cal F} $ and aspect ratios $ {\cal
    A} $  for these structures.  }
{Our data are based on {\it Planck} FIR and  HI4PI \hi\ 
  observations. Filaments have  previously been extracted by
  applying the Hessian operator. We trace individual filamentary
  structures along the plane of the sky and determine $ {\cal A} $ and $
  {\cal F} $.}
{ Filaments in the diffuse interstellar medium (ISM) are seldom
  isolated structures, but are rather part of a network of filaments with a well-defined, continuous distribution in $ {\cal A} $ and $ {\cal F} $. This
  distribution is self-replicating, and the merger or disruption of individual
  filamentary structures leads only to a repositioning of the filament
  in $ {\cal A} $ and $ {\cal F} $ without changing the course of the
  distribution. }
{ FIR and \hi\ filaments identified at high Galactic latitudes are a close match
  to model expectations for narrow filaments with approximately
  constant widths. This distribution is continuous without clear upper
  limits on the observed aspect ratios. Filaments are associated with
  enhanced column densities of CO-dark \h2.  Radial velocities along the
  filaments are coherent and mostly linear with typical dispersions of
  $\Delta v_{\mathrm{LSR}} = 5.24 $ \kms.  The magnetic field strength
  in the diffuse turbulent ISM scales with hydrogen volume density as $B
  \propto n_{\mathrm{H}}^{0.58} $. At high Galactic latitudes, we
  determine an average turbulent magnetic field strength of $\langle \delta B
  \rangle = 5.3 ~\mu$G and an average mean strength of the magnetic
  field in the plane of the sky of $\langle B_{\mathrm{POS}} \rangle =
  4.4 ~\mu$G. }


  \maketitle
%
%________________________________________________________________

\section{Introduction}
\label{Intro}

A significant fraction of the interstellar medium (ISM) is filamentary.
Predominantly within the last two decades, evidence has been accumulating from
high-resolution all-sky surveys that the dust emission, stellar
reddening, and \hi\ line emission in large parts of the sky are
correlated and shaped in filamentary structures. Moreover, it appears
that ridges in the far-infrared (FIR) emission ---observable with {\it
  Planck} at 857, 545, and 353 GHz--- are aligned with the magnetic field
measured on the structures \citep{Planck2016}. Evidence for magnetically
aligned \hi\ fibers at high Galactic latitudes that extend for many
degrees was found by \citet{Clark2014} based on observations with the
Galactic Arecibo L-Band Feed Array survey
(GALFA-\hi\ \citealt{Peek2018}). \citet{Kalberla2016} used the HI4PI
survey \citep{HI4PI2016} to demonstrate that filamentary structures in
\hi\ are correlated with FIR emission at 353 GHz. Filamentary structures
are also important in dense molecular gas regions, setting the initial
conditions for star formation; see e.g. \citet{Hennebelle2019} and
\citet{Hacar2022} for a recent reviews. Our investigations are 
restricted to the diffuse ISM with column densities $N_H \la
10^{21.7}\,{\rm cm^{-2}}$; gravitational instabilities are unexpected in
this range and the magnetic field is in general aligned parallel to the
filaments \citep{Hennebelle2019}.

Our current contribution is a follow-up to the work of \citet[][Paper
  I]{Kalberla2021}, where we studied the coherence between FIR emission
at 857 GHz and \hi\ on angular scales of 18\arcmin. A Hessian analysis
was applied to the large-scale FIR distribution observed with {\it
  Planck}, and to the \hi\  observed with HI4PI. Such an analysis
determines local maxima in the FIR and \hi\ distributions
from second-order partial derivatives. The data reveal that structures,
or ridges, in the intensity map have counterparts in the Stokes Q and/or
U maps.  Upon tracing structures in orientation angles $\theta$ along the
maxima, a tight agreement between FIR and \hi\ was found in narrow
velocity intervals of 1 \kms. The FIR filaments are organized in
coherent structures with well-defined radial velocities. Accordingly,
radial velocities can be assigned to FIR filaments. Orientation angles
$\theta$ along the filaments ---projected perpendicular to the line of
sight--- are fluctuating systematically. The derived distribution of
filament curvatures was found to be characteristic for the distribution
generated by a turbulent small-scale dynamo. 

Here, we continue the analysis from \citetalias{Kalberla2021}, 
generating a complete census of filamentary structures in the diffuse ISM
at high Galactic latitudes. Our intention is to provide observational
constraints for aspect ratios of the fluctuation dynamo. To obtain better
insight into the nature of filamentary structures, we use 2D Minkowski
functionals. We trace individual structures and measure their surfaces
and perimeters to derive their filamentarity. From an independent
determination of the filament widths, we also derive
aspect ratios of FIR/\hi\ filaments  for the first time and relate these to their
filamentarities. For each individual structure, we also determine the
average radial velocity and its dispersion. In Sect. \ref{Obs}, we give
an overview of the observations and the data processing. Data analysis
and results on the filamentarity of FIR/\hi\ structures are presented in
Sect. \ref{Data_analysis}.  We estimate the turbulent magnetic field
strength in Sect. \ref{field} and the mean field strength in the plane
of the sky (POS) in Sect. \ref{rimLB_mean}. We then discuss our results in
Sect. \ref{Discussion} and provide a summary in Sect. \ref{Summary} .

\section{Observations and data processing}
\label{Obs}

Our analysis is based on pre-processed data and results from
\citetalias{Kalberla2021}. We used HI4PI \hi\ observations
\citep{HI4PI2016}, combining data from the Galactic all-sky survey
(GASS; \citet{Kalberla2015}) measured with the Parkes radio telescope
and the Effelsberg-Bonn \hi\ Survey (EBHIS; \citealt{Winkel2016}) with
data from the 100 m telescope.  The \hi\ data were correlated with FIR
emission maps from the Public Data Release 4 (PR4) \citep{Planck2020} at
frequencies of 857, 545, and 353
GHz\footnote[1]{HFI\_SkyMap\_857\_2048\_R4.00\_full.fits and
  HFI\_SkyMap\_545\_2048\_R4.00\_full.fits, and
  HFI\_SkyMap\_353-field-IQU\_2048\_R4.00\_full.fits from
  \url{https://irsa.ipac.caltech.edu/data/Planck/release_3/ancillary-data/HFI_Products.html}}.
The aim of the analysis of  \citetalias{Kalberla2021} was to study the velocity-dependent coherence
between FIR filaments at 857 GHz and \hi; here we briefly describe
their basic data processing.

\subsection{Preliminaries: Hessian analysis}
\label{Hessian}

In \citetalias{Kalberla2021}, the Hessian operator $H,$ which is based on
partial derivatives of the intensity distribution, was used as a tool to
classify structures as filament-like:
\begin{equation}
     \label{eq:hessI} 
        H(x,y)\, \equiv \, \left ( \begin{array}{cc} H_{xx} & H_{xy }\\
            H_{yx} & H_{yy} \end{array} \right ).
\end{equation}  
Here {\it x} and {\it y} refer to the true angle approximations in
longitude and latitude as defined in Sect. 4.1 of
\citet{Gorski2005}. The second-order partial derivatives are
$H_{xx}=\partial^2 I / \partial x^2$, $H_{xy}=\partial^2 I / \partial x
\partial y$, $H_{yx}=\partial^2 I / \partial y \partial x$, and
$H_{yy}=\partial^2 I / \partial y^2$.

The eigenvalues of H, 
\begin{equation}
\label{eq:lambda}
\lambda_{\pm}=\frac{(H_{xx}+H_{yy}) \pm \sqrt{(H_{xx}-H_{yy})^2+4H_{xy}H_{yx}}}{2},
\end{equation}
describe the local curvature of \hi\ and FIR features; $\lambda_- < 0 $
is in the direction of least curvature and indicates filamentary structures
or ridges. The 5 × 5 pixel matrix of the Hessian operator implies a
spatial filtering that corresponds to a scale of 18\arcmin. Taking the
resolution limits of the HI4PI survey into account, this is the highest
possible spatial resolution that can be used to search  the whole sky for
filamentary FIR/\hi\ structures.  An
adapted filtering was applied to improve sensitivity in the FIR; see \citetalias{Kalberla2021} for
details.

The local orientation of filamentary structures relative to the Galactic
plane is given by the angle
\begin{equation}\label{eq:theta}
\theta =
\frac{1}{2}\arctan\left[\frac{H_{xy}+H_{yx}}{H_{xx}-H_{yy}}\right],
\end{equation}
in analogy to the relation
\begin{equation}\label{eq:theta2}
\theta_S =\frac{1}{2}\arctan \frac{U}{Q},
\end{equation}
which can be derived from polarimetric observations that provide the 
Stokes parameters $U$ and $Q$.

The Hessian analysis, including the determination of local orientation
angles along the filaments, was applied to FIR data at 857 GHz and
repeated in \hi\ for all velocity channels at $|v_{\mathrm{LSR}}| < 50 $
\kms. Applying an appropriate level $\lambda_- < -1.5\ \mathrm{K
  deg}^{-2}$ at 857 GHz, a HEALPix bit map was derived that defines
positions with significant filamentary structures. These data exceed an
average signal-to-noise ratio (S/N) level of 68 at high Galactic
latitudes. Positions with $\lambda_- > -1.5\ \mathrm{K deg}^{-2}$ were
flagged as undefined and excluded from the further analysis.
  
As the Hessian matrix operator was applied independently to all
channels of the \hi\ database, it is possible to consider velocity-dependent orientation angles $\Theta$ for the \hi. Spurious structures
from low-S/N data were excluded by applying an \hi\ threshold of $\lambda_-
= -50\ \mathrm{K deg}^{-2}$.  Each individual position that was
previously determined to belong to a FIR filament was searched for the
velocity with the best agreement in FIR and \hi\ orientation
angles. These velocities were encoded in a single nside=1024 HEALPix
database that is used for the current analysis. We emphasize that
matching \hi\ and FIR orientation angles locally (at individual
positions, not within channel maps) allows us to measure random motions and
velocity gradients along the filaments.

\hi\ data with different velocity widths were considered in
\citetalias{Kalberla2021}. The best correlation with typical alignment
errors of 3\fdg1 was found for narrow velocity intervals of $\delta
v_{\mathrm{LSR}} = 1 $ \kms. Velocities along the filaments are coherent
but with typical fluctuations of $\Delta v_{\mathrm{LSR}} = 5.5 $
\kms. A Gaussian analysis shows that the \hi\ that is associated
with the FIR filaments is cold with typical line widths of $\delta
v_{\mathrm{LSR}} = 3 $ \kms, implying that the velocity fluctuations
along the filaments are caused by supersonic turbulence.  The
FIR/\hi\ correlation at latitudes $|b| \ga 20\deg$ is dominated by
single narrow components belonging to the cold neutral medium
(CNM). Broad lines from the warm neutral medium (WNM) have very little effect on
the correlation. Blends from multiple independent CNM components may
disqualify the data analysis that was described above. Fortunately, such
blends from overlapping unrelated structures are rare at
Galactic latitudes $|b| \ga 20\deg$ (see Sect. \ref{plane}).

The dominance of the CNM within filaments that are exposed to supersonic
turbulence has important consequences for the data analysis. Selecting
channel maps at constant velocities with narrow velocity intervals as
low as $\delta v_{\mathrm{LSR}} = 3 $ \kms\ in the presence of velocity
gradients and turbulent motions with $\Delta v_{\mathrm{LSR}} = 5.5 $
\kms \ leads to an artificial fragmentation of filamentary
structures. Turbulent motions would partly shift structures to
neighboring channels and the results depend on the selected velocity
grid and the channel width.  Broadening the velocity channels or using
total column densities degrades the FIR--\hi\ correlation (see Table 1
and Fig. 3 of \citetalias{Kalberla2021}). In addition, using a coarse
velocity resolution implies limitations to the observed velocity
coherence (see there Sect. 2.6 and Figs. 4 and 5) because filaments
cannot be properly identified. Furthermore, the
  FIR--\hi\ correlation with velocity coherence along the filaments
  cannot be generated by random processes from unrelated
  \hi\ distributed along the line of sight. Such velocity caustics with
  prevailing density fluctuations in broad velocity intervals,
  predicted by \citet{Lazarian2000} and \citet{Lazarian2018}, would be
  in conflict with our interpretation of three-dimensional density
  structures in the ISM.

\subsection{Minkowski functionals and filamentarity}
\label{Minkowsi}

The aim of \citetalias{Kalberla2021} was to determine velocity
coherence, local orientation angles, and the curvature distribution along
the filaments. Here we restrict the analysis to the probability density
distribution (PDF) of filamentary axis ratios. The easiest way to perform a
parameterization is to consider aspect ratios $ {\cal A} = L / W $ by
measuring the lengths $L$ and widths $W$ of filaments. A more general way is
to consider geometrical descriptors for the morphology of filamentary
structures in terms of Minkowski functionals.  These functionals derive
from the theory of convex sets and generalize curvature integrals over
smooth surfaces. A fairly complete introduction to the basic
morphological measures can be obtained from \citet{Mecke1994}, \citet{Mecke2000},
and \citet{Kerscher2000}. \citep{Sahni1998} and \citet{Bharadwaj2000}
apply Minkowski functionals to study the degree of filamentarity in the
Las Campanas Redshift Survey.

In three dimensions, the four Minkowski functionals characterizing the
morphology of a compact manifold are the volume $V_\mathrm{3D}$, the
surface area $S_\mathrm{3D}$, the integrated mean curvature
$C_\mathrm{3D}$, and the integrated Gaussian curvature $G_\mathrm{3D}$.
These functionals can be used to define measures for the length $L =
C_\mathrm{3D}/ 4 \pi$ , width $W = S_\mathrm{3D}/C_\mathrm{3D}$ , and
thickness $T = 3V_\mathrm{3D}/S_\mathrm{3D}$ of objects under
consideration. With a normalization $L=W=T = r$ for a sphere of radius
$r$, a spherical object would be characterized by $ L \simeq W \simeq T
$, a pancake as $ L \simeq W \ggg T$, a ribbon as $ L \ggg W \ggg T $,
and a filament as $ L \ggg W \simeq T $. Based on the three different
length-measures $L$, $W$, and $T$, object shapes can be characterized
with two dimensionless shapefinders: the planarity $P =(W-T)/(W+T)$ and
filamentarity $F=(L-W)/(L+W)$.

Here, we are concerned with two dimensions: the thickness (hence also the
planarity) cannot be measured but the filamentarity is still well
defined. \citet{Bharadwaj2000} found it useful  in this
case to define
the filamentarity ${\cal F}$ by applying slightly different conventions,
as mentioned above, using perimeter $P$ and surface area $S$,
\begin{equation}
{\cal F} = \frac{P^2 - 4\pi S}{P^2 + 4\pi S}.
\label{eq:F}
\end{equation}
By definition, $ 0 \le {\cal F} \le 1$.  We obtain $ {\cal F} = 0 $ for a
filled circle and $ {\cal F} = 1 $ in the limit for a
line segment of infinate length. As an example, and for a different shape, we mention $ {\cal F}
= (4-\pi )/(4+\pi )\sim 0.12 $ for a filled square. Therefore, by calculating
$\cal F,$ it is possible to characterize the shape of an object with a
single number and this is particularly useful if we want to quantify the extent to which the shape of an object can be described as a filament.
\citet{Makarenko2015} applied this concept for the first time to
characterize the shapes of \hi\ cloud structures in the Milky Way.

\subsubsection{Pixel counting}
\label{Data}

The Hessian analysis results in a HEALPix map with the best-fit radial
velocities for each position on the filament. To determine $P$ and $S$
for individual structures, we need to trace the individual structures on this
map. The surface area $S$ can be determined simply by counting all
the pixels $n_{\mathrm{S}}$ that belong to the filament. For an nside=1024
HEALPix database, a single pixel covers an area of $4 \pi /12~582~912
\sim 10^{-6} $ sr or $ 3.3~10^{-3} \mathrm{deg}^2$.  The perimeter $P$
can be determined in several ways using the Cauchy-Crofton formula
\citep{Legland2011}.  However, we are faced with a time-critical
computational problem \citep{Lehmann2012}: we need to analyze $5.4~10^6$
boundary pixels. We decided to solve this by pixel counting in two ways,
counting pixels $n_{\mathrm{Pout}}$ just outside the structure, and
alternatively counting pixels $n_{\mathrm{Pin}}$ along the rim, but inside it; using
the average of the two counts approximates the set of curves that can go through
adjacent regions; see Fig. 2c of \citet{Stawiaski2007}.   In
addition,  for each position inside the filament, we determine the number of nearest
neighbors $n_{\mathrm{Neighbor}}$ that are also inside. The HEALPix
tessellation allows up to eight neighbors\footnote[2]{There are eight
  HEALPix positions on the sky with only seven neighbors}. Positions
with $n_{\mathrm{Neighbor}} = 8 $ can be considered as deep inside the
filament.

\subsubsection{Simulations}
\label{Simulations}

A basic check for numerical applications of Eq. (\ref{eq:F}) is to
ensure that the calculations result in $ {\cal F} = 0 $ for a filled
circle.  \citet{Bharadwaj2000} notice a further simple relation that
allows us to test the accuracy of calculating $ {\cal F} $ for filamentary
structures. The region between two concentric circles with radii $R_2 <
R_1$ has a filamentarity of
\begin{equation}
{\cal F} = \frac{R_2}{R_1}.
\label{eq:R}
\end{equation}

To test our numerical procedure, we generated an nside=1024 HEALPix
database,  first defining a filled circular disk centered on the
Galactic plane with a radius of $ r = 25\deg$. We then generated a series
of pseudo-filaments, that is, regions between two concentric circles at radii
$R_1 = r + \delta r$ and $R_2 = r - \delta r$. These simulate filaments
with widths of $2 \delta r$, surfaces of $ \pi (R_1^2 - R_2^2), $ and
perimeters of $2 \pi (R_1 + R_2)$. The filamentarity can be determined
according to Eq. (\ref{eq:R}).

The direct numerical application of Eq. (\ref{eq:F}), using
$P_{\mathrm{Pout}}$ or $P_{\mathrm{Pin}}$ to determine the perimeter and
$n_{\mathrm{S}}$ to determine the surface led immediately to an inconsistent
result, namely $ {\cal F} = 0.25 $ for the filled circle. Equation (\ref{eq:F})
is defined for the ideal case of a distribution that can be represented
with continuous contours.  \citet{Bharadwaj2000}
pointed out that this definition should be modified for data on a grid. These authors propose
different shapefinder versions for a rectangular grid. However, using the version
with their Eq. (6) leads only to a slight improvement, that is, $ {\cal
  F}_1 = 0.13 $ for the filled circle. Next we tested another
shapefinder relation according to Eq. (8) of \citet{Bharadwaj2000} and
found that with $ {\cal F}_2 = 0.23 $ for the filled circle, this
relation is also not appropriate for our application.

We decided to continue along the definition in Eq. (\ref{eq:F}) but to
consider the properties of the HEALPix database in detail. The HEALPix
tessellation provides the same surface area for all pixels but slightly
different shapes \citep{Gorski2005}. The surface is therefore well
defined when counting pixels $n_{\mathrm{S}}$ inside the filament, but the
HEALPix grid is variable,  locally approximating a rectilinear
equal-area grid that is rotated against the Galactic coordinate system
by 45\deg\ \citep[][Figs. 4 and 5]{Gorski2005}.  The direction weights
for rectangular grids that are needed for the Crofton formula in
discrete images \citep[][Fig. 1]{Lehmann2012} are position
dependent, which implies that distances between individual pixels
along the perimeter of a filament are variable. Direct distance measures
along jagged filament borders using $n_{\mathrm{Pout}}$ or
$n_{\mathrm{Pin}}$ are biased in any case. We take this into account by
applying an average correction using a bias factor $f_{\mathrm{b}}$,
such that the outer and inner perimeters are approximated by
$P_{\mathrm{Pout}} = n_{\mathrm{Pout}} f_{\mathrm{b}}$ or
$P_{\mathrm{Pin}} = n_{\mathrm{Pin}} f_{\mathrm{b}}$, respectively.

Generating a series of pseudo-filaments at a constant radius of 25\deg
with variable widths $2 \delta r$, we calculated the expected
filamentarity ${\cal F_{\mathrm{model}}}$ according to Eq. (\ref{eq:R})
and varied the geometric correction factor $f_{\mathrm{b}}$. From about
3~600 model calculations, the best fit with ${\cal F}/ {\cal
  F_{\mathrm{model}}} = 1.00005 \pm 1.\ 10^{-3}$ was found for
$f_{\mathrm{b}} = 0.77$\footnote[3]{Errors given throughout this paper
  are dispersions rounded to a significant digit. In this case, we
  derived  ${\cal F}/ {\cal
    F_{\mathrm{model}}} = 0.999974 \pm 8.097 \ 10^{-4}$ from 114 samples
  with variable widths with a different realization.}. For the filled circle, we determine in this case
${\cal F} = -1.16\ 10^{-3}$, and also the minimal deviation from zero that
we found during the simulations. We emphasize that the correction with a
factor $f_{\mathrm{b}} = 0.77$ is a statistical correction only. The
simulations described above used a well-defined test bed with a uniform
distribution of orientation angles. We expect that the correction factor
$f_{\mathrm{b}} = 0.77$ is unbiased, but the application to complex
filamentary structures is probably far less accurate than the average
0.1\% error level that could be achieved during the
simulations. In any case, measurements of the perimeters of pixelized structures always come with uncertainties, which depend on the size and orientation of
the object. Using computationally more elaborate Crofton methods, one
would expect relative errors of up to 1\% or 5\% depending on the algorithm
used; see Fig. 5 of \citet{Lehmann2012}.

We like to note here that outer and inner perimeter counts show only
minor deviations as long as the filaments are extended and well defined
(as in the simulations).  For short filaments, significant differences
are notable and we therefore use the average $P = f_{\mathrm{b}} ~
(n_{\mathrm{Pout}} + n_{\mathrm{Pin}}) /2 $ as the best estimate for the
perimeter. Perimeters around surfaces with $n_{\mathrm{S}} < 10$ are ill
defined. From Fig. 4 of \citet{Lehmann2012}, we expect that errors,
which are independent of the numerical method used, would exceed
20\% in these
cases. Such objects are disregarded.

\begin{figure}[th] %%  1
   \centering
   \includegraphics[width=9cm]{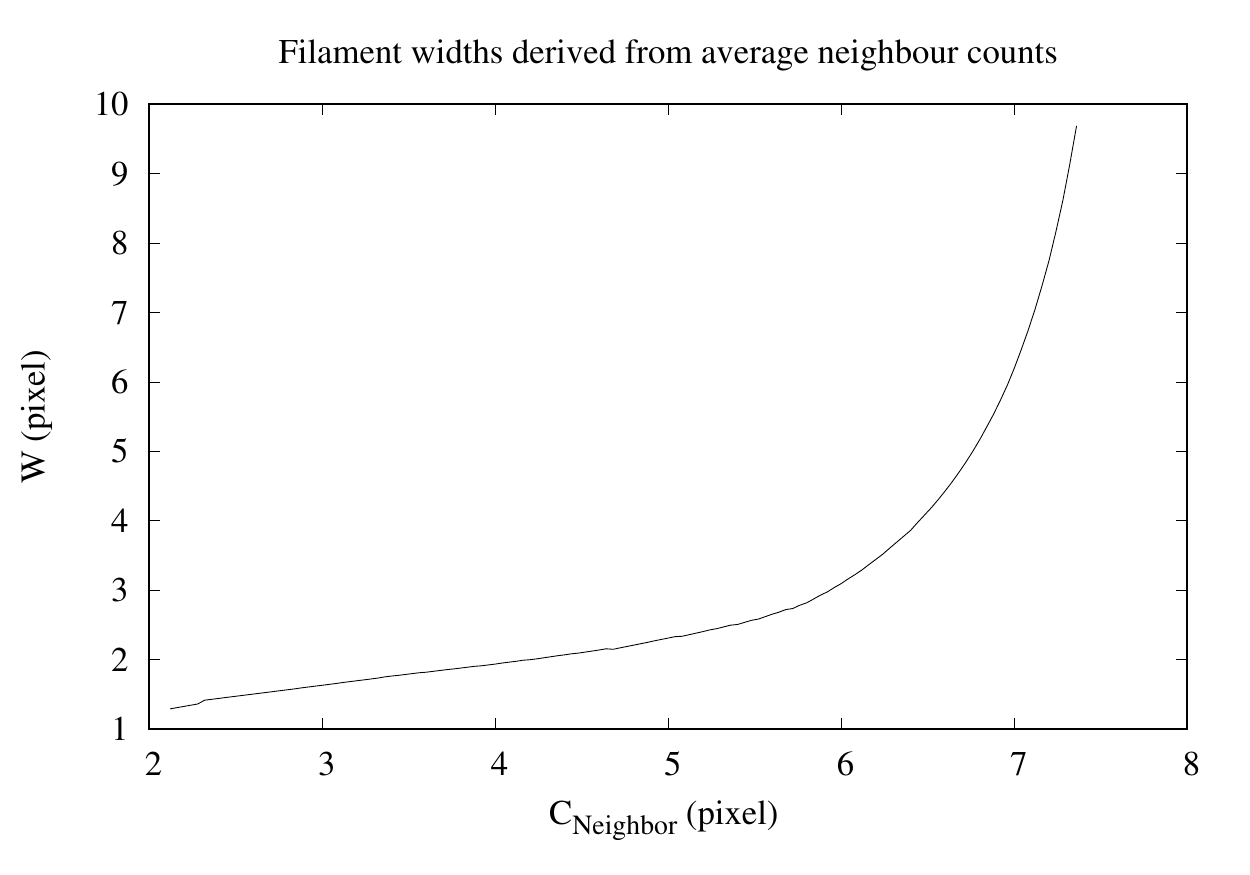}
   \caption{Average filament widths derived from neighbor counts during
     simulations. }
   \label{Fig_width}
\end{figure}

Our simulations were also used to estimate the filament widths, which are defined
as the average width along the filament.  For a filament with a simple
geometry one would first determine the skeleton (or center) of the
filament and measure  ---for each position along the skeleton--- the width
perpendicular to the direction defined by neighboring skeleton
elements. This procedure can become
complicated when faced with complex filament geometries or filament
branching,  with ambiguities at sharp turns.

We used a different method that is independent of filament shape.
For each pixel inside the filament, we counted the number of neighbors
$n_{\mathrm{Neighbor}}$. We then determined the average neighbor count
$C_{\mathrm{Neighbor}} = \sum n_{\mathrm{Neighbor}} / n_s $.  During our
simulations, we tabulated $C_{\mathrm{Neighbor}}$ as function of the
filament widths $W$ known from the model. This table is later used to
infer the average widths $W$ from the $C_{\mathrm{Neighbor}}$ counts;
Fig. \ref{Fig_width} shows the calibration curve. This method was found
to produce robust results up to a width of 10 pixels. We show below
that we only need to consider filaments with an average width of up to 5.5
pixels.  The width counts $W$ measures average distances (corrected for
geometrical sampling effects by application of the factor
$f_{\mathrm{b}}=0.77$), for an nside=1024 HEALPix database in units of
$10^{-3}$ rad or $5.7~10^{-2} $ deg.
     
In closing this section, we need to point out that the path from the
theoretical definition of filamentarity in Eq. (\ref{eq:F}) to a
meaningful derivation of parameters is hampered by a number of
uncertainties. Naive methods for the surface, perimeter, and width
determination can lead to huge systematic errors
\citep{Lehmann2012}. Our numerical recipes are valid for filamentary
objects with a uniform distribution of orientation angles and have been
tested against objects with known filamentarity and width.  It is
impossible to estimate uncertainties in cases of observed filament
geometries, which can be highly complex and variable. We therefore refer
to Sect. \ref{F_A}, where we compare derived aspect ratios and
filamentarities with the expected model distribution.  Uncertainties and
systematical errors should show up in this instance.

\section{Data analysis and results}
\label{Data_analysis}

Using the results from the Hessian analysis in
\citetalias{Kalberla2021}, we trace whether or not each position of the nside=1024
HEALPix database is part of a filament. For each position
inside a filament, we determine the number of neighbors
$n_{\mathrm{Neighbor}}$. By tracing individual filamentary structures, we
calculate surface $S$, perimeter $P$, average width $W$, and average
position in Galactic coordinates. We also calculate the average radial
velocity $v_{\mathrm{LSR}}$ with the associated dispersion $\Delta v$
along the filaments. In Sect. \ref{rimLB}, we use distances to the wall
of the magnetized Local Bubble (LB) from dust data modeled by
\citet{Pelgrims2020} and determine average filament distances, assuming
that these might be associated with the LB
wall\footnote[4]{\url{https://dataverse.harvard.edu/dataverse/Pelgrims}}.

\begin{figure}[th] %%  2
   \centering
   \includegraphics[width=9cm]{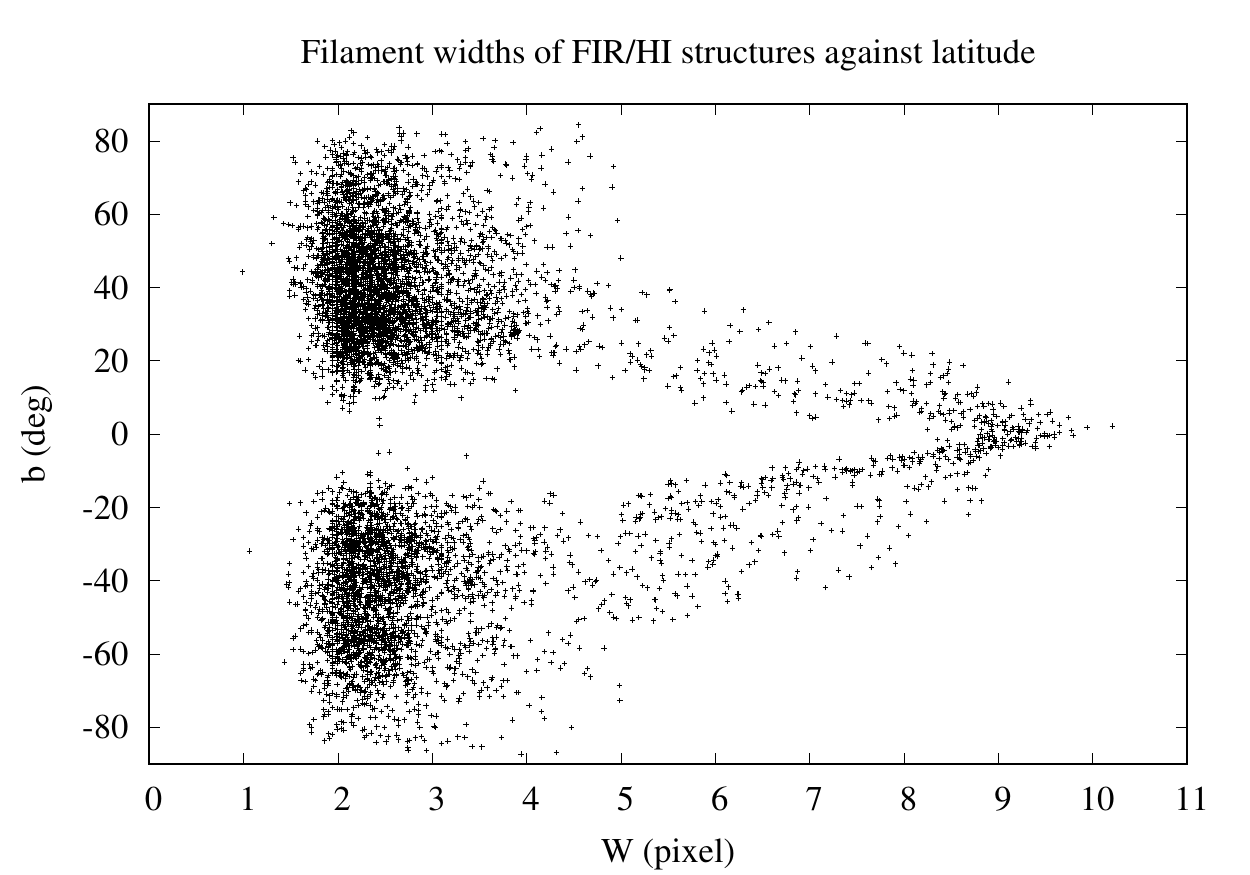}
   \caption{Spatial distribution of filaments with width $W$ as function
     of latitude $b$. }
   \label{Fig_all_sky}
\end{figure}

\subsection{Limitations: Confusion in the Galactic plane}
\label{plane}

To study the global distribution of filaments in the Milky Way, we first
show the latitude distribution of filament widths $W$ in
Fig. \ref{Fig_all_sky}. At high Galactic latitudes, the derived widths
are typically in the range of $1.5 \la W \la 4.5 $ pixels. However,
approaching the Galactic plane, we find that $W$ becomes strongly
biased, up to $W \sim 10$. This distribution is caused by
confusion. Numerous filaments overlay each other at velocities $-50 <
v_{\mathrm{LSR}} < 50 $ \kms\ and appear as one in projection. This
leads to spurious structures that cannot be reasonably interpreted with
our current approach.  The low latitude bias is not unexpected, because
the analysis in \citetalias{Kalberla2021} focused on high latitudes.  We
also note that all spurious structures in the Galactic plane have huge
surfaces.

Inspecting the results of our analysis, we find that large filaments at
latitudes $|b| \ga 30\deg$ extend to somewhat less than 20~000 pixels,
and those at low latitudes to far larger dimensions, propagating in
position to lower latitudes. As an example, for a reliable filament
extraction, we mention the structure at $l = 44\fdg1,~ b = 41\fdg1$,
which is discussed later in Sect. \ref{network}, with a surface of $
n_{\mathrm{S}} = 18~563$ pixels. Extrapolating this result, we use a
fiducial upper limit of $ n_{\mathrm{S}} = 20~000$ pixels for the
analysis of filamentary structures, corresponding to 0.02 sr or 65.6
deg$^2$.  Such extended structures are not traced for further extensions
and are also eliminated from the statistical analysis. This simple
recipe has proven to be quite successful, because it allows the user to
eliminate all structures that show obvious signs of originating from
blending. At the same time, features with a width of $W \ga 5.5$ are
removed. The final constraint for our analysis is that surfaces must be
in the range of $ 10 < n_{\mathrm{S}} < 20~000$ pixels. This restriction
leads to a total number of 6~568 filaments that can be used for our analysis. 

\begin{figure}[th] %%  3
   \centering
   \includegraphics[width=9cm]{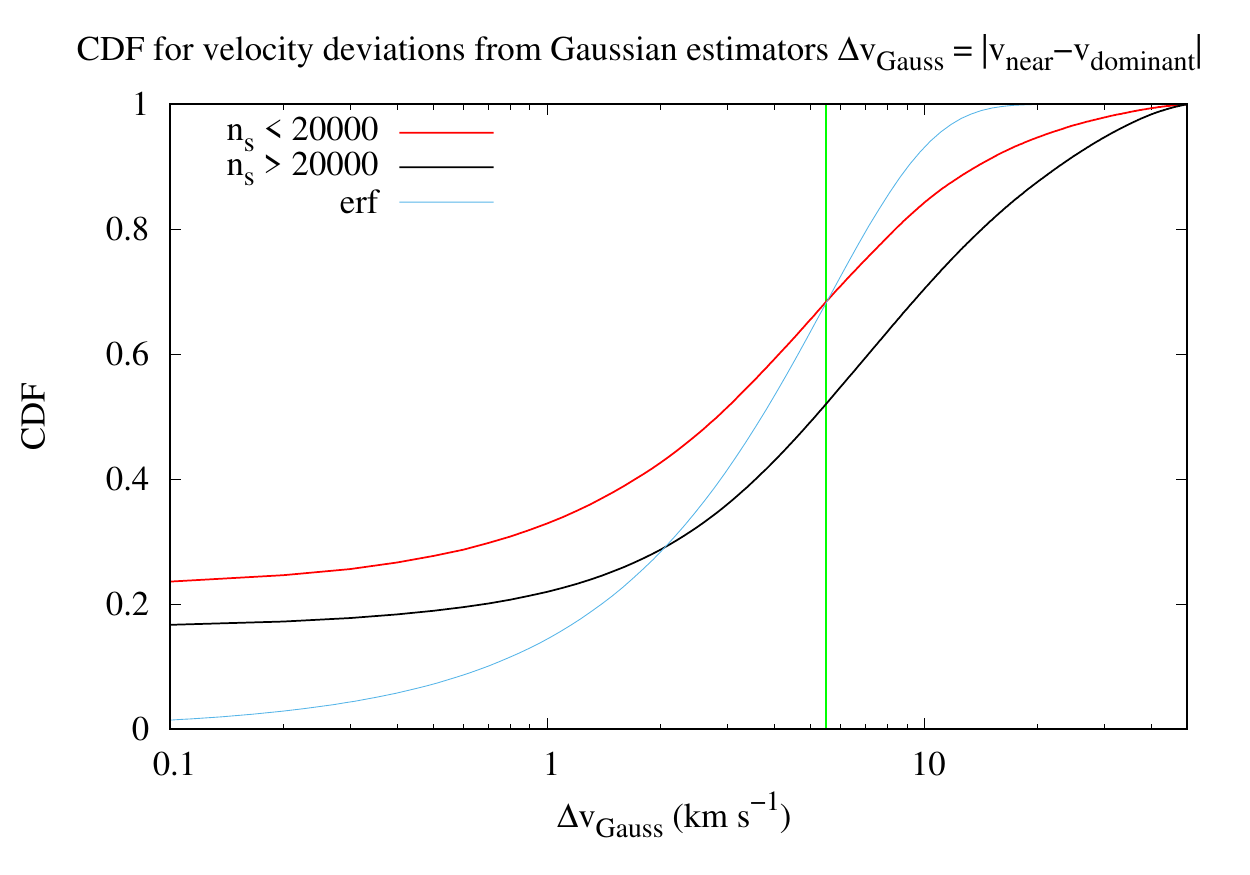}
   \caption{Cumulative distribution function of velocity separations
     $\Delta v_{\mathrm{Gauss}} = | v_{\mathrm{near}} -
     v_{\mathrm{dominant}}| $ between Gaussian components associated
     with \hi\ filaments that are nearest in velocity and dominant in
     column density.  }
   \label{Fig_Gauss}
\end{figure}

Our constraint of $ n_{\mathrm{S}} < 20~000$ is somewhat arbitrary; we
experimented with limits in the range $ 15~000 < n_{\mathrm{S}} <
20~000$ and found that our results are not significantly affected by a
particular choice. However, a limit of this kind turned out to be very
useful in identifying \hi\ regions that are affected by
confusion. Using the Gaussian decomposition from
\citetalias{Kalberla2021},   we compared the
best-fit \hi\ velocity for each filament position (from matching FIR/\hi\ orientation angles) with
the center velocities of the Gaussian components.  We selected the nearest
component in velocity at $v_{\mathrm{near}}$ and also ---independently--- the
dominant component with the largest column density in the spectrum at
$v_{\mathrm{dominant}}$.  Assuming that the FIR/\hi\ correlation in
orientation angles from \citetalias{Kalberla2021} is caused by
\hi\ density enhancements, both selections may point to
\hi\ counterparts of the FIR filaments.  For filaments with $
n_{\mathrm{S}} < 20~000,$ we find that in 25\% of all cases, identical
components are selected. Component velocities deviate in 66\% of all
cases by $\Delta v_{\mathrm{Gauss}} = | v_{\mathrm{near}} -
v_{\mathrm{dominant}}| < 5.5 $ \kms, which is equal to the velocity dispersion along the
filaments determined in \citetalias{Kalberla2021}. Figure
\ref{Fig_Gauss} shows the cumulative distribution function (CDF) for the
occurrence of a component separation, $\Delta v_{\mathrm{Gauss}}$. In
comparison to a normal distribution with a velocity dispersion of $ 5.5
$ \kms, we find a clear observational preference for component
separations of $\Delta v_{\mathrm{Gauss}} < 5.5 $ \kms. The CDF for
filaments with $ n_{\mathrm{S}} > 20~000$ is for all $\Delta
v_{\mathrm{Gauss}}$ values significantly below the selection with $
n_{\mathrm{S}} < 20~000$.

\begin{figure}[th] %%  4
   \centering
   \includegraphics[width=9cm]{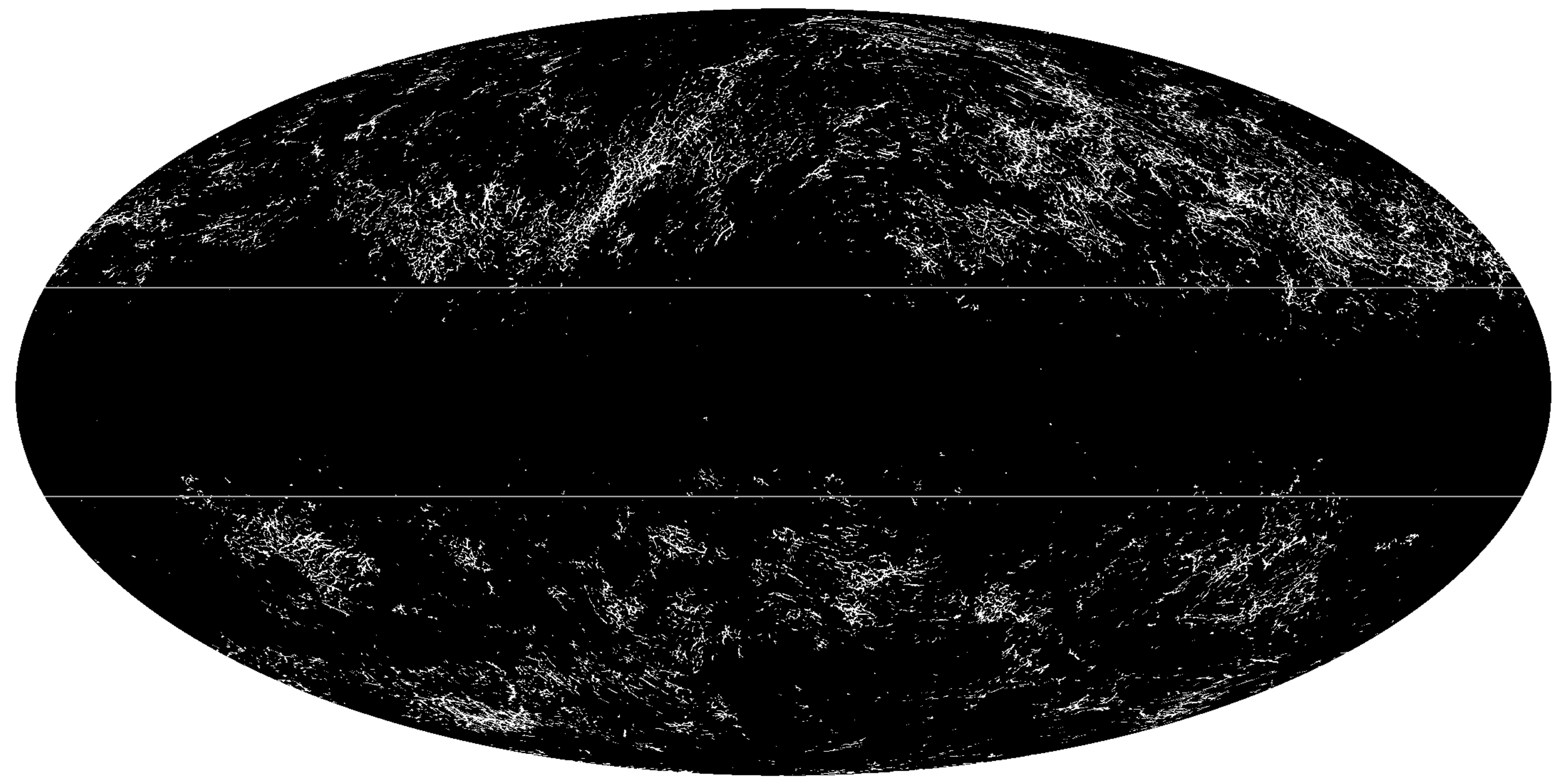}
   \caption{Distribution of the 6~568 FIR/\hi\ filaments in Mollweide
     projection.  The horizontal lines indicate Galactic latitudes of $ b =
     20\degr$ and $ b = -20\degr$. }
   \label{Fig_Sky_dist}
\end{figure}

Figure \ref{Fig_Sky_dist} shows the sky distribution of
FIR/\hi\ filaments for the selection $ n_{\mathrm{S}} < 20~000$ that is
used in the following. As opposed to the distribution shown in Fig. 1 of
\citetalias{Kalberla2021}, most of the filamentary structures at Galactic
latitudes $ | b | \la 20\degr$ are excluded because of confusion.  Such
a selection is consistent with the empirical assumption shared by many
observers that only the range $ | b | \ga 20\degr$ may be considered as
``high Galactic latitudes'', and therefore unperturbed by confusion.

\subsection{Filament widths}
\label{widths}

\begin{figure}[th] %%  5
   \centering
   \includegraphics[width=9cm]{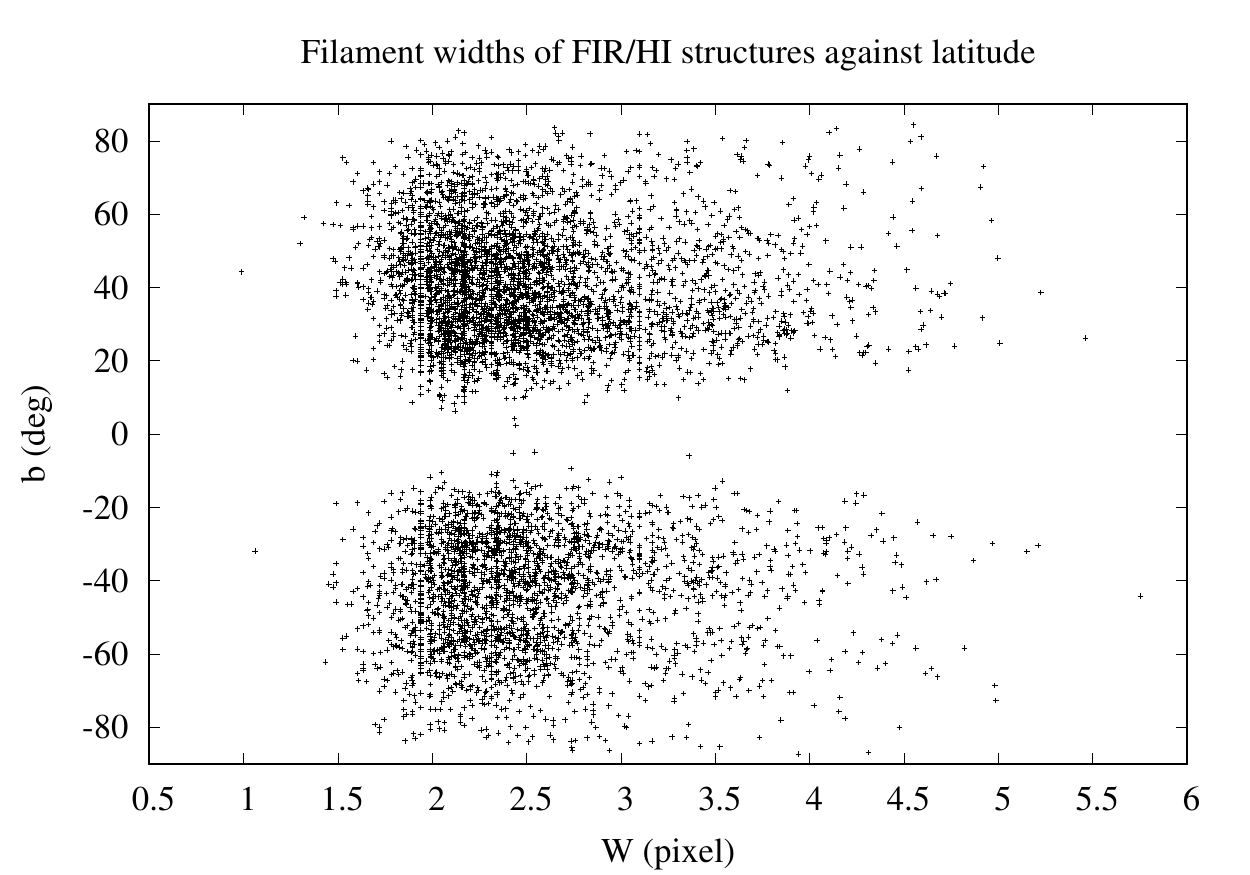}
   \includegraphics[width=9cm]{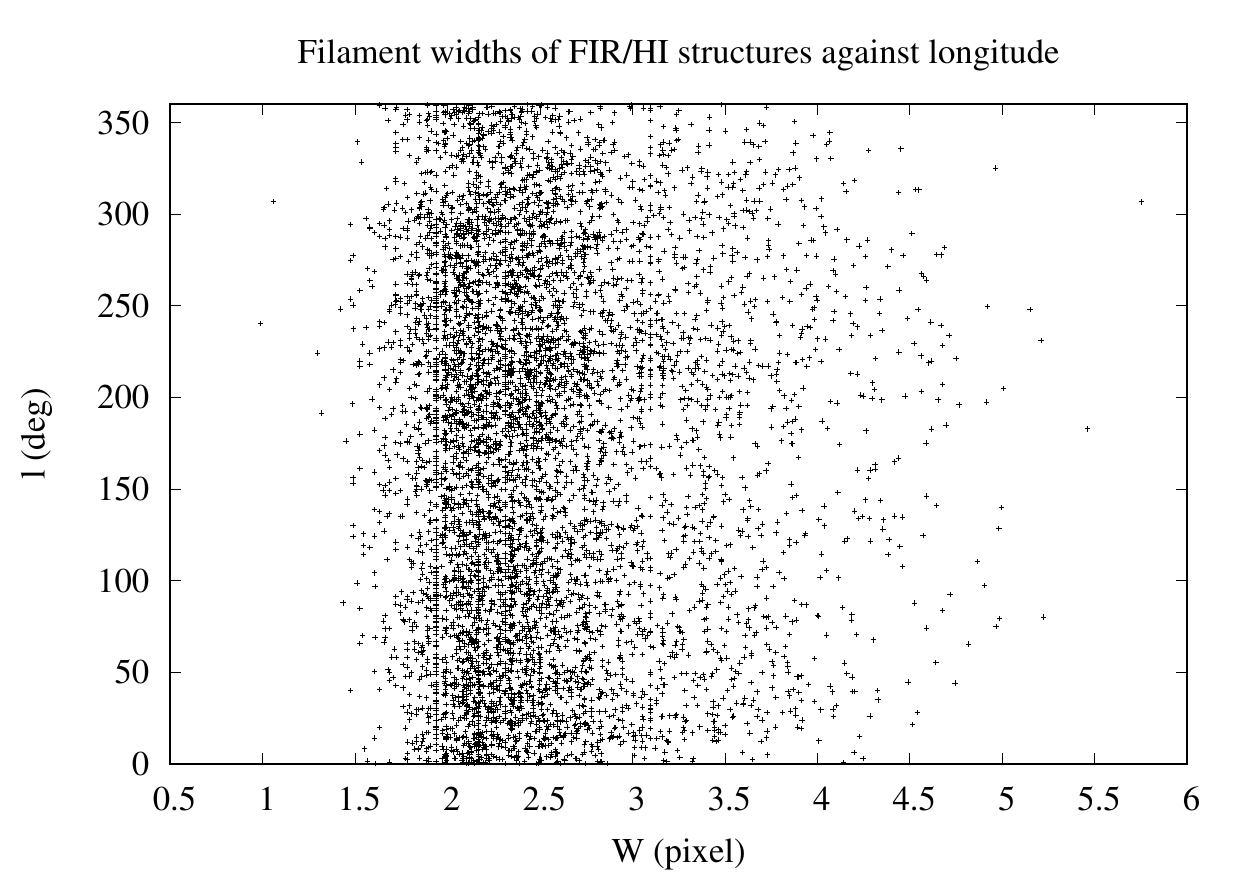}
   \caption{Spatial distribution of filaments with width $W$ for $ 10 <
     n_{\mathrm{S}} < 20~000$, as a function of latitude $b$ (top) and longitude $l$ (bottom).  }
   \label{Fig_W_dist}
\end{figure}

Figure \ref{Fig_W_dist} displays the spatial distribution of the
filaments with their derived widths after constraining the surface
area. On average, we find a width of 2.53 pixels. For a single pixel in
a nside=1024 HEALPix database, the angular resolution is $
\Theta_{\mathrm{pix}} = 3\farcm44 $ \citep{Gorski2005}. Assuming
filament distances of 100 pc \citep{Sfeir1999}, we estimate an average
filament width of 0.25 pc.

\begin{figure}[th] %%  6
   \centering
   \includegraphics[width=9cm]{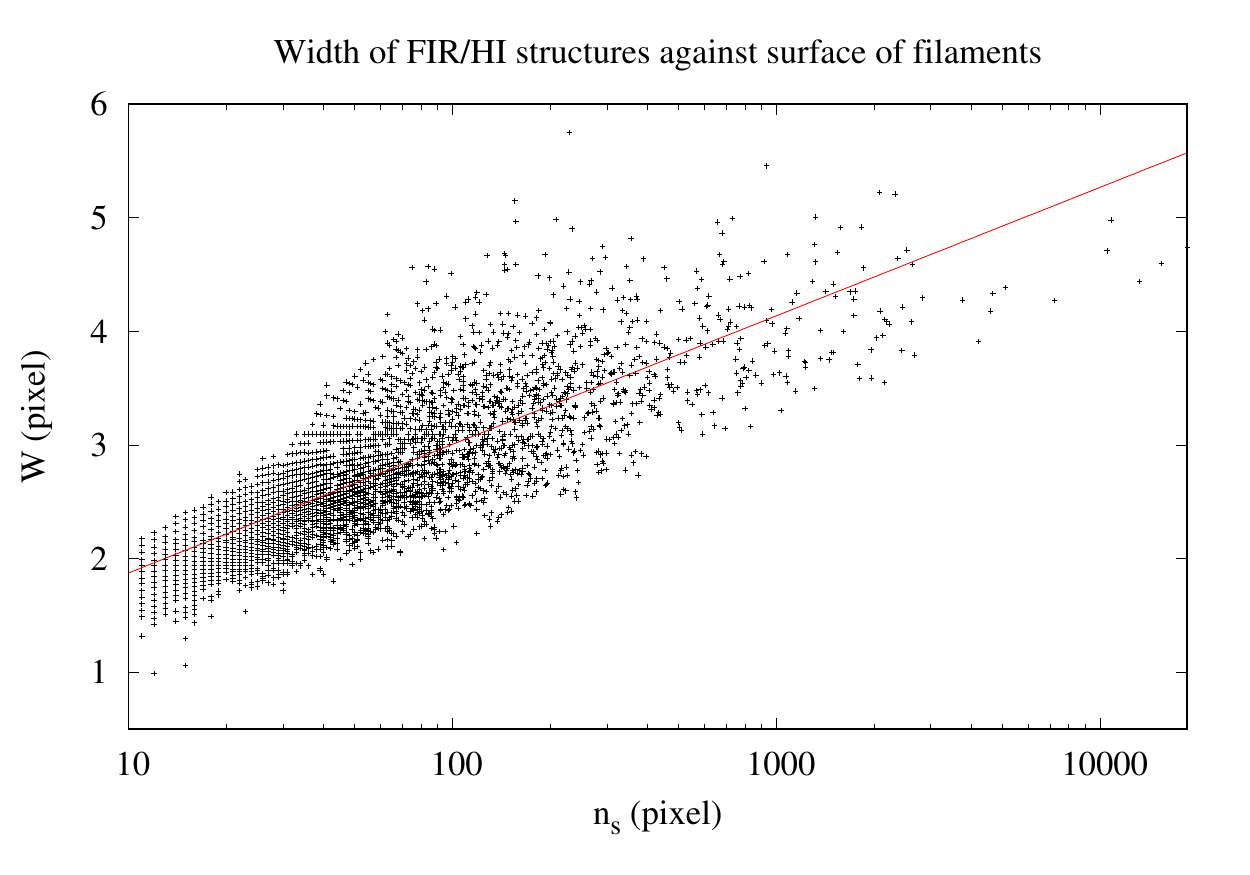}
   \caption{Average filament width $W$ as a function of filament surface,
     measured by counting surface pixels $n_{\mathrm{S}}$. The red line
     represents the fit $ W = 0.74 \pm .01 + (0.491 \pm .004)
     \ln(n_{\mathrm{S}}) $ }
   \label{Fig_Widths}
\end{figure}

\begin{figure}[th] %%  7
   \centering
   \includegraphics[width=9cm]{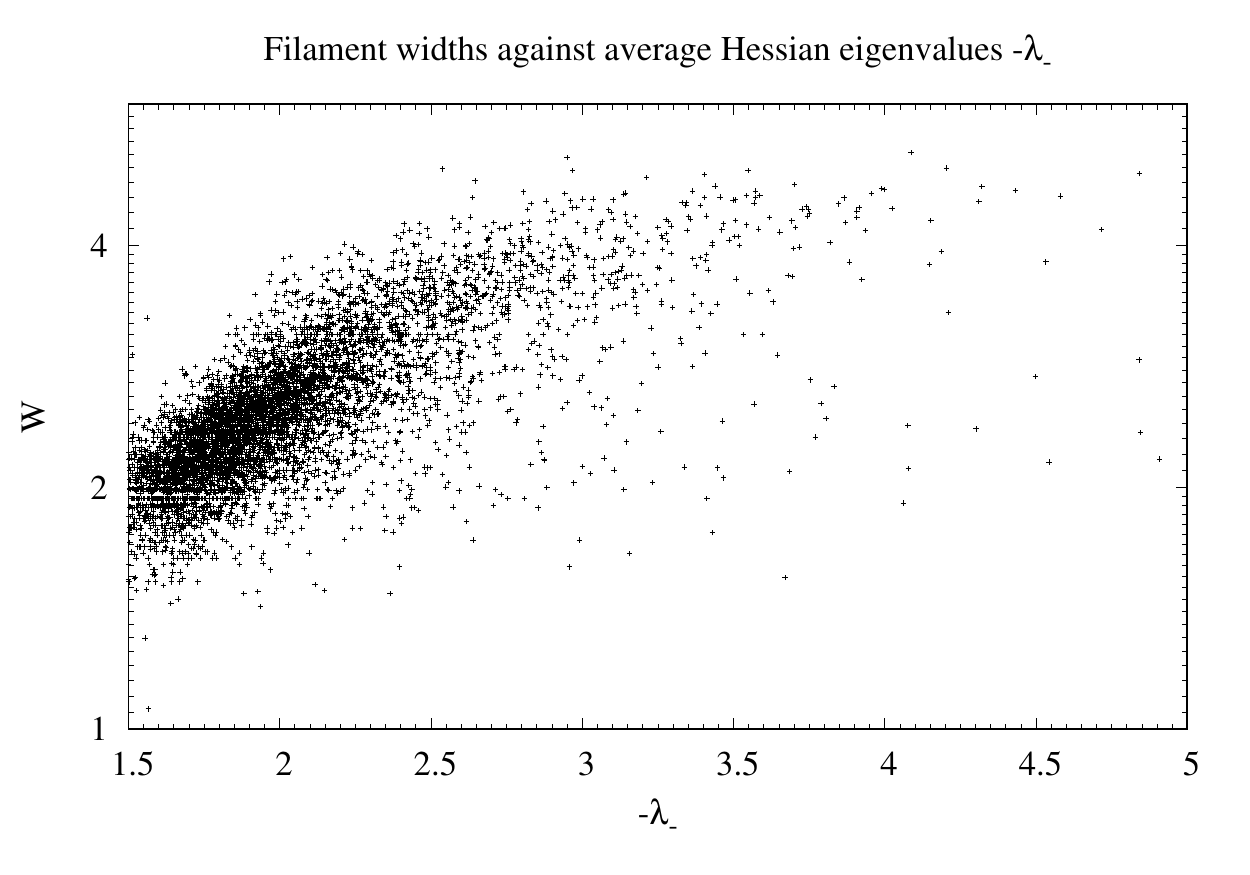}
   \caption{Average filament width $W$ as a function of the average
     Hessian eigenvalue $-\lambda_-$ from 857 GHz data.  }
   \label{Fig_Widths_lam}
\end{figure}

We find that the average filament widths increase with the
filament surfaces, $ W = 0.74 \pm .01 + (0.491 \pm .004)
\ln(n_{\mathrm{S}}) $. The data and fit\footnote[5]{Throughout the paper, for fitting a set of data points, we
  use an implementation of the
  nonlinear least-squares Marquardt-Levenberg algorithm provided by
  Gnuplot and cite rounded formal asymptotic standard errors of the
  fit.} are shown in Fig. \ref{Fig_Widths}. Stripy structures in this
plot come from the fact that $n_{\mathrm{S}}$ as a pixel sum is an
integer number, while averages over pixel counts can be treated as
real numbers. Also, perimeter counts are, by definition, quantized and some of
the later presentations are affected by integer arithmetic due to pixel
counts. The Hessian operator is selective and most sensitive to
FIR/\hi\ data with local S/N maxima (constant multiple rule). This
implies that prominent ridges have stronger eigenvalues $-\lambda_-$,
and therefore the average filament widths $W$  increase depending on the significance
level $\lambda_- < -1.5\ \mathrm{K deg}^{-2}$ at 857 GHz.
This effect is reproduced in Fig. \ref{Fig_Widths_lam} and discussed in
Sect. \ref{F_A} in more detail. We interpret such structures with
enhanced column densities in gas and dust.

\subsection{Filamentarity and aspect ratios}
\label{F_A}

For each of the 6~568 structures, we calculate the filamentarity ${\cal F}$
according to Eq. (\ref{eq:F}) by determining the
pixel counts $n_{\mathrm{S}}$   for the surfaces $S$ within the filaments. The perimeters are
derived from the average of the inner and outer pixel counts along the
rims of the filaments by applying the correction factor $f_{\mathrm{b}}$
as determined in Sect. \ref{Simulations}, that is, $ P =
(P_{\mathrm{Pout}}+P_{\mathrm{Pin}})/2 = f_{\mathrm{b}} 
(n_{\mathrm{Pout}} + n_{\mathrm{Pin}})/2 $.

\begin{figure}[th] %%  8
   \centering
   \includegraphics[width=9cm]{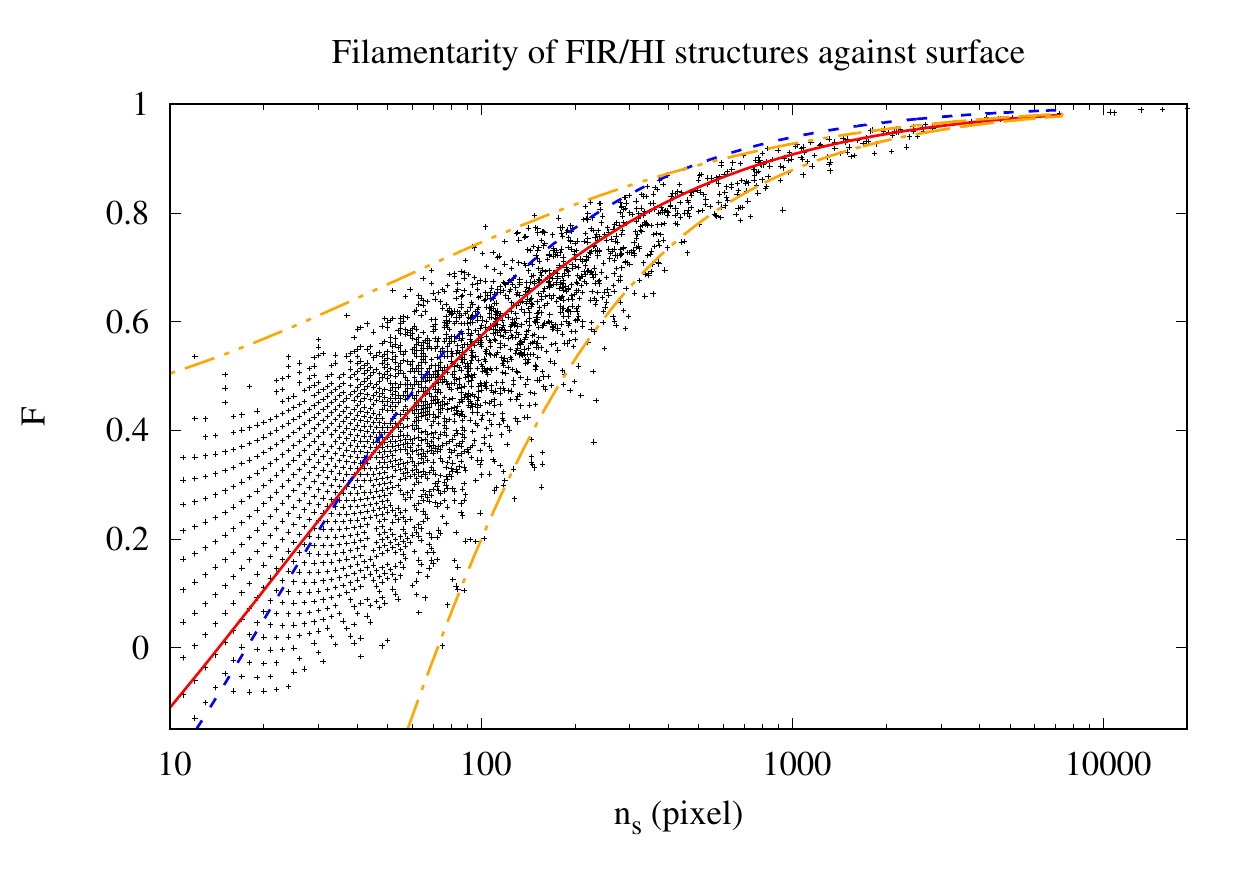}
   \caption{Filamentarity of 6~568 structures as a function of filament
     surface measured by counting filament surface pixels
     $n_{\mathrm{S}}$. The dashed blue line shows a model of the
     filamentarity in the case of a constant width of $W = 2.53 $ pixels, and the
     red line shows the model for the broadening fit from
     Fig. \ref{Fig_Widths}. Formal numerical $3 \sigma$ uncertainties
     from perimeter counts are indicated by the dash-dotted orange lines.  }
   \label{Fig_F_S}
\end{figure}

The distribution of the resulting filamentarities as a function of the
surface pixel count $n_{\mathrm{S}}$ is shown in Fig. \ref{Fig_F_S}
together with two simple filamentarity models. The first model (blue
dashed line) assumes a common width of $W = 2.53 $ pixels. For the
second model (red line), we use the fit $ W = 0.74 + 0.491 \ln(S) $.
This model shows obvious improvements for large surfaces, and we use it
in the following.  We find a rather homogeneous and continuous
distribution that covers three orders of magnitude in surface. The
scatter can be explained by the uncertainties in the parameter
determination. Most significant is the statistical error $ \Delta P =
\sqrt{ f_{\mathrm{b}} (n_{\mathrm{Pout}} + n_{\mathrm{Pin}})/2 } $ in
the perimeter determination. Neglecting independent errors in $W,$ we
estimate $3 \sigma$ uncertainties, which are plotted with dash-dotted
orange lines in Fig. \ref{Fig_F_S}.  Outliers can be traced back to
filaments with large widths in Fig. \ref{Fig_Widths}. We find no
evidence for other systematical biases.

Aspect ratios are usually estimated visually as ${\cal A} = L / W $ from
the measured filament lengths $L$ and widths $W$. Here, we generalize
this concept. The filamentarity ${\cal F}$, which depends on the Minkowski
functionals $P$ and $S$ only, is sensitive to the shape of a structure,
but not to its size; it is further invariant against translation or
rotation and describes the filamentary shape of a class of objects
whenever the $P$ and $S$ values lead to the same value ${\cal F}$
defined by Eq. (\ref{eq:F}). In a similar way, we may consider all
objects with the same $ L / W $ ratio as equivalent. Filamentarity and
aspect ratio can then be linked. First, we generalize the length as $ L =
P/2 $ and obtain
\begin{equation}
{\cal A} = \frac{P}{2 W}.
\label{eq:A}
\end{equation}

\begin{figure}[th] %%  9
   \centering
   \includegraphics[width=9cm]{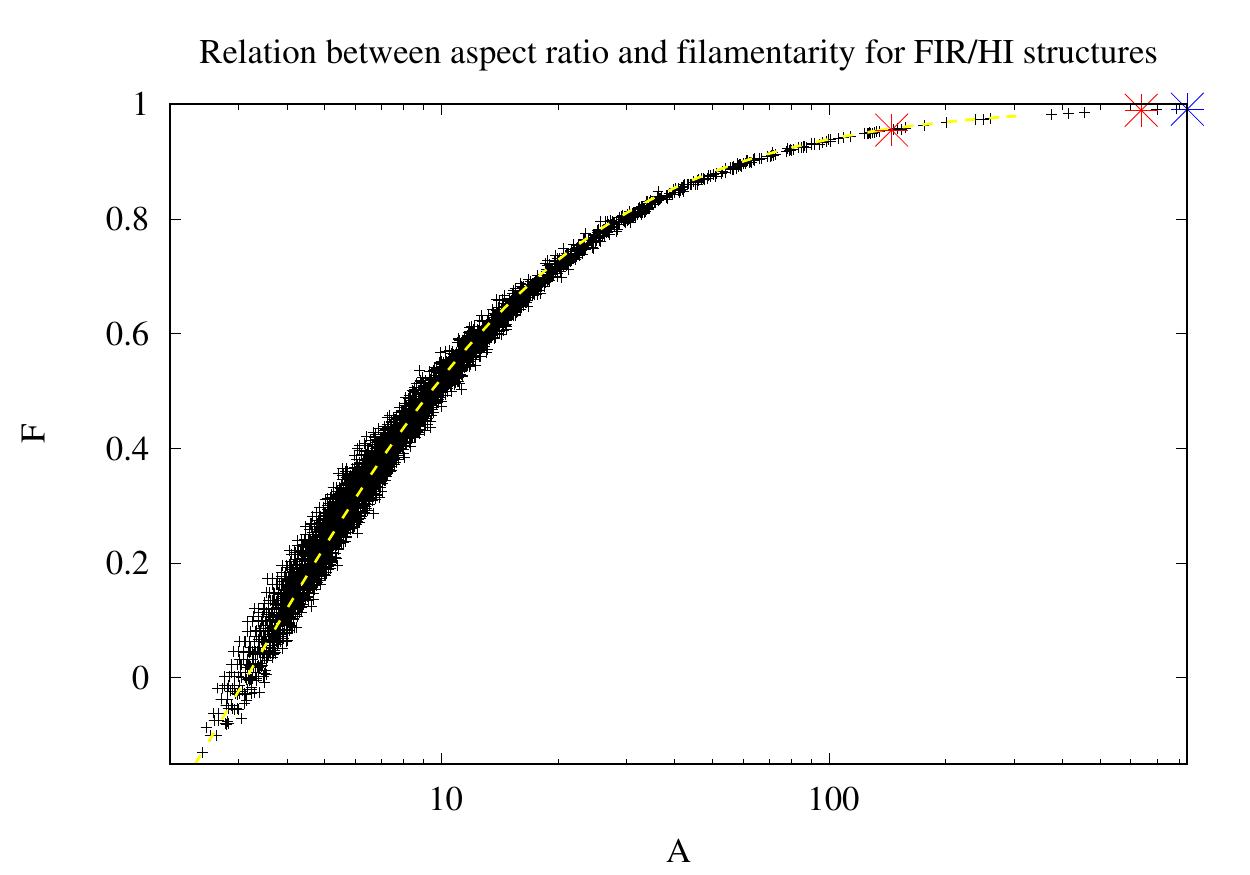}
   \includegraphics[width=9cm]{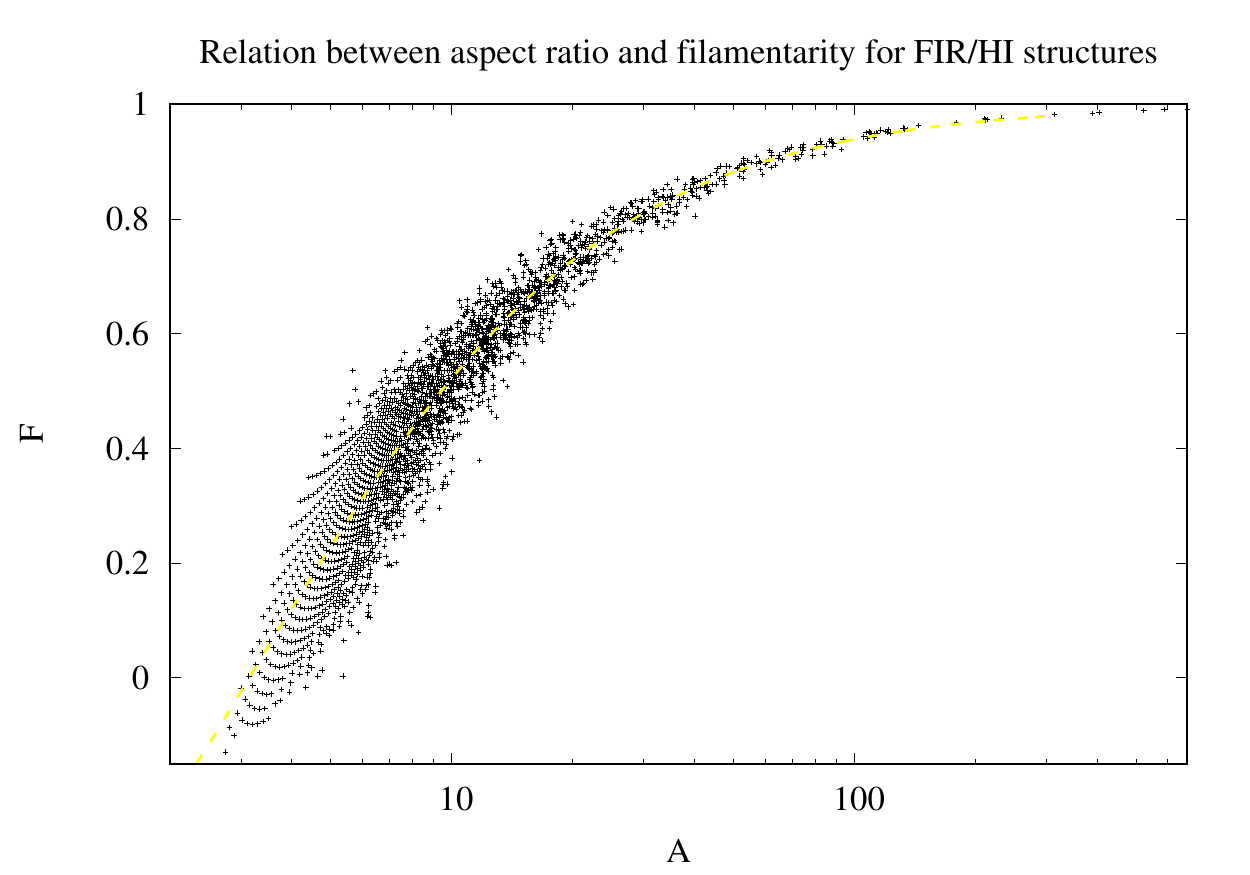}
   \caption{Distribution of filaments covering the parameter space in
     aspect ratio ${\cal A}$ and filamentarity ${\cal F}$. The
     distribution in the top panel was derived using the
     width $W$ as described in Sect. \ref{Simulations}  for each filament. For the distribution in the bottom panel, we use widths as determined from the fit described in
     Sect. \ref{widths}.  The dashed yellow line is derived from
     Eq. (\ref{eq:F_A}) using aspect ratios according to
     Eq. (\ref{eq:A}). Positions marked in red in the upper plot belong
     to the two filaments described in Sect. \ref{network}; merging these
     filaments results in a single structure, marked with blue.}
   \label{Fig_A_F12}
\end{figure}

The top panel of Fig. \ref{Fig_A_F12} shows the relation between ${\cal F}$ and
${\cal A}$ for all of our objects, using independent determinations of
surface $S$, perimeter $P,$ and average width $W$. For the lower plot, we
use the relation $ W = 0.74 + 0.491 \ln(n_{\mathrm{S}}) $ from
Fig. \ref{Fig_Widths} to estimate the width from the surface. This
statistical estimate can be compared with the measurements. In case of
independent measures for $P$ and $W$ (Fig. \ref{Fig_A_F12} top), we
obtain a very tight and clean relation between ${\cal F}$ and ${\cal
  A}$, suggesting a close relation of $ W = 2 S / P $  for the filaments,
and therefore\begin{equation}
{\cal A} = \frac{P^2}{4 S}
\label{eq:A2}
,\end{equation}
and we obtain the ${\cal A}$-to-${\cal F}$ relation,
\begin{equation}
{\cal F} = \frac{1 - \pi / {\cal A}}{1 + \pi / {\cal A}}.
\label{eq:F_A}
\end{equation}

A low filamentarity observed for an isolated structure must be formally considered as an indication that this structure does not represent
a filament. The implication of Fig. \ref{Fig_A_F12} is that low
filamentarity may not in all cases be interpreted this way. Our sample
needs to be considered as a homogeneous class of objects. Short
filaments with low aspect ratios are characterized by a low
filamentarity but these objects do not separate from the sample shown in
Fig. \ref{eq:F_A}.  Objects belonging to
this low end of the filament distribution may suffer from observational
uncertainties, in some cases even leading to negative filamentarities, but are definitively part of the filament population. It is remarkable that the upper plot
in Fig. \ref{eq:F_A} with aspect ratios from Eq. (\ref{eq:A2}) shows less scatter than the plot below with the $W$ fit described in
Sect. \ref{widths}. This means that the individually measured
filament widths $W$ are well defined, and a significant fraction of the
scatter in the top panel of Fig. \ref{Fig_Widths}  must reflect filament properties.

\subsection{Probability density distributions }
\label{Probability}

So far we have considered predominantly the distribution of filamentarities
and aspect ratios for individual structures. Here, we focus also on
the total surface covered by filaments, which is characterized by $\Sigma
n_{\mathrm{S}}$, the sum over all filaments in a given range.

\begin{figure}[th] %%  10
   \centering
   \includegraphics[width=9cm]{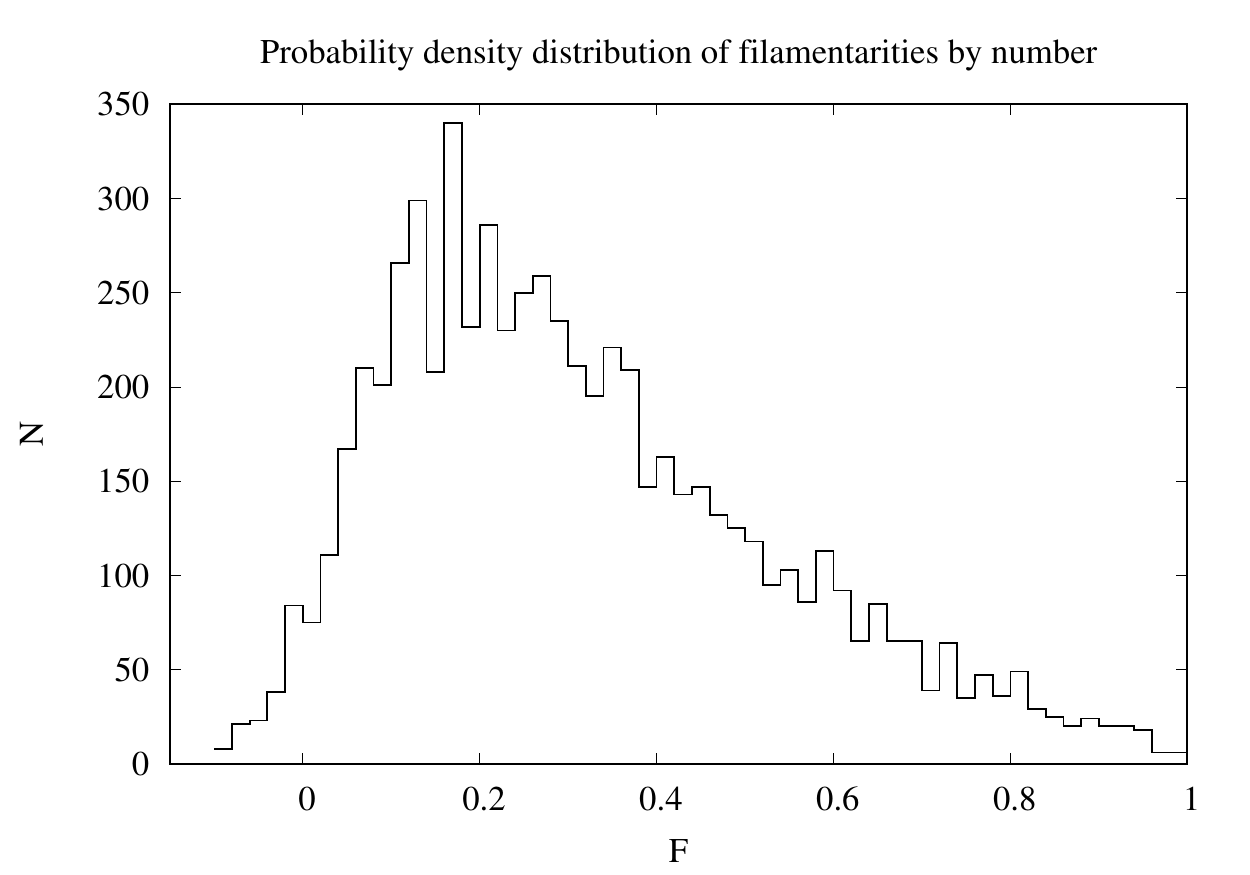}
   \includegraphics[width=9cm]{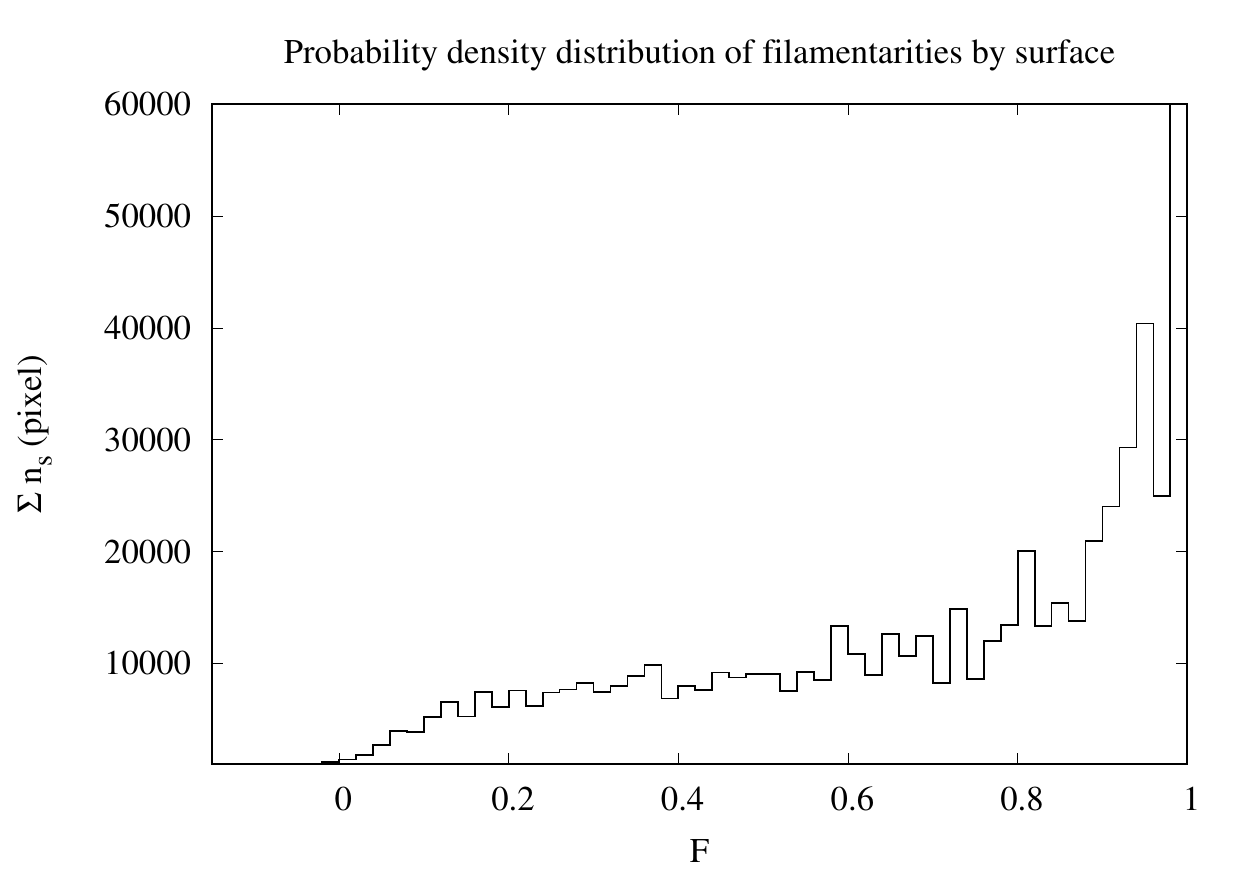}
   \caption{Probability density distributions of filamentarities
     ${\cal F}$. Top: Distribution by number. Bottom: Distribution
     weighted by the total surface covered by  filaments. }
   \label{Fig_F_hist}
\end{figure}

\subsubsection{Filamentarity}

Figure \ref{Fig_F_hist} shows that the PDF of filamentarity by number is
fundamentally different from that obtained from the distribution
obtained by weighting with the total surface of the filaments. The upper part
of Fig. \ref{Fig_F_hist} may be compared with the PDF derived by
\citet[][Fig. 7]{Makarenko2015}. These authors considered isocontours of
the fluctuations in the gas number density obtained from the
\hi\ distribution at a constant Galactocentric radius of 16 kpc,
obtaining a peak of the PDF at ${\cal F} \sim 0.15$ with a power-law tail
at larger values with a truncation at ${\cal F} \sim 0.75 $. Their
distribution resembles that of triaxial ellipsoids and
\citet{Makarenko2015} argue that the form of their PDF indicates that
the \hi\ distribution is indeed filamentary in 3D but truncated at an
aspect ratio of length/thickness of about 20. In an extensive analysis,
\citet{Soler2020} and \citet{Soler2022} studied the filamentary
structure in atomic hydrogen emission toward the Galactic plane and
confirmed that a significant fraction of the \hi\ distribution is in
filaments; however, these authors did not determine aspect ratios or
filamentarity.
 
We do not observe a truncation at high ${\cal F}$ values, and the
weighted PDF of filament surfaces, displayed in the bottom panel of
Fig. \ref{Fig_F_hist}, shows rising surfaces for increasing ${\cal
  F}$. Filaments with values ${\cal F} \ga 0.9 $   contribute significantly to the total surface covered by the filamentary
\hi\ and FIR distribution. We note that filaments with surfaces of $ n_{\mathrm{S}} >
20~000$ pixels are excluded and do not contribute to Fig. \ref{Fig_F_hist}.

\subsubsection{Aspect ratio}

\begin{figure}[th] %%  11
   \centering
   \includegraphics[width=9cm]{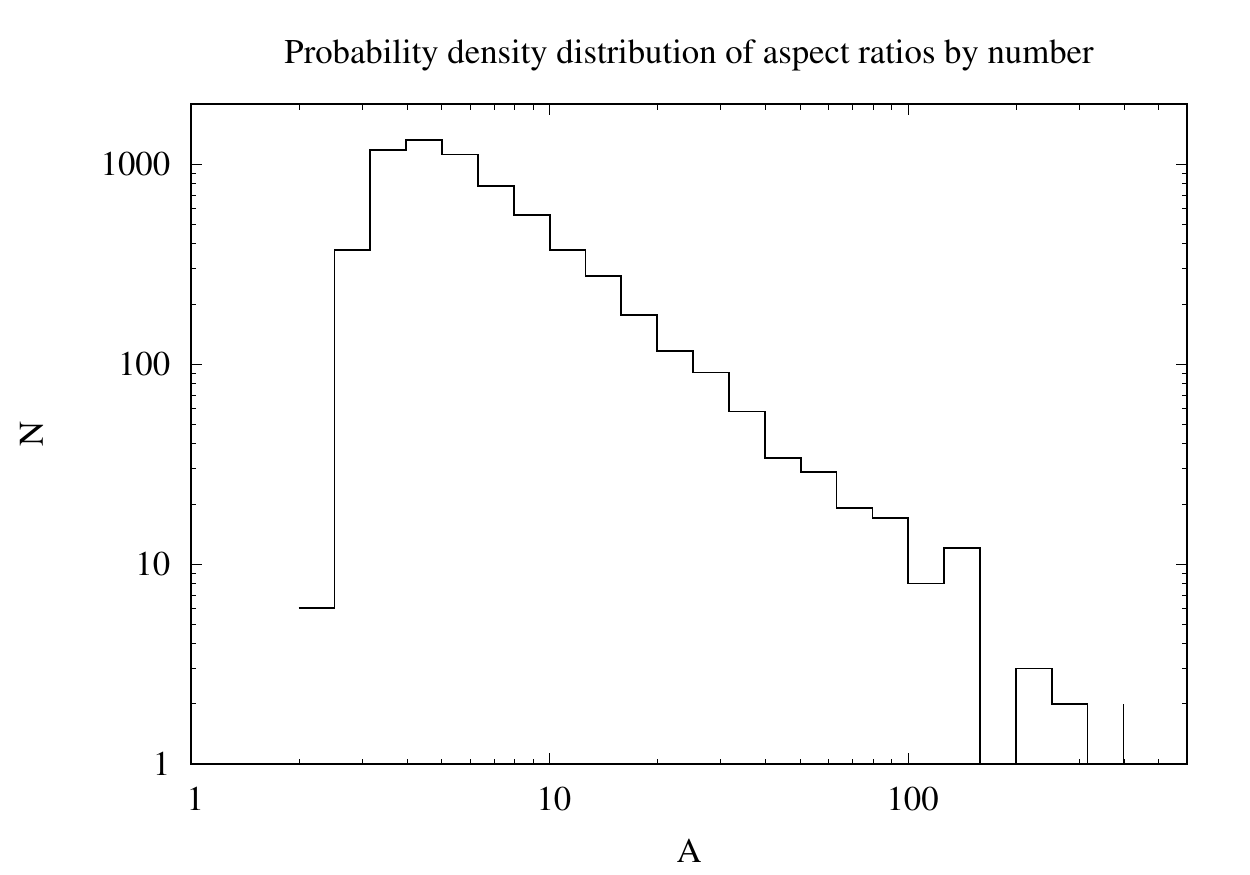}
   \includegraphics[width=9cm]{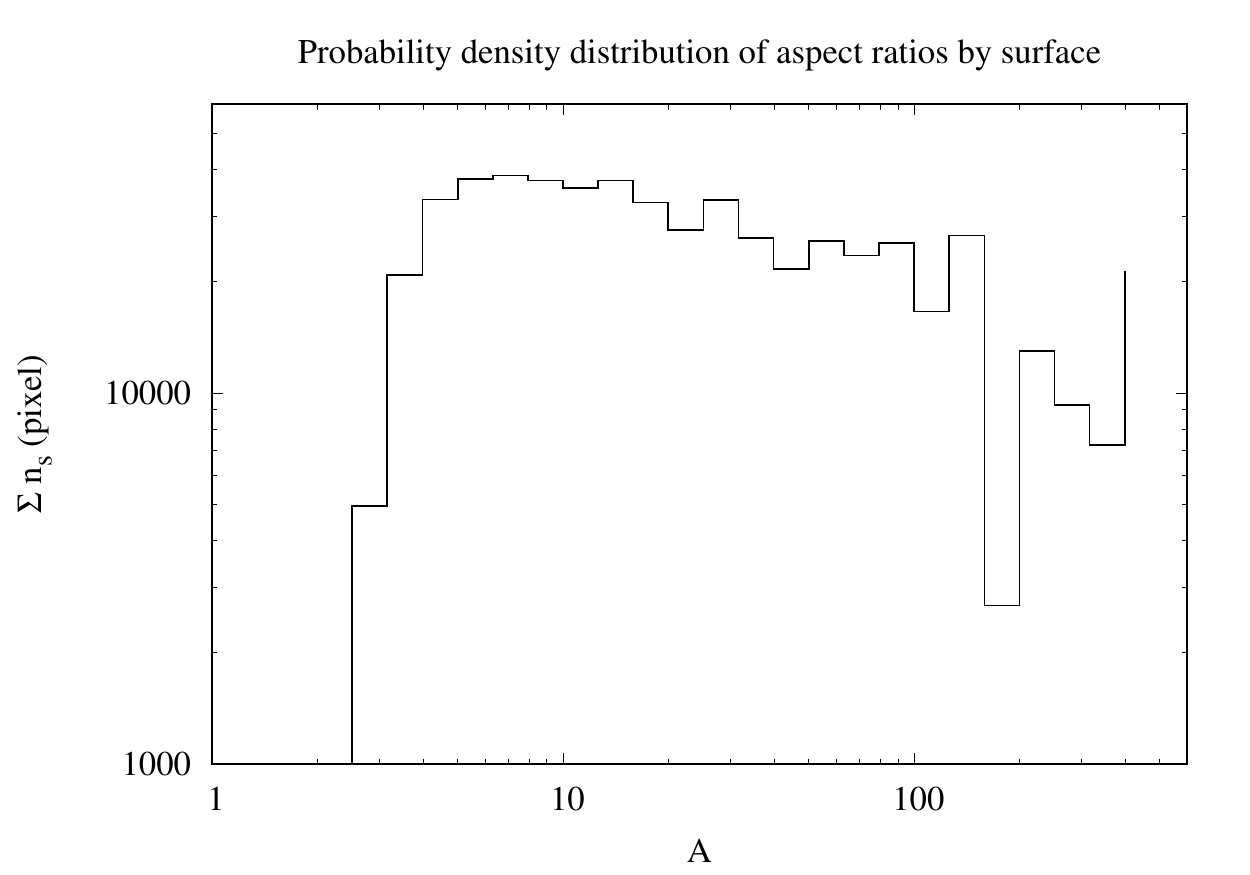}
   \caption{Probability density distributions of aspect ratios ${\cal
       A}$. Top: Distribution by number. Bottom: Distribution weighted
     by the total surface covered by  filaments.  }
   \label{Fig_A_hist}
\end{figure}

Figure \ref{Fig_A_hist} presents the PDFs for the
aspect ratios  for comparison. Most of the analyzed filamentary
structures have low aspect ratios. The distribution is continuous, at
least up to ${\cal A} \sim 125$. The PDF of filament surfaces indicates
that even aspect ratios of $200 \la {\cal A} \la 600$  exist, but these
contribute less to the PDF at the bottom of Fig. \ref{Fig_A_hist}, which is
weighted by surface. The exclusion of filaments with surfaces of $ n_{\mathrm{S}} >
20~000$ pixels implies that large aspect ratios are under-represented
because of confusion limitations.
Our results deviate significantly from those of \citet{Makarenko2015},
who considered \hi\ cloud complexes in the Galactic plane at a constant
Galactocentric radius of 16 kpc and found only moderate aspect ratios.

\subsection{Velocity coherence revisited}
\label{velocity}

One of the most striking results from the analysis in
\citetalias{Kalberla2021} is that FIR filaments at 857 GHz are
associated with \hi\ at well-defined radial velocities. Paying attention
to the constraints discussed in Sect. \ref{Hessian}, it is possible to
assign radial velocities to the FIR filaments.  The average velocity
dispersion along the filaments is defined in Sect. 2.6 of
\citetalias{Kalberla2021} from a two-point function within a radius of
1\degr. Velocity fluctuations of $\Delta v_{\mathrm{LSR}} \sim 5.5 $
\kms\ along the filaments were found to be representative for turbulent
motions within the ISM. Some strong velocity deviations ---exceeding the
expected three-sigma limit for a normal distribution--- are found from
visual inspection but such deviations appear not to disrupt the
filamentary structure. As opposed to
\citetalias{Kalberla2021}, we wish to consider here individual filaments. Having extracted these, we intend
to check whether the velocity coherence is constrained by the
filamentarity of the structures.  First, we
determine the average radial velocity  of each filament, followed by the
standard deviation of the velocity fluctuations along the filament.

\begin{figure}[th] %%  12
   \centering
   \includegraphics[width=9cm]{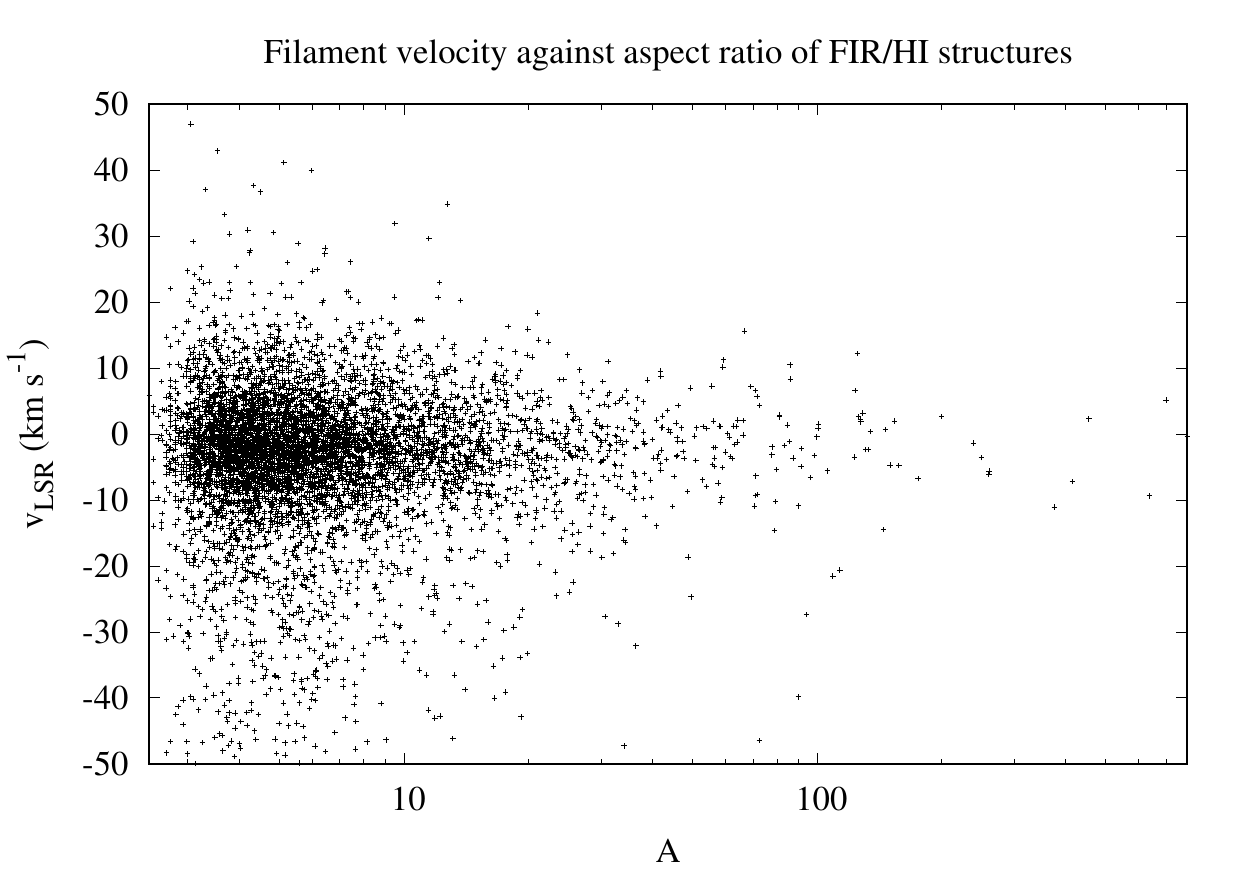}
   \includegraphics[width=9cm]{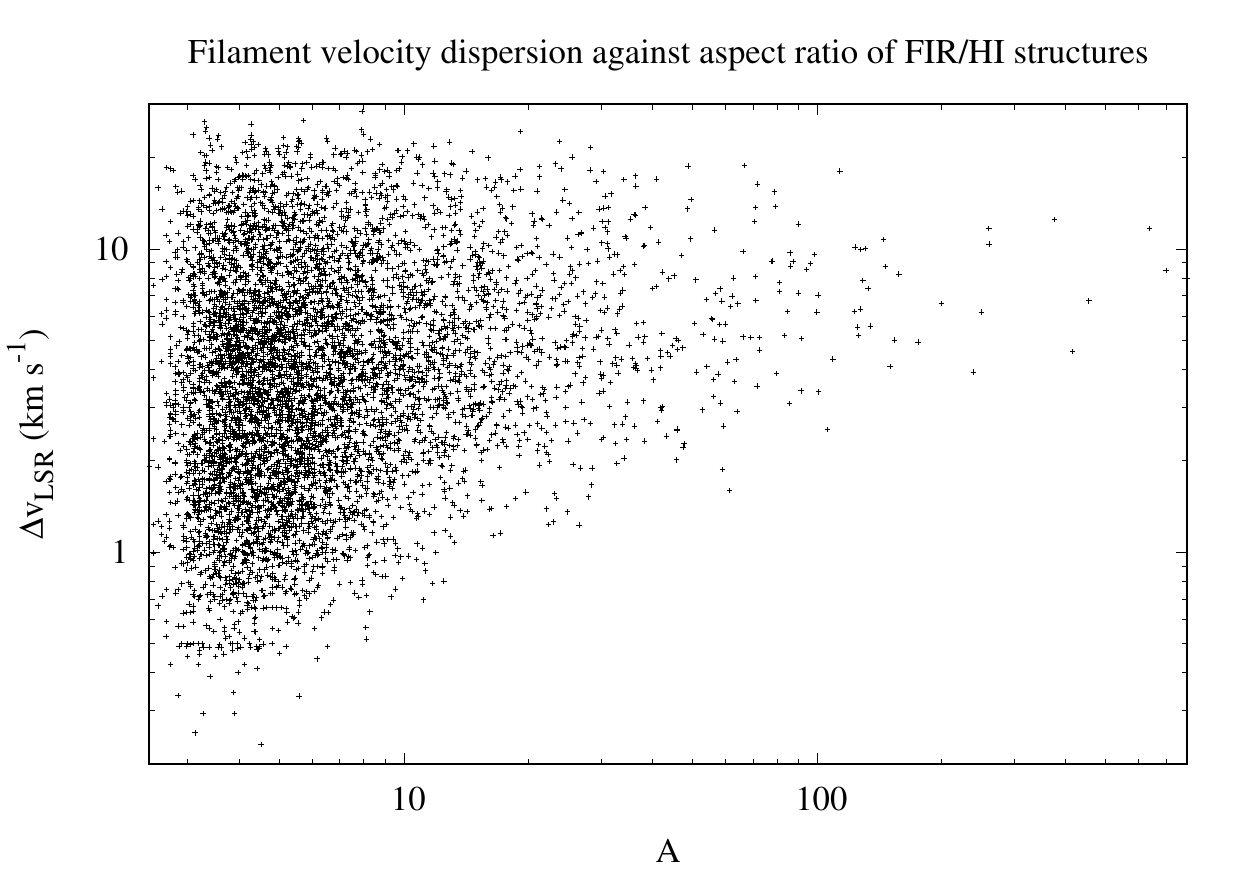}
   \caption{Velocity distribution as a function of aspect ratio.  Top:
     Average filament \hi\ radial velocities. Bottom: Velocity
     dispersions within the filaments. }
   \label{Fig_Vel}
\end{figure}

In the top panel of Fig. \ref{Fig_Vel}, we plot the average radial velocities of the
filaments and below the corresponding velocity dispersions as a function
of aspect ratio. The largest scatter in radial velocity and
associated velocity dispersion exists for ${\cal A} \la 20 $.  A
fraction of the filaments at velocities $ v_{\mathrm{LSR}} \la -20 $
\kms\ can be attributed to intermediate-velocity clouds (IVCs) with low
velocity dispersions.  The radial velocities of the filaments are mostly close to zero for
large aspect ratios, implying that these filaments
must be predominantly local. The average velocity dispersion for our
sample of filaments is $\Delta v_{\mathrm{LSR}} = 5.24 $ \kms.  While
this dispersion may be considered as representative for internal
turbulent motions within the filaments, a slight increase for large
filaments can be caused by velocity gradients along the
filaments. Larger surfaces imply larger aspect ratios. The dispersions
in Fig. \ref{Fig_Vel} appear to increase slightly with aspect ratio, but
this impression is biased by low dispersions from IVCs at low aspect
ratios; the correlation coefficient is 0.08. In a similar way, the
correlation coefficient between velocity dispersion and surface is
0.06. There is no clear evidence that large filaments must also have
large velocity dispersions.

\begin{figure}[th] %%  13
   \centering
   \includegraphics[width=9cm]{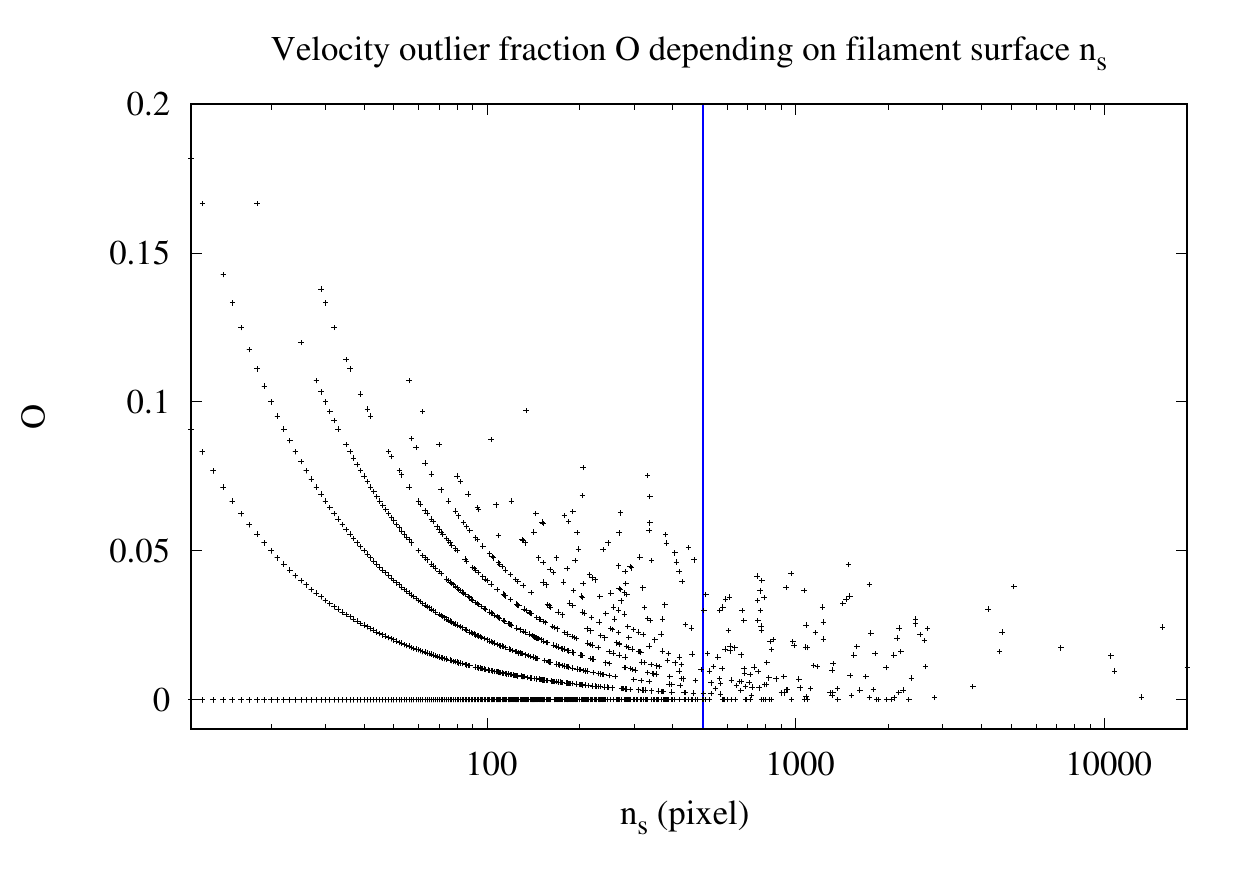}
   \caption{Fraction $O$ of outlier positions depending on filament
     surface $n_{\mathrm{S}}$. Outliers are defined as positions with
     velocities deviating from the average filament velocity by more
     than three times the velocity dispersion.  The presentation is
     strongly quantized, showing, from bottom to top, outlier counts of
     zero, one, two, and so on. The blue vertical line divides the sample at
     $n_{\mathrm{S}}= 500$. }
   \label{Fig_Outlier}
\end{figure}

\begin{figure}[th] %%  14
   \centering
   \includegraphics[width=9cm]{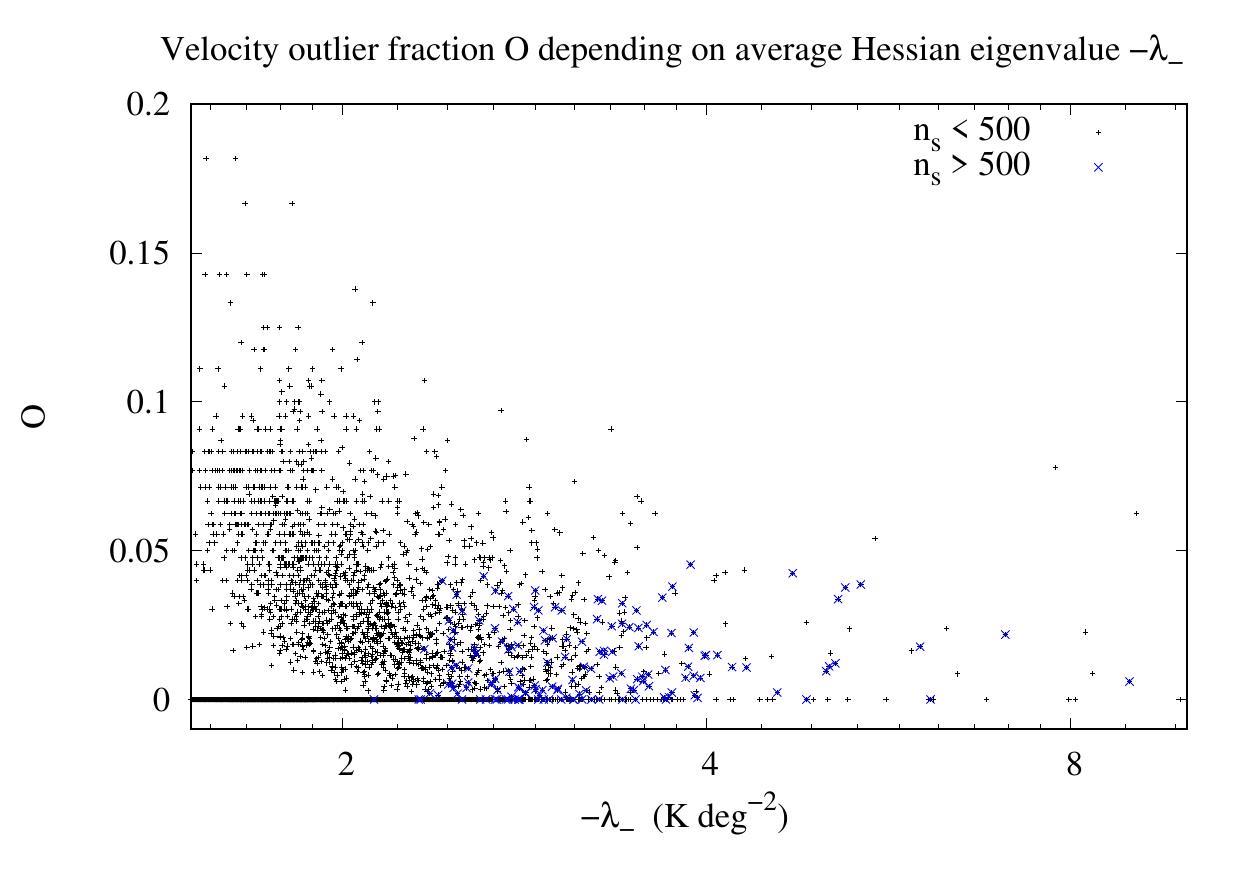}
   \caption{Fraction $O$ of outlier positions as  a function of the average
     Hessian eigenvalue $-\lambda_-$ from the FIR data. Blue data points
     are from prominent filaments with $n_{\mathrm{S}} > 500$.  }
   \label{Fig_outlier}
\end{figure}

Outliers, which are defined at positions with velocities that deviate from the
average filament velocity by more than three times the velocity
dispersion, may be caused by statistical uncertainties, by turbulence-driven deviations, or alternatively by systematic effects, indicating
blending from unrelated structures. We define the outlier fraction $O$
as the fraction of positions with outliers relative to the number of
positions of the total filament surface. Figure \ref{Fig_Outlier} shows
this outlier fraction as a function of filament surface $ n_{\mathrm{S}}$.
The presentation is dominated by integer arithmetic, from bottom up with
outlier counts of zero, one, two, and so on. There is no evidence that
filaments with large surfaces $n_{\mathrm{S}}$ suffer from a large
outlier fraction. The blue vertical line in Fig. \ref{Fig_Outlier}
divides the filament distribution at $n_{\mathrm{S}}= 500$ in two
samples containing approximately the same total column density. We use this threshold
in the following analysis to distinguish filaments that, visually,
appear to be prominent (for $n_{\mathrm{S}} \ga 500$) from those appearing less prominent
($n_{\mathrm{S}} \la 500$). Prominent structures appear to be less  affected
by outliers. Figure \ref{Fig_outlier} shows that the outlier
fraction is largest at positions that are less significant and have low
average Hessian eigenvalues $-\lambda_-$. Filaments were defined in
\citetalias{Kalberla2021} according to FIR distribution, and the radial velocities were then determined by matching \hi\ structures. We conclude that most of the velocity
outliers are probably caused by sensitivity limitations.  Visual
inspection shows many cases of outliers along continuous structures on
small scales. At high Galactic latitudes, we find no direct evidence that
a significant number of outliers is caused by confusion with unrelated
structures. Filament positions are derived from the Hessian analysis of
the FIR at 857 GHz. We use the \hi\ data to determine the velocity field
but we do not flag FIR filament positions as invalid in the case of outliers.

\subsection{A network of filaments }
\label{network}

\begin{figure}[th] %%  15
   \centering
   \includegraphics[width=8cm]{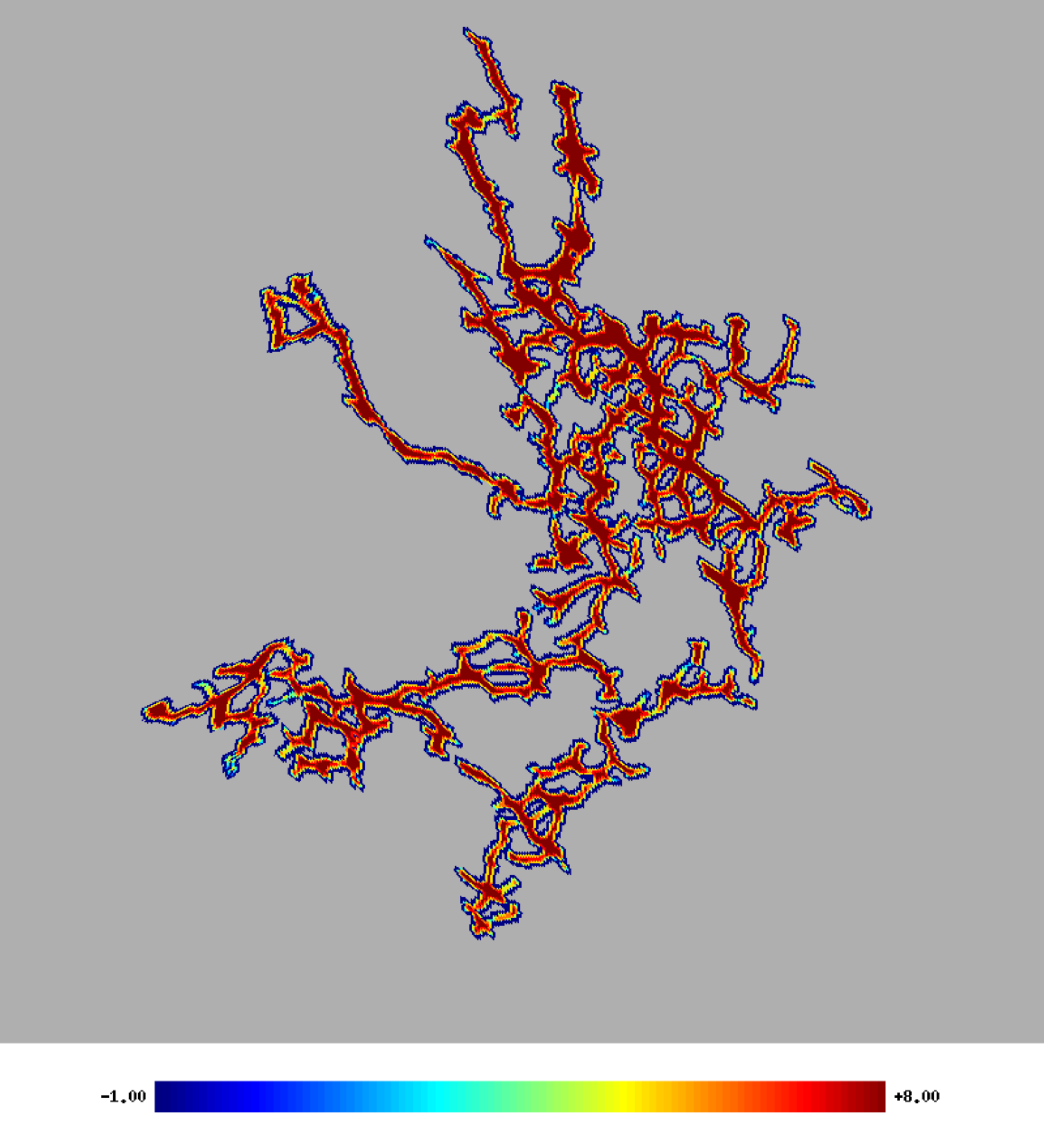}
   \includegraphics[width=8cm]{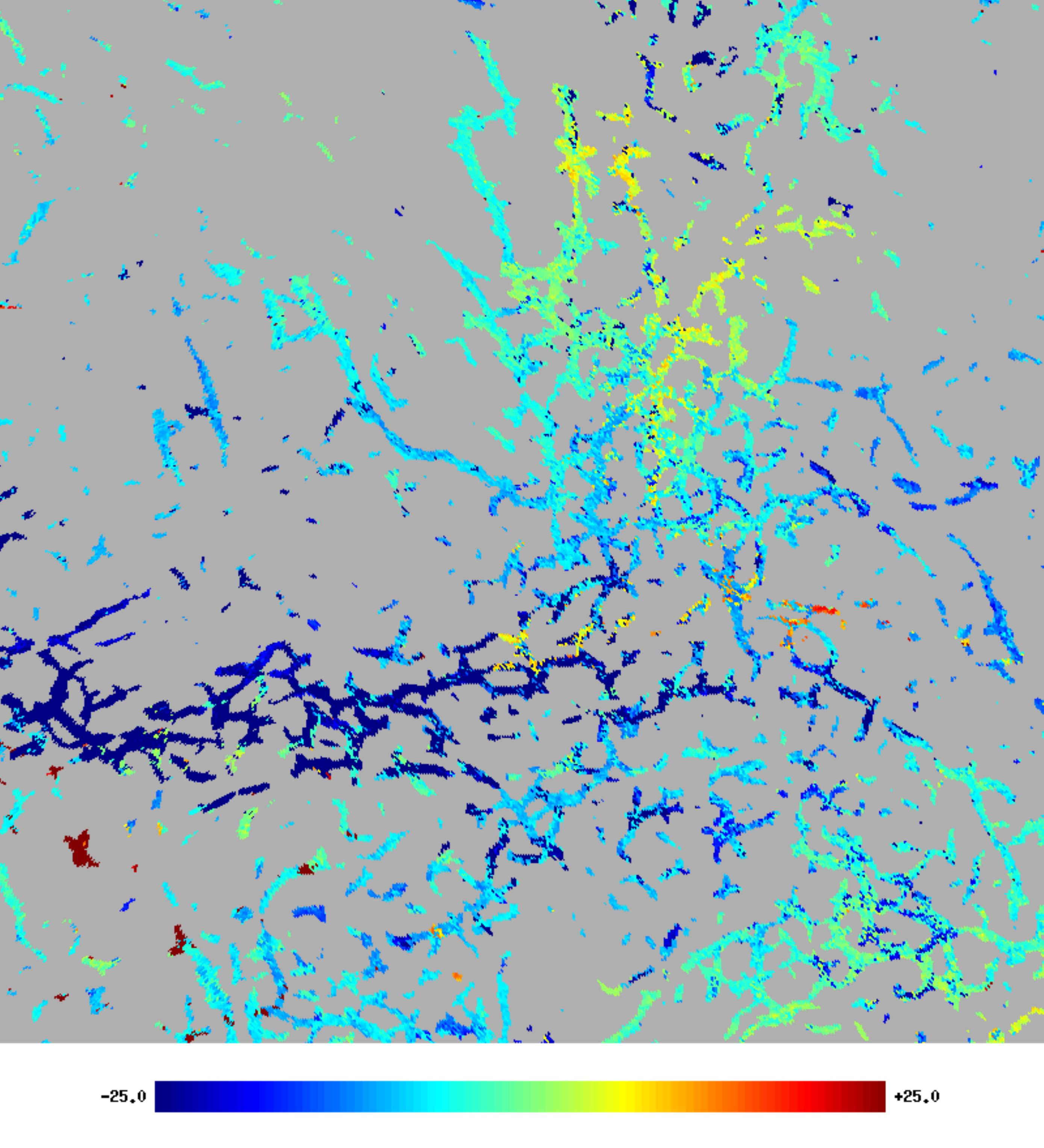}
   \caption{Spatial distribution of filamentary structures in gnomonic
     projection. The field size is 27\fdg7. Top: Isolated filamentary
     structures discussed in the text.  For each
     position, the color coding shows weights according to the number of neighbors
     $n_{\mathrm{Neighbor}}$ inside the filament. Boundary pixels
     $n_{\mathrm{Pout}}$ just outside the filaments have weight -1 and
     are coded in blue. Bottom: Velocity field of the filamentary
     structures, including all other filaments in the field. The center
     position is $ l = 260\deg$, $b = 58\deg $. }
   \label{Fig_fil_1}
\end{figure}

Filaments do not only exist as distinct, thin elongated structures; they
appear also as part of an extended network. To enable the presentation
of individual filamentary structures, we extracted two such filamentary
networks that contain elongated features, which are shown in the top
panel of Fig. \ref{Fig_fil_1}. The color coding shows the number of
neighbors $n_{\mathrm{Neighbor}}$ that are used to characterize the
thickness of the filament. Boundary pixels $n_{\mathrm{Pout}}$ just
outside the filaments are shown in blue. The display appears to show a
single structure but a closer look at the lower structure in latitude
shows that this filament is disconnected. Two of the boundary pixels
touch each other, giving the impression of a connection.

The northern filament covers a surface of 13~198 pixels, resulting in a
width of $W = 4.4 $ in ${\cal A} = 637$, ${\cal F} =0.9897$. Velocities are
centered on $v_{\mathrm{LSR}} = -9.4 $ \kms\ with a dispersion of
$\Delta v_{\mathrm{LSR}} = 11.7 $ \kms. For the southern part, we measure $n_{\mathrm{S}} = 1~967$, $W = 3.6 $, ${\cal A} = 145$, ${\cal F} =
0.9553$, $v_{\mathrm{LSR}} = -14.4 $ \kms, and $\Delta v_{\mathrm{LSR}}
= 10.7 $ \kms, 
respectively.  The
filament velocity field is shown in the lower part of Fig. \ref{Fig_fil_1}, including other structures in the
vicinity. Some of the smaller fragments appear to be aligned with larger
filaments, suggesting a common flow. The color code allows us to estimate
velocity gradients; in most cases these show smooth transitions but there are  rapid changes in a
few cases.

\begin{figure}[th] %%  16
   \centering \includegraphics[width=8cm]{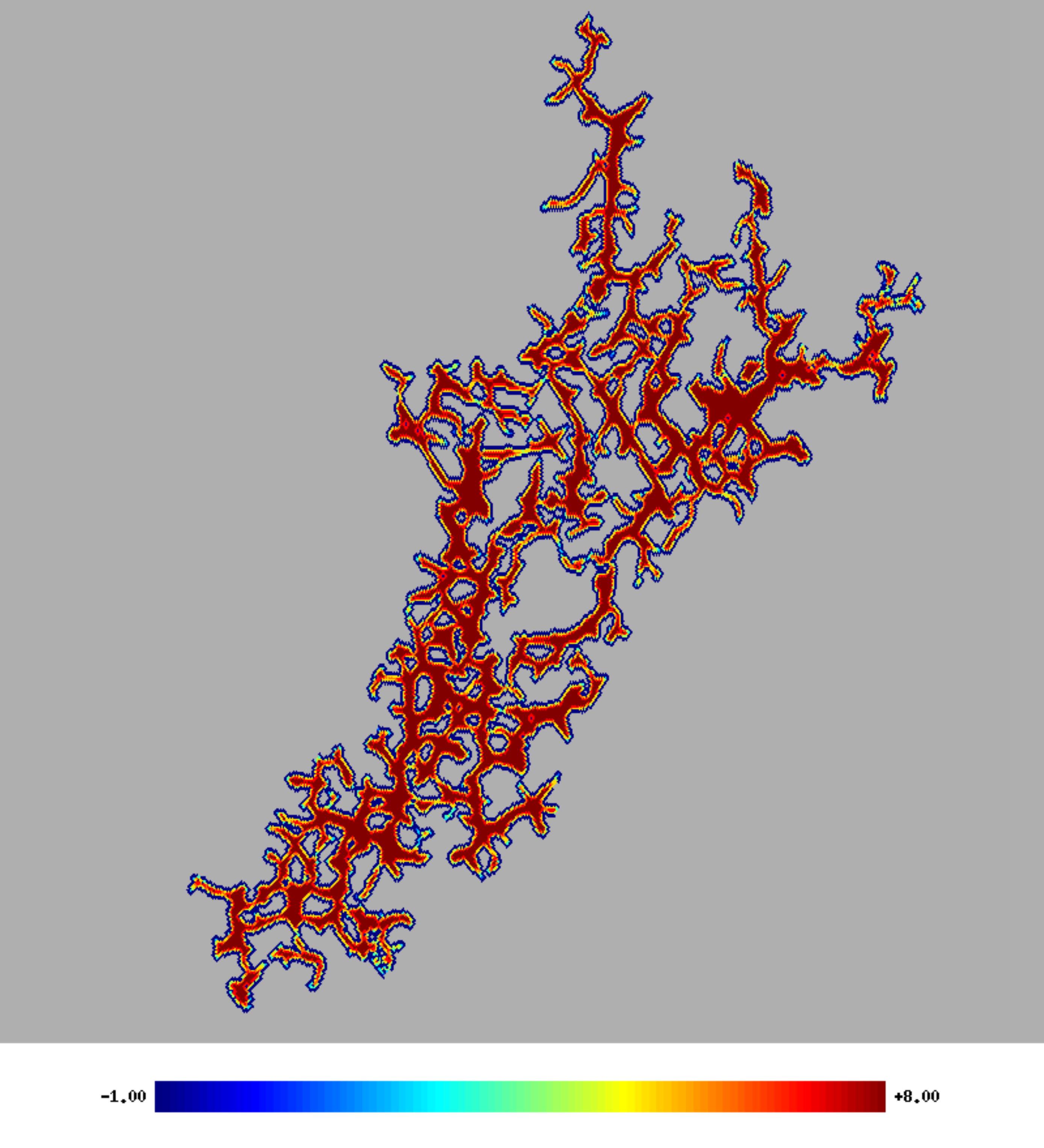}
   \includegraphics[width=8cm]{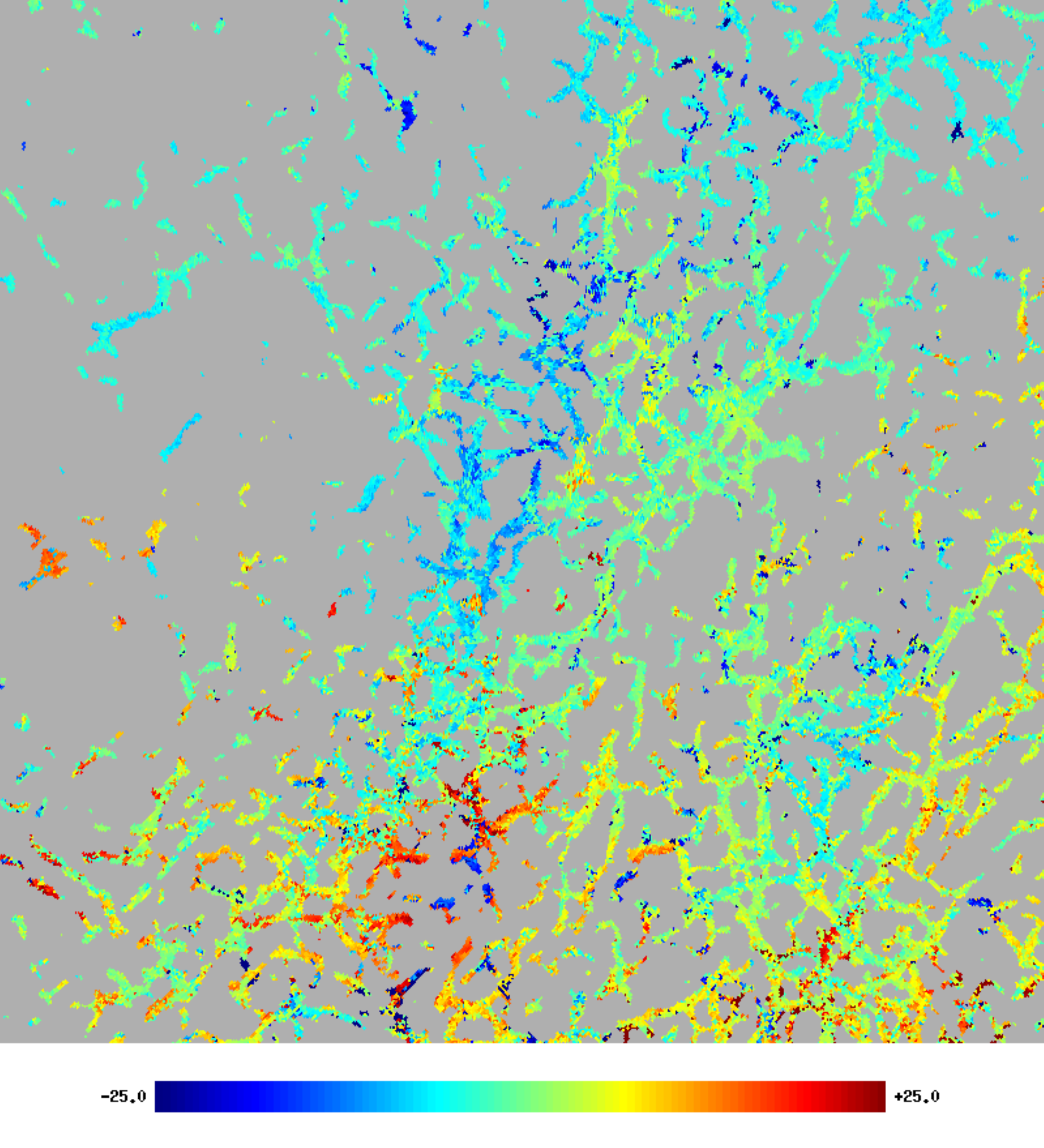}
   \caption{As in Fig. \ref{Fig_fil_1} but centered at $ l = 45\deg$, $b =
     42\deg $.  }
\label{Fig_fil_2}
\end{figure}

Surprisingly, concerning the characterization of filaments as being
located on the ${\cal A}$-to-${\cal F}$ track (Fig. \ref{Fig_A_F12} and
Eq. (\ref{eq:F_A})), whether or not these two filaments are connected is
irrelevant. The filaments are part of a hierarchy of filaments, defined
by ${\cal A}$-to-${\cal F}$ relation.   Coalescence or breaking of
filamentary structures leads to a repositioning of the filament
positions along the ${\cal A}$-to-${\cal F}$ track shown in
Fig. \ref{Fig_A_F12}. This can be demonstrated in the case of the
filaments shown in Fig. \ref{Fig_fil_1}. We indicate the positions of
the two observed filaments with red markers in the ${\cal A}$-to-${\cal
  F}$ plot in the top panel of Fig. \ref{Fig_A_F12}.  Merging both
filaments replaces these entries with a new one on the ${\cal
  A}$-to-${\cal F}$ track indicated in blue. In general, merger or
disruption of filaments leads to a repositioning of structures on the
path defined by Eq. (\ref{eq:F_A}) but these structures do not leave the
${\cal A}$-to-${\cal F}$ track. Thus, the distribution is
self-replicating. This property mitigates the observational limitations
to some extent. Sensitivity limitations imply that we are limited in our
analysis to the strongest features; there are likely to be many more,
weaker features. As long as observational gaps due to sensitivity
limitations are not too large, we expect the ${\cal A}$-to-${\cal F}$
distribution (Fig. \ref{Fig_A_F12}, top) to be representative.

As pointed out by \citet{Mecke1994}, Minkowski functionals are global
and additive measures (see their Eq. (12)).  Additivity allows us to
calculate these measures by summing up local contributions, as in our
example.  We also note that filaments that are affected by projection
effects remain on the ${\cal A}$-to-${\cal F}$ track as long as the
measured filament width $W$ remains unaffected.  The aspect ratio is
largest for filaments that are oriented perpendicular to the line of
sight. Assuming for example a simple linear structure, a rod. Tilting
such a filament by an angle $\theta_{\mathrm{POS}}$ against the plane of
the sky reduces $S$ and $P$ by the same factor of
$\cos{(\theta_{\mathrm{POS}})}$; this also modifies the aspect ratio but also affects ${\cal F}$ ---according to Eq. (\ref{eq:F_A})---such that the filament
stays on the ${\cal A}$-to-${\cal F}$ track. The additivity property of
Minkowski functionals allows us to generalize this simple example in the case
of complex structures by considering a piece-wise decomposition in local
structures with subsequent addition. However, an observational difficulty is
that,  with increasing misalignment, it becomes increasingly
difficult to accurately determine surface, perimeter, and width. We
relate some of the scatter for low aspect ratios in
Fig. \ref{Fig_A_F12} to observational uncertainties of this kind and note
that, during our simulations (Sect. \ref{Simulations}), inaccuracies in
the width determination were found to be most critical at low aspect
ratios. As a counter example, objects with large planarity (pancakes)
may appear in projection as filaments. For an isotropic distribution,
members of this class should also appear as circular structures in
the plane of the sky with ${\cal F} \sim 0 $. This is not observed within the
uncertainties shown in Fig. \ref{Fig_F_S}.

\begin{figure}[th] %%  17
   \centering
   \includegraphics[width=8cm]{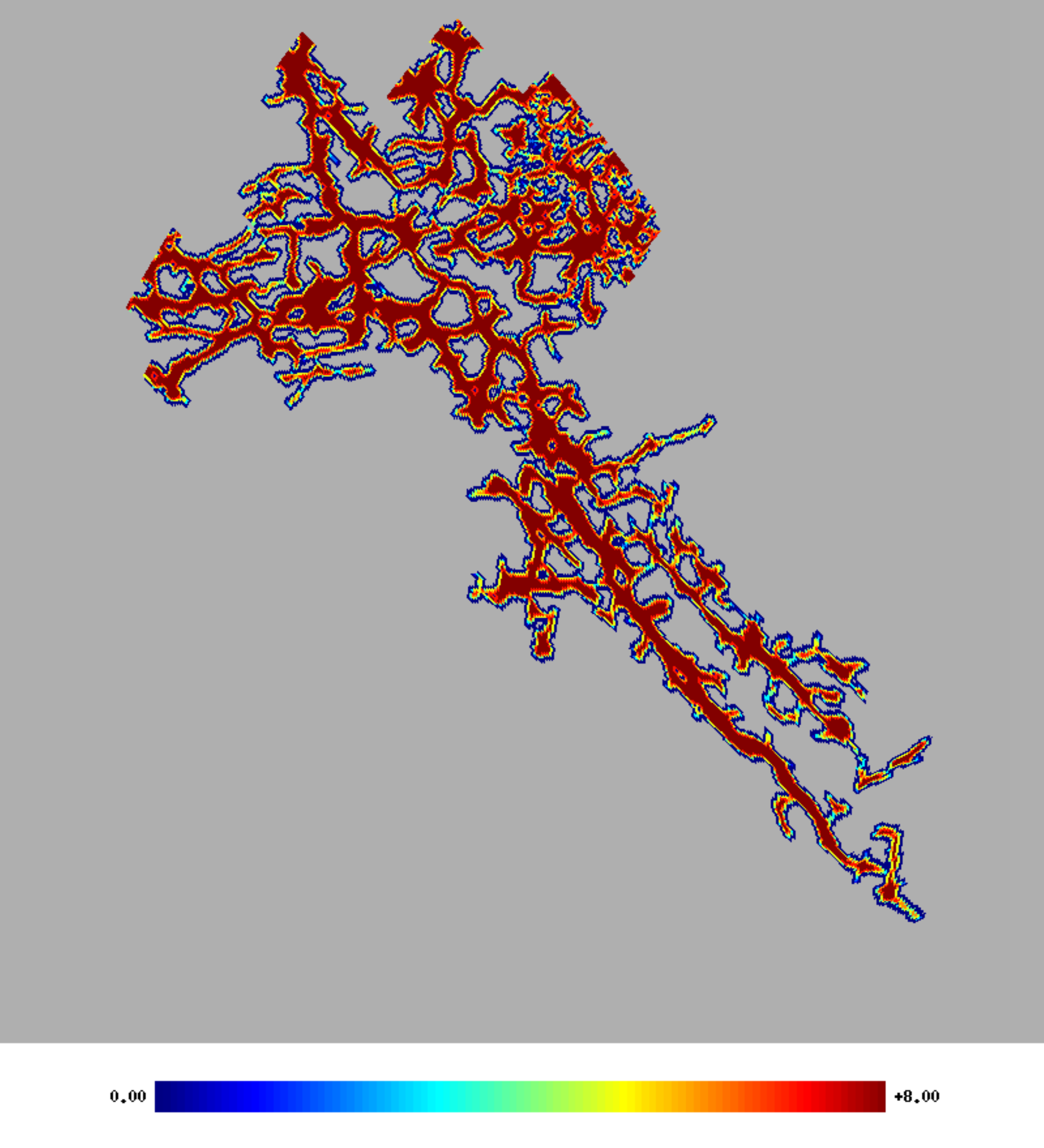}
   \includegraphics[width=8cm]{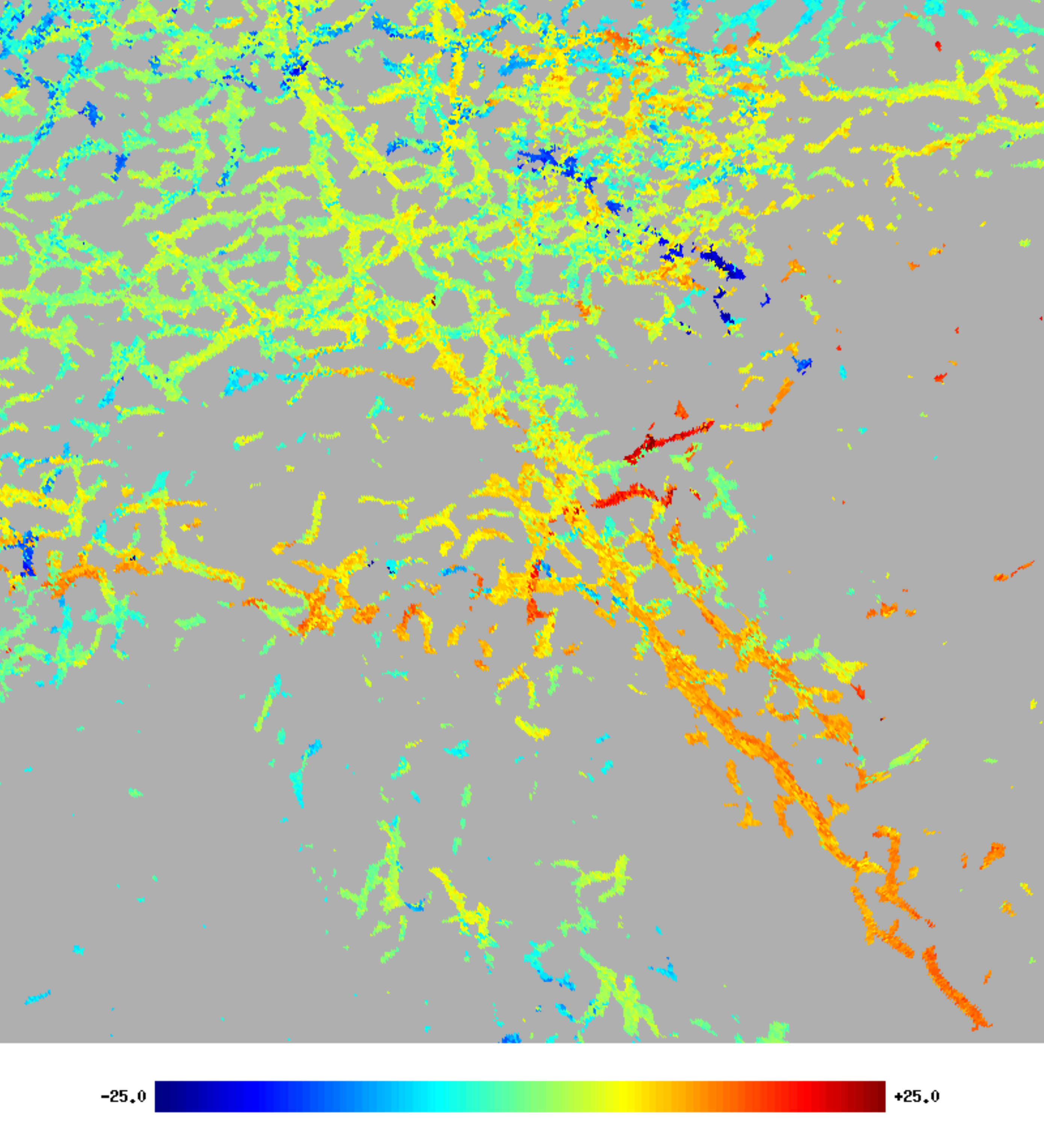}
   \caption{As in Fig. \ref{Fig_fil_1} but centered at $ l = 283\deg$, $b =
     -43\deg $.   }
   \label{Fig_fil_3}
\end{figure}

Figure \ref{Fig_fil_2} shows another prominent and well-defined
filamentary network at high Galactic latitudes. We determine in this
case $ n_{\mathrm{S}} = 18~563$, $W = 4.7 $, ${\cal A} = 787$, ${\cal F} = 0.9917,$
and $v_{\mathrm{LSR}} = -0.4 $ \kms\ with $\Delta v_{\mathrm{LSR}} = 8 $
\kms. In comparison to Fig. \ref{Fig_fil_1}, we find less pronounced
linear structures but more branches and loops.

We close this subsection with an example of a complex, confusion-limited
structure that we excluded from our analysis. This structure,
shown in Fig. \ref{Fig_fil_3}, is the famous dusty
filament\footnote[6]{\url{
 https://www.esa.int/ESA_Multimedia/Images/2015/09/The_Magellanic_Clouds_and_an_interstellar_filament}},
aligned with the galaxy’s magnetic field in the foreground of the
Magellanic clouds \citep[][Figs. 4 and 18]{Planck2016}. Albeit a prime example of a filament, it cannot be
separated from the complex
  northern part of the \hi\ distribution, which appears to be strongly
blended. During data processing, this structure was truncated
automatically after reaching a limit of $n_{\mathrm{S}} > 20~000$ (see
Sect. \ref{plane}).  In this case, we determine formal parameters of
$W = 5.2 $, ${\cal A} = 706$, ${\cal F} = 0.9906$, and $v_{\mathrm{LSR}}
= 4.7 $ \kms\ with $\Delta v_{\mathrm{LSR}} = 5.7 $ \kms. These
parameters are consistent with those of other filaments but we
emphasize that this outstanding source did not enter our statistics.

\subsection{The dark connection }
\label{dark}

Associated with the FIR distribution at 857 GHz are filamentary
\hi\ structures that are considerably colder than the surrounding ISM
\citepalias{Kalberla2021}. Adopting the hypothesis of a constant
$E(B-V)/N_{\mathrm{H}}$ ratio,  \citet{Kalberla2020}
suggested that these structures must be associated with CO-dark molecular
hydrogen. We use column densities in $N_{\mathrm{H}}$ and
$N_{\mathrm{H2}}$ determined by these authors.  To test whether there
might be an enhancement of the total hydrogen column densities
$N_{\mathrm{H}}$ within the filaments, we first calculated the pixel-based
average hydrogen column density ${N_{\mathrm{Hoff}}}$ for all positions
that are not covered by FIR filaments. To take spatial fluctuations of
${N_{\mathrm{Hoff}}}$ into account, we generated an average distribution
$\overline{N_{\mathrm{Hoff}}(l,b)}$ on an nside=128 HEALPix grid,
implying a spatial averaging over a pixel size of 27\farcm5. We then
determined the average column density $\overline{N_{\mathrm{Hfil}}}$ for
each filament and the average
$\overline{N_{\mathrm{Hcenter}}}$ for the center positions with
$n_{\mathrm{Neighbor}} = 8$, separately. The ratios $\overline{N_{\mathrm{Hfil}}} /
\overline{N_{\mathrm{Hoff}}}$ and $\overline{N_{\mathrm{Hcenter}}} /
\overline{N_{\mathrm{Hoff}}}$ can be used to characterize an excess in
hydrogen column density.

\begin{figure}[th] %%  18
  \centering
  \includegraphics[width=9cm]{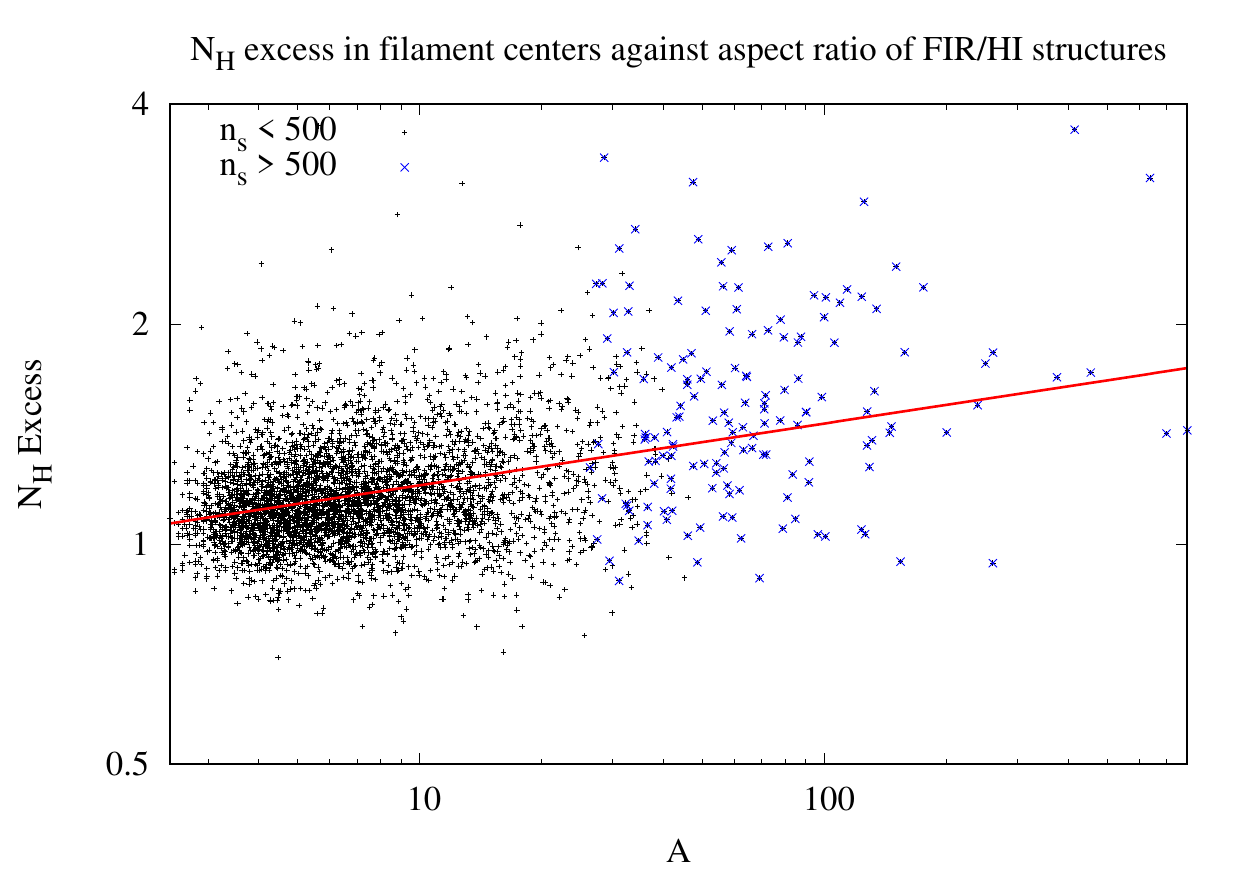}
   \caption{Excess in total hydrogen column density
     $\overline{N_{\mathrm{Hcenter}}} / \overline{N_{\mathrm{Hoff}}}$ in
     the filament centers as a function of aspect ratio ${\cal A}$. The
     fit $ \ln{(\overline{N_{\mathrm{Hcenter}}} /
     \overline{N_{\mathrm{Hoff}}})} =  -0.0095 \pm 0.008 + (0.085 \pm
     .004) \ln ({\cal A})  $ is indicated by the red line.  }
\label{Fig_NH}
\end{figure}

\begin{figure}[th] %%  19
  \centering
  \includegraphics[width=9cm]{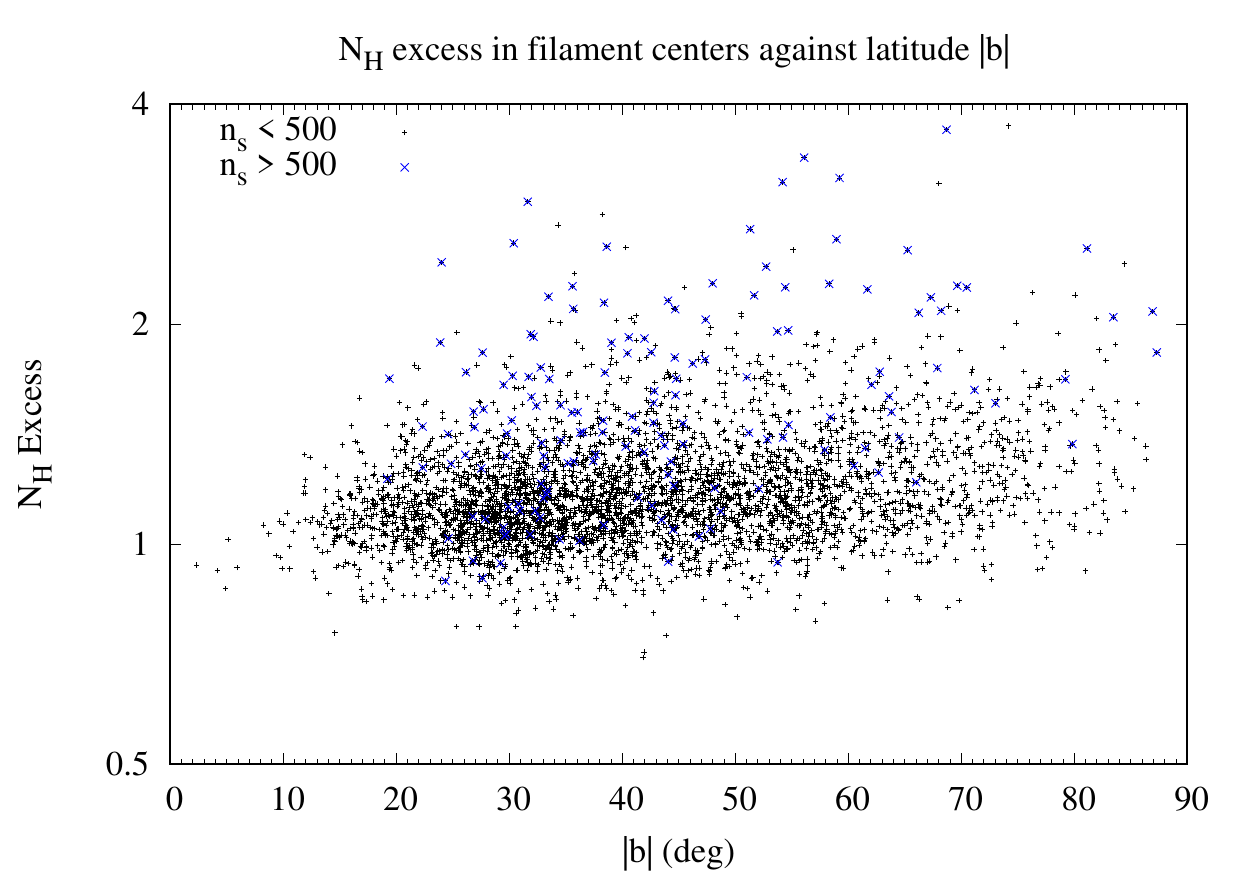}
   \caption{Excess in total hydrogen column density
     $\overline{N_{\mathrm{Hcenter}}} / \overline{N_{\mathrm{Hoff}}}$ in
     the filament centers as a function of Galactic latitude $|b|$. }
\label{Fig_NH_b}
\end{figure}

Summarizing over all filaments from our sample, we determine
$\overline{N_{\mathrm{Hfil}}} / \overline{N_{\mathrm{Hoff}}} = 1.43 $
(1.71) and $\overline{N_{\mathrm{Hcenter}}} /
\overline{N_{\mathrm{Hoff}}} = 1.56 $ (1.86).  These calculations can be
repeated using the CO-dark \h2\ content.  The average \h2\ excess in
filaments is in this case 4.32 (5.60) and 5.29 (6.67), respectively. The
values in brackets are obtained when only prominent filaments are selected and
indicate a further increase of $\sim 23$\%. Figure \ref{Fig_NH} confirms
the trend that prominent filaments with $n_{\mathrm{S}} > 500$ tend to
have a larger $N_{\mathrm{H}}$ excess.  Figure \ref{Fig_NH_b} shows the
latitude dependence of the ${N_{\mathrm{H}}}$ excess. There is no
significant dependence on Galactic latitude (or on longitude, but
not shown); most of the ${N_{\mathrm{H}}}$-excess is contained within
filaments with large aspect ratios or surfaces. The \h2\ distribution
caused by condensations in the colder part of the CNM is strongly
organized in filaments (see Fig. 11 of \citet{Kalberla2020}).

In principle, dependencies between column densities and aspect ratios
can be used to infer the extent to which filaments are tilted away from the plane of
the sky. The aspect ratio is in this case reduced by a factor
$\cos{(\theta_{\mathrm{POS}}),}$ but the total column density of the
filament remains constant. The apparent surface also decreases by a
factor of $\cos{(\theta_{\mathrm{POS}})}$ and we should observe an
average column density excess for low aspect ratios; however, this is not
observed.  On the contrary, Fig. \ref{Fig_NH} indicates a weak tendency
---with a correlation coefficient of 0.34--- for an increase in the column
density excess with increasing aspect ratio, hence with increasing surface.

\begin{figure}[th] %%  20
  \centering
  \includegraphics[width=9cm]{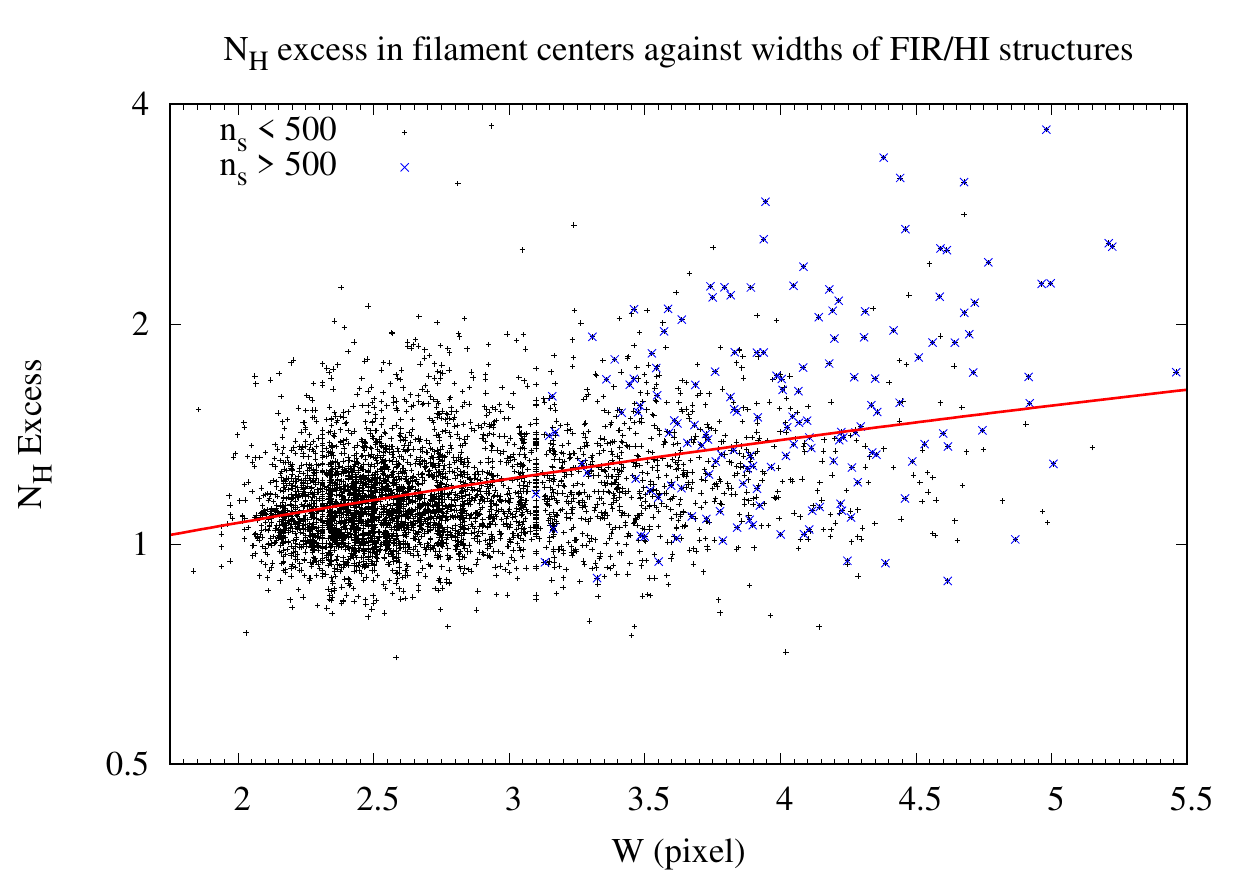}
   \caption{Excess in total hydrogen column density
     $\overline{N_{\mathrm{Hcenter}}} / \overline{N_{\mathrm{Hoff}}}$ in
     the filament centers as a function of filament width $W$. The
     fit $\overline{N_{\mathrm{Hcenter}}} /
     \overline{N_{\mathrm{Hoff}}} = 0.75 \pm 0.02 + (0.16 \pm
     .006) W $ is indicated by the red line.  }
\label{Fig_NH_W}
\end{figure}

Figure \ref{Fig_NH_W} shows that there is also a weak correlation between $N_{\mathrm{H}}$ column density excess and
width,  with a
coefficient of 0.36. We conclude that denser FIR/\hi\ filaments tend to have an
increased width $W$. This tendency cannot be explained by projection
effects.  We note in Sect. \ref{widths} that filament width increases
with total filament surface area (Fig. \ref{Fig_Widths}), and also that the filamentarity
is affected by an increase in $W$ (Sect. \ref{F_A} and
Fig. \ref{Fig_F_S}).  Figure \ref{Fig_Widths_lam} indicates that such
structures tend to have a high significance level
$-\lambda_-$. Consolidating this information, prominent filaments tend to
have increased width and to be associated with $N_{\mathrm{H}}$ enhancements,
and it is tempting to speculate that such filaments may also be associated with
stronger magnetic fields.

\begin{figure}[th] %%  21
  \centering
  \includegraphics[width=9cm]{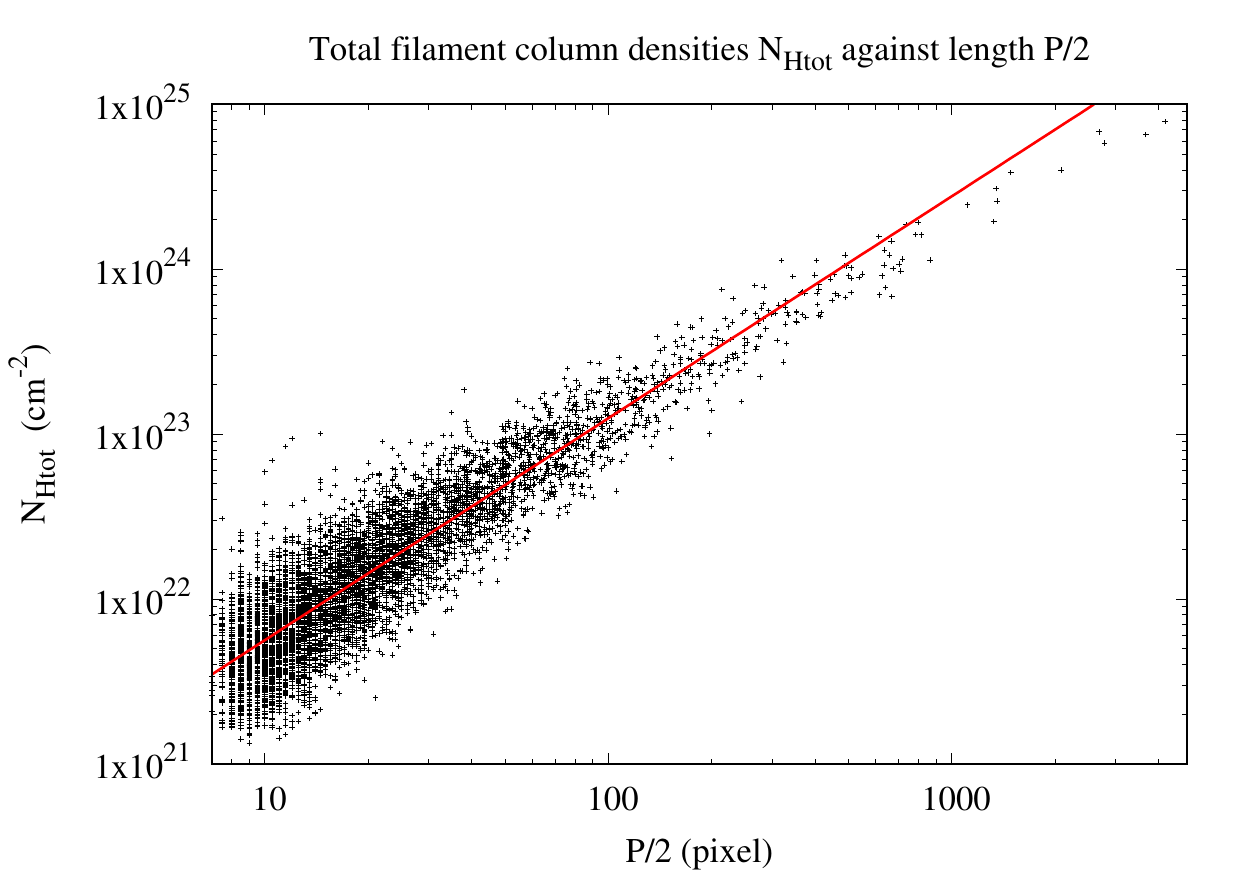}
  \includegraphics[width=9cm]{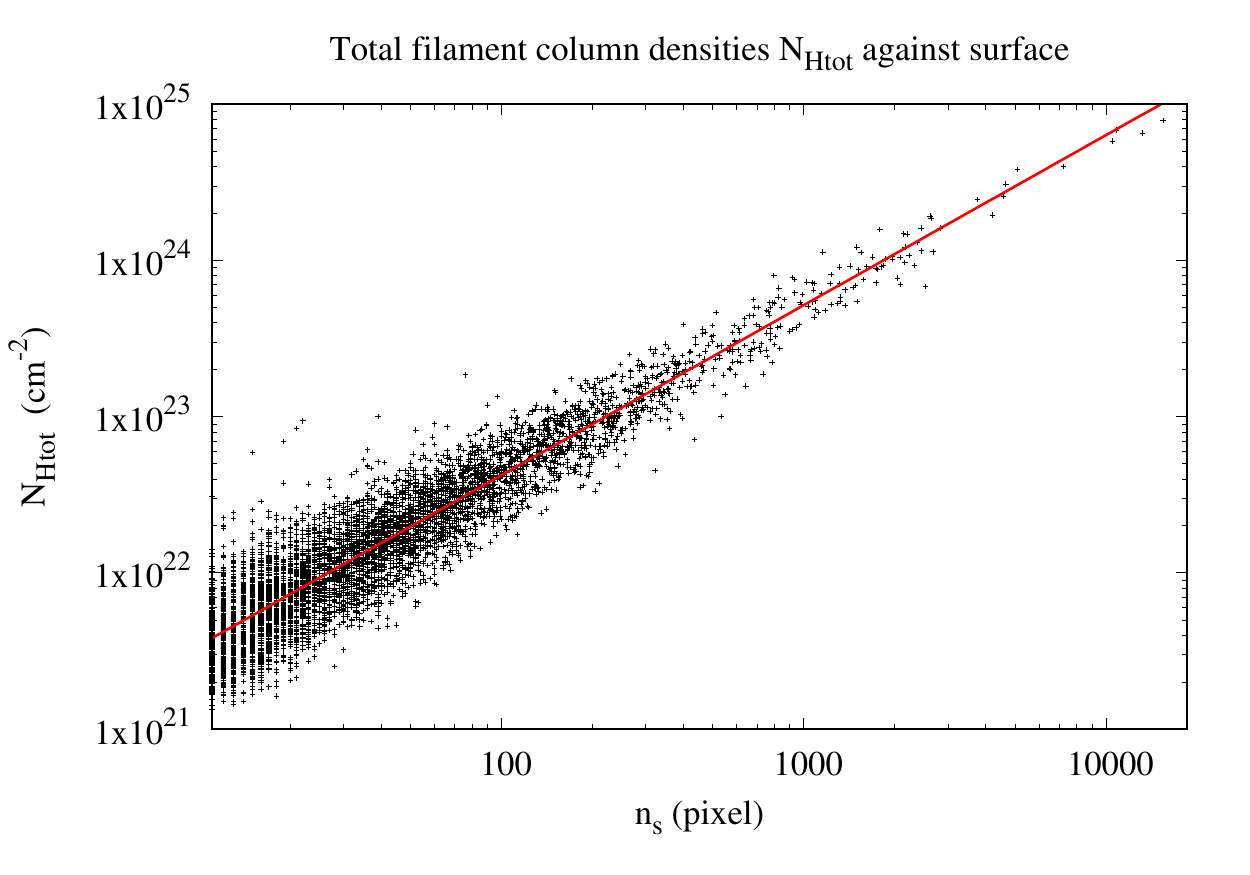}
   \caption{Distribution of total hydrogen column densities in
     filaments. Top: ${N_{\mathrm{Htot}}}$ as a function of
     length $P/2$. The fit $\ln(N_{\mathrm{Htot}}) = 47.0 \pm 0.02 +
     (1.345 \pm 0.008) \ln(P/2) $ is indicated by the red line.  Bottom:
     ${N_{\mathrm{Htot}}}$ as a function of surface $n_{\mathrm{s}}$ and with the fit
     $\ln(N_{\mathrm{Htot}}) = 47.1 \pm 0.02 + (1.089 \pm 0.005)
     \ln(n_{\mathrm{s}})$ shown as a red line.  }
\label{Fig_NH_ns}
\end{figure}

We do not know the extent to which projection effects have led to enhancements in
the observed hydrogen column densities. However, the additional increase
in the filament centers cannot be explained by mere projection effects.
The observed effects are relatively strong in comparison to the results by 
\citet{Koch2015}, who obtain enhancements in brightness temperature by a factor of
1.3  for filaments in the {\it Herschel} Gould
Belt Survey. We conclude that FIR filaments in the diffuse ISM are associated
with cold \hi\ gas and CO-dark molecular hydrogen that is concentrated
toward the filament centers.

For a complete evaluation of the distribution of hydrogen column
densities in filaments, we calculated the total column density content
$N_{\mathrm{Htot}}$ for each individual filament. Figure \ref{Fig_NH_ns}
shows that the FIR/\hi\ filaments are part of a well-defined homogeneous
population of structures. For the filament length $ L = P/2, $ we fit
$N_{\mathrm{Htot}} \propto (P/2)^{1.35}$ and for the surface
$N_{\mathrm{Htot}} \propto n_{\mathrm{s}}^{1.09}$. Assuming that
filaments at different scales effectively sample similar gas densities
in the ISM, we expect an increase of $N_{\mathrm{Htot}} \propto
n_{\mathrm{s}}$. The observed enhancement of $N_{\mathrm{Htot}}$ at
large surfaces is consistent with the increase in $W$ for large
surfaces, also with increasing eigenvalues $-\lambda_-$
(Sect. \ref{widths}), implying stronger FIR/\hi\ ridges. The
$N_{\mathrm{Htot}}$ distribution is partly self-replicating, and merger
or disruption of individual filamentary structures leads to a
repositioning of the filaments within the distribution of
Fig. \ref{Fig_NH_ns}. Projection effects decrease the projected surface
without changing $N_{\mathrm{Htot}}$ and therefore cause an increase in
the observed scatter.  \citet{Hacar2022} relate the mass $M$ in
filaments to their length $L$ and find that the entire distribution of
filaments (except \hi\ filaments) roughly follows the distribution $L
\propto M^{1/2}$.  Figure \ref{Fig_NH_ns} top is incompatible with
such a distribution.

\subsection{Fractals: The surface--perimeter relation}
\label{fractal}

\begin{figure}[th] %%  22
  \centering
  \includegraphics[width=9cm]{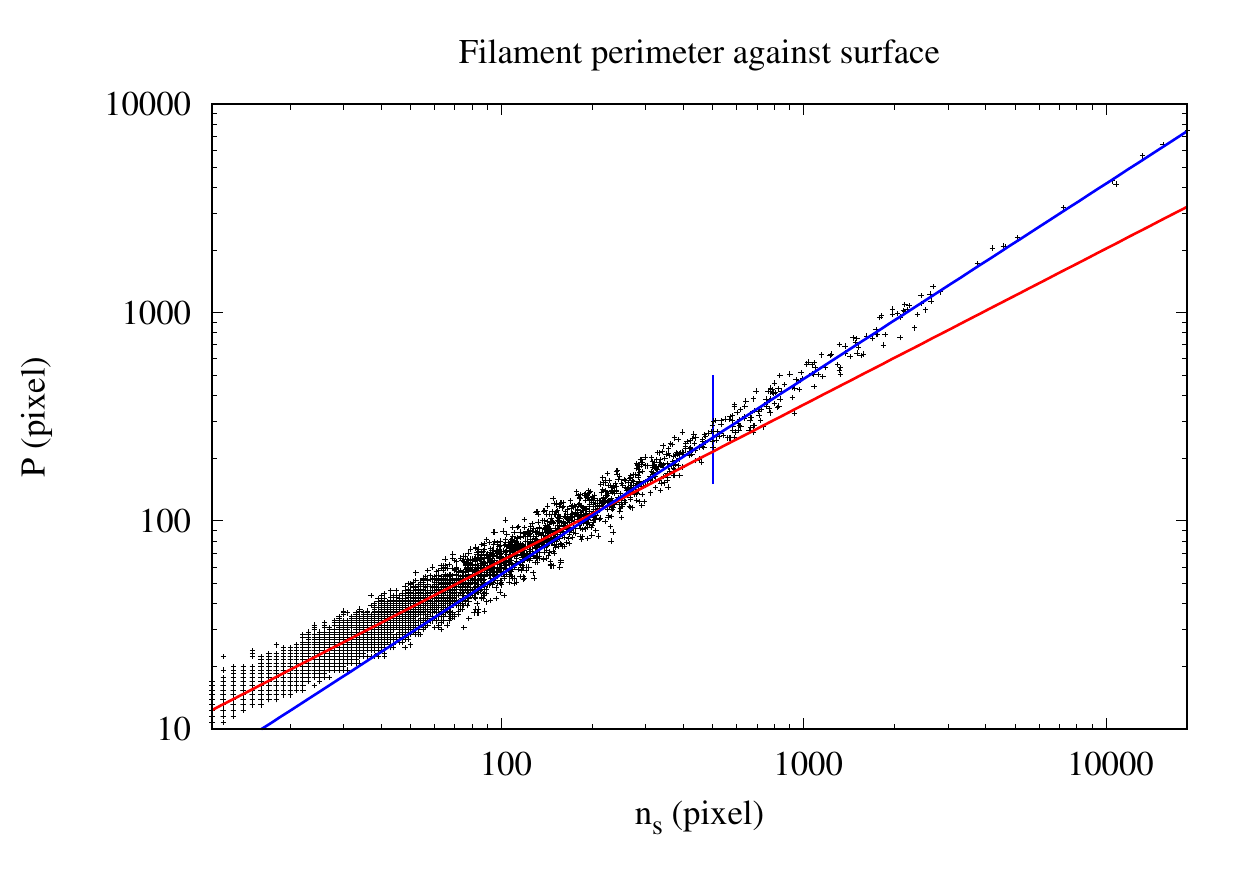}
  \caption{Filament perimeters $P$ as a function of surface counts 
    $n_{\mathrm{s}}$.  The fits $\ln(P) = 0.716 \pm 0.007 + (0.749 \pm
    0.002) \ln(n_{\mathrm{s}}) $ for $n_{\mathrm{s}} < 500$ (red) and
    $\ln(P) = -0.30 \pm 0.08 + (0.94 \pm 0.01) \ln(n_{\mathrm{s}}) $ for
    $n_{\mathrm{s}} > 500$ (blue) are indicated, which correspond to fractal
    dimensions $D = 1.5$ and $D = 1.9,$  respectively. }
\label{Fig_S_P}
\end{figure}

Filaments represent hierarchical configurations with spatial structures
on all scales. According to \citet{Mandelbrot1983}, self-similar
structures are ubiquitous in nature and are often linked to a fractal
geometry, which characterizes the self-similar scaling of surface
structures.  \citet{Mandelbrot1984} define the fractal dimension $D$
(sometimes referred to as the Hausdorff dimension) with the surface--perimeter
relation $ P \propto S^{D/2}$. Here, $D$ can range from circular sources with
$D = 1$ to highly convoluted structures with $D = 2$.  In an extensive
analysis, \citet{Falgarone1991} applied this approach to molecular cloud
structures in the range 0.1 to 100 pc. These authors obtained a common fractal
dimension $D = 1.36 \pm 0.02$ independent of the different scales considered, and concluded that the molecular gas is organized in a
self-similar distribution of sizes continuing down to a threshold
estimated to be smaller than 2000 AU.

To compare these results with fractal dimensions of FIR/\hi\ filaments,
we calculate the surface--perimeter relation shown in
Fig. \ref{Fig_S_P}. This distribution is clearly bent and is inconsistent
with a single fractal dimension for all scales.  In order to quantify changes of the fractal dimension, we divided our database in
two parts, that is, filaments with small and large surfaces. We find $D = 1.5 $
for $n_{\mathrm{s}} \la 500$ and $D = 1.9 $ for $n_{\mathrm{s}} \ga
500$. The difference is highly significant.
  
 \citet{Mandelbrot1984} noted that some of their data
indicated that the central region of the log-log plots splits into two
distinct subregions characterized by different values for
$D$. \citet{Criscuoli2007} studied the fractal nature of magnetic
features in the solar photosphere and found that the fractal dimension
of bright magnetic features in CaIIK images ranges between values of 1.2
and 1.7 for small and large structures, respectively. Their Fig. A.1
shows a clear trend that is strikingly similar to our
Fig. \ref{Fig_S_P}. Systematic changes in $D$ have been known about since
\citet{Meunier1999}. Indeed, their Figs. 3, 7, and 8 show a similar trend for
magnetograms observed by the SOHO spacecraft and the authors note that the fractal
dimension increases with the area of the active regions and mention that
differences in the granulation patterns might be due to different
magnetic-field strengths.

Searching in our case for an explanation for variations of $D$ with
surface area, we consider Eqs. (\ref{eq:A2}) and (\ref{eq:A}), which for
filaments result in $ P = \sqrt{{4\cal A}S} = (2/W)~S $. Therefore, for
filaments, $D = 2 \log{P}/\log{S} $ depends in a nonlinear way on ${\cal
  A}$ or $W$. Following the relation (\ref{eq:F_A}), filamentary
structures are incompatible with a constant fractal dimension.  In
Sect. \ref{widths}, we demonstrate that the filament widths tend to
increase with increasing surface (see Fig. \ref{Fig_Widths}); we
interpret this trend with enhanced column densities in gas and dust for
prominent filaments.  Here, we find a steepening of the fractal
dimension for large surface areas. The question arises as to whether
such a steepening could be related to an enhanced magnetic field
strength, as in the case of solar magnetograms.

\section{Estimating the turbulent magnetic field strength }
\label{field}

The magnetization of the ISM can be estimated by assuming that the
energy of magnetic field fluctuations is in equipartition with the
kinetic energy of turbulence \citep{Alfven1949},
\begin{equation}
\frac{\delta B^2}{8  \pi} = \frac{\rho \Delta v_{\mathrm{LSR}}^2} {2} ,
\label{eq:B}
\end{equation}
where $\delta B$ is the strength of the fluctuating part of the magnetic
field and $\rho$ is the volume density of the gas phase that is supplying
the kinetic energy. Here, we need to consider the total amount of gas,
including hydrogen and helium. We use the estimates of the molecular
hydrogen content of \hi\ clouds according to \citet{Kalberla2020} (see
also Sect. \ref{dark}) and include a 34\% correction for the helium mass
fraction \citep{Steigman2007}.  It is assumed that the true velocity
perturbations are isotropic, and therefore the dispersion in the velocity
transverse to the filament spines is equal to the rms line-of-sight
(LOS) velocity \citep{Chandrasekhar1953} and $\Delta v_{\mathrm{LSR}}$
is the observed 1D LOS velocity dispersion along the filaments.  Below,
we discuss two cases, assuming filaments at a constant average distance
and alternatively assuming the filaments to be located at the LB rim
with distances as determined by \citet{Pelgrims2020}.

\subsection{Distance dependencies}
  \label{withinLB}

A generally accepted model assumption is that winds and supernova
explosions inflated the LB of gas and dust. The cavity wall was
stretched to create the aligned filaments and magnetic field lines that
we observe. For many years, the distance to the LB wall remained unknown
and it became common to use 100 pc as a fiducial number
\citep{Sfeir1999}. Below we describe how we followed this assumption
first in order to derive results that are comparable to previously
published findings, and then explored how the results change with
distance.

\begin{figure}[th] %%  23
  \centering
  \includegraphics[width=9cm]{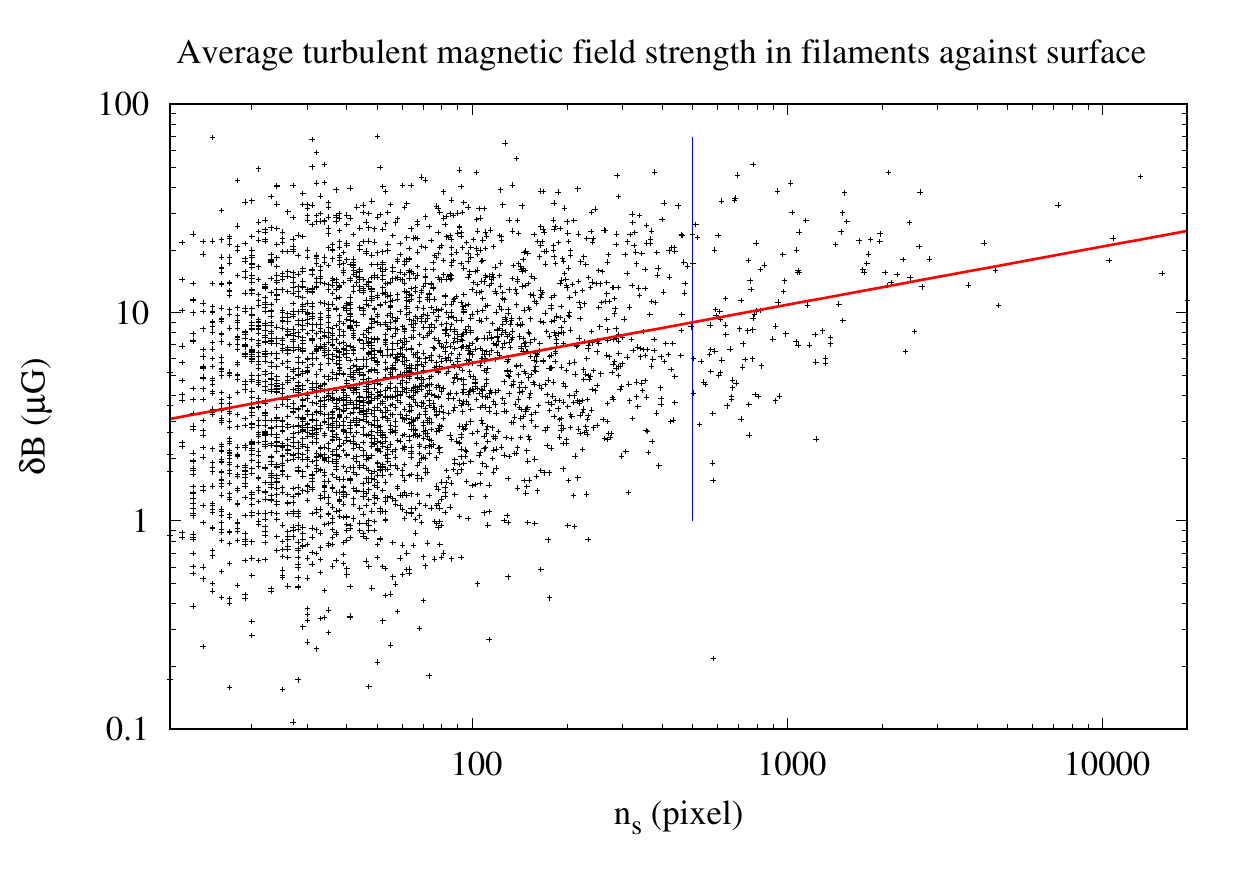}
  \includegraphics[width=9cm]{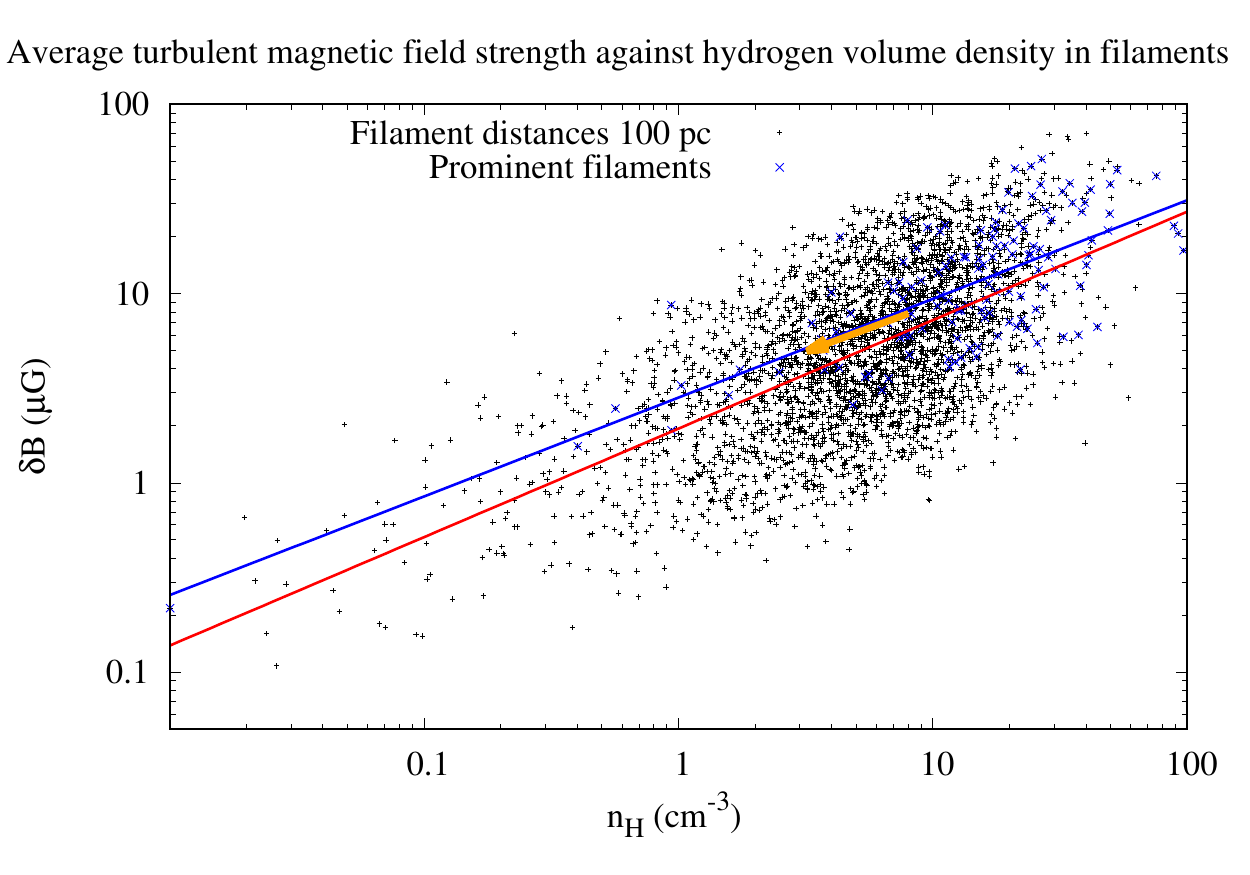}
  \caption{Average turbulent magnetic field strength $\delta B$ in
    filaments. Top: $\delta B$ depending on filament surface
    $n_{\mathrm{s}}$. The fit
    $\ln(\delta B) = 0.46 \pm 0.07 + (0.28 \pm 0.02) \ln(n_{\mathrm{s}})
    $ is indicated by the red line. Bottom: Average turbulent magnetic
    field strength in filaments $\delta B$ as a function of the average
    hydrogen volume density.  The fit $\ln(\delta B) = 0.66 \pm 0.02 +
    (0.57 \pm 0.01) \ln(n_{\mathrm{H}})$ for all 6~568 filaments is
    indicated in red. Fitting only prominent filaments with surfaces for
    $n_{\mathrm{s}} > 500$ (right from the vertical blue line in the top
    figure), we get $\ln(\delta B) = 1.0 \pm 0.1 + (0.52 \pm 0.04)
    \ln(n_{\mathrm{H}}),$ as indicated by the blue line. The orange arrow
    points from the center of mass of the point cloud ($\langle
    n_{\mathrm{H}} \rangle = 8.0~{\rm cm^{-3}}$, $\langle \delta B
    \rangle = 7.9~\mu$G)--- assuming a distance of 100 pc--- to the center
    of mass ($\langle n_{\mathrm{H}} \rangle = 3.2~{\rm cm^{-3}}$,
    $\langle \delta B \rangle = 5.0 ~\mu$G) for a distance of 250 pc. }
\label{Fig_B}
\end{figure}

Using the parameters from the previous section, we derived an average magnetic field strength of $\langle \delta B
\rangle = 7.9~\mu \mathrm{G }$  for an average
distance of 100 pc. The distribution of $\delta B$ for all
filaments is shown in Fig. \ref{Fig_B}. The top panel displays $\delta B$ as
a function of surface $n_{\mathrm{s}}$. As expected from previous
discussions, filaments with large surfaces tend to be associated with
higher magnetic fields; the correlation coefficient here is 0.28.  For
$n_{\mathrm{s}} > 500,$ we find the average $\langle \delta B \rangle = 14.2~\mu
\mathrm{G}$. The lower plot in Fig.  \ref{Fig_B} displays $B$ as
a function of $n_{\mathrm{H}}$. We find a general trend of $\delta B \propto
n_{\mathrm{H}}^{0.57} $, which is indicated with the red line. The blue line
shows a similar trend for prominent filaments with $n_{\mathrm{s}} >
500$.

According to recent distance determinations, which are discussed in the following
subsection in more detail, most of the LB wall is located at distances
of between 200 and 300 pc \citep{Pelgrims2020}. To estimate distance biases,
we considered dependencies on our estimates for the averages of $\langle
n_{\mathrm{H}} \rangle = 8.0~{\rm cm^{-3}}$, $\langle \delta B \rangle = 7.9
~\mu$G, at a distance of 100 pc, by changing the assumed distance to 250
pc. We obtain $\langle n_{\mathrm{H}} \rangle = 3.2~{\rm cm^{-3}}$,
$\langle \delta B \rangle = 5.0 ~\mu$G. The resulting average bias is indicated
in Fig. \ref{Fig_B} with an orange arrow. For filamentary structures
with variable distances, we expect that the distribution shown in this
figure can be stretched by shifting individual data points up to the
amount indicated by the orange arrow.

\begin{figure}[th] %%  24
  \centering
  \includegraphics[width=9cm]{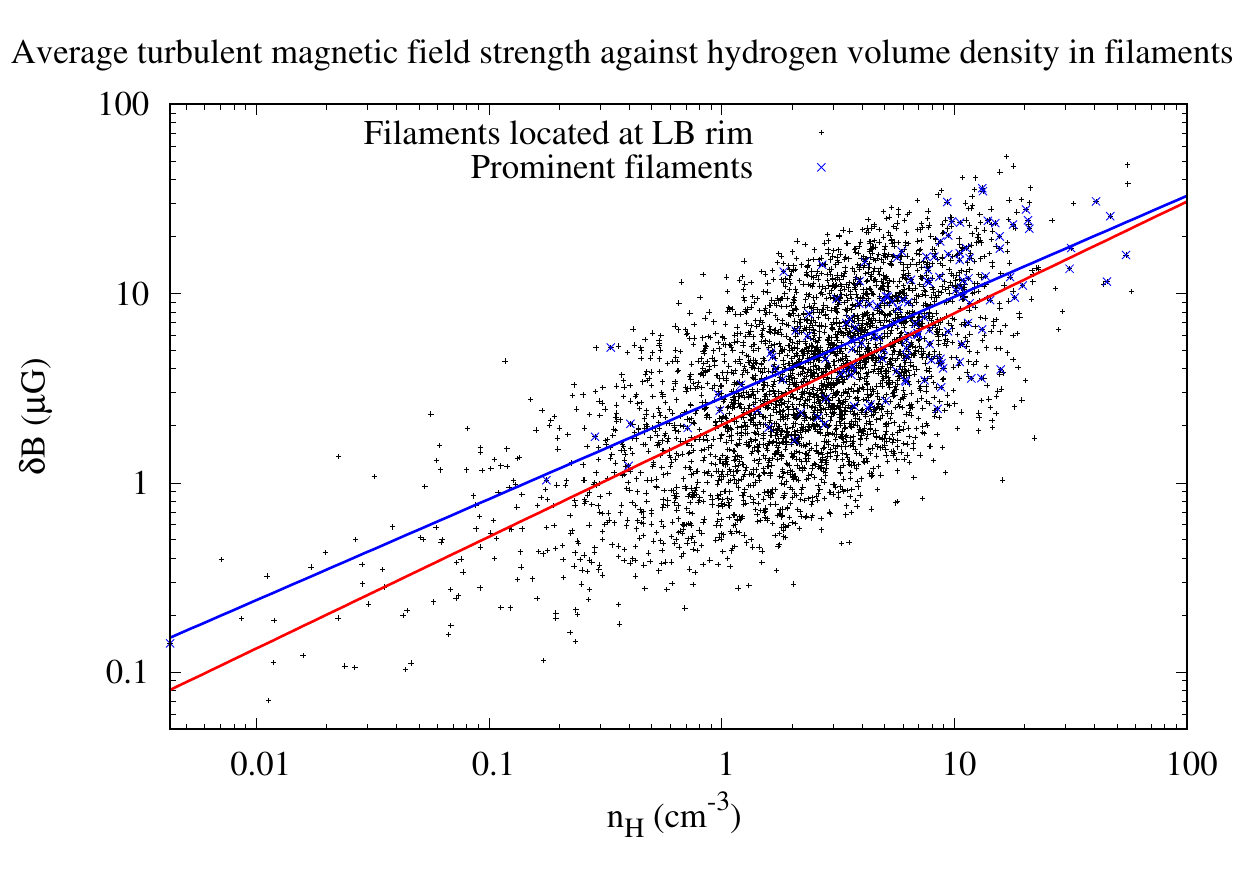}
  \caption{Average turbulent magnetic field strength in filaments
    $\delta B$ as a function of the average hydrogen volume density for
    filaments located at the LB wall.  The fit $\ln(\delta B) = 0.71 \pm
    0.02 + (0.59 \pm 0.01) \ln(n_{\mathrm{H}})$ for all 6~568 filaments
    is indicated in red. When fitting only prominent filaments with surfaces
    for $n_{\mathrm{s}} > 500$ (right from the vertical blue line in
    Fig. \ref{Fig_B} top), we get $\ln(\delta B) = 1.03 \pm 0.09 + (0.53
    \pm 0.04) \ln(n_{\mathrm{H}})$ as indicated by the blue line.  }
\label{Fig_B_LB}
\end{figure}

\subsection{Filaments at the rim of the the Local Bubble}
\label{rimLB}
Figure \ref{Fig_B_LB} shows the distribution of the turbulent magnetic
field strength $\delta B$ using recent distance estimates.
\citet{Pelgrims2020} determined the geometry of the LB shell from 3D
dust extinction maps. We use distances for the inner surface of the LB
shell as extracted from the \citet{Lallement2019} extinction map shown
in the top panel of Fig. 4 in \citet{Pelgrims2020}. The spread in distance is $ 81
< D < 366$ pc. We verified that distances within the filaments are
homogeneous, and find no significant scatter in distance within
individual filaments. Furthermore, we find no distance dependencies for
filament width or velocity.

As expected, Fig. \ref{Fig_B_LB} shows a somewhat stretched data
distribution.  We derive $\delta B \propto
n_{\mathrm{H}}^{0.59} $  on average  and $\delta B \propto
n_{\mathrm{H}}^{0.53} $  for prominent filaments. The average turbulent magnetic field strength
is $\langle \delta B \rangle = 5.3 ~\mu$G at an average filament volume
density of $\langle n_{\mathrm{H}} \rangle = 3.8~{\rm cm^{-3}}$. In the case of
prominent filaments, we obtain $\langle \delta B \rangle = 9.5 ~\mu$G and
$\langle n_{\mathrm{H}} \rangle = 8.6~{\rm cm^{-3}}$. The average
magnetic field strength is only 6\% higher than the estimates in
Sect. \ref{withinLB} in the case of a constant distance of 250 pc. Our
exercise in Sect. \ref{withinLB} demonstrates that regional distance
uncertainties cannot affect the average turbulent magnetic field
strength significantly. We repeated our analysis, this time also using the model
extinction map with a spherical harmonic expansion up to
$l_{\mathrm{max}} = 6$, shown at the bottom of Fig. 4 in
\citet{Pelgrims2020}. We find changes in the scatter diagram but no
significant changes in the derived averages.

All filaments considered here have $n_{\mathrm{H}} < 100~{\rm
  cm^{-3}}$.  \citet{Crutcher2010} and \citet{Crutcher2012} note that in
the range $n_{\mathrm{H}} < 300~{\rm cm^{-3}}$, which is characteristic for the
diffuse ISM, $\delta B$ does not scale with density. These authors conclude that
diffuse clouds are assembled by flows along magnetic field lines, which
would increase the density but not the magnetic field strength. Our case
is different; we explicitly study filamentary structures driven by a
turbulent small-scale dynamo. These structures are in direct interaction
with turbulence-induced magnetic fields. Therefore, we consider a selective
sample, occupying only 25\% of the diffuse ISM.  \citet{Pelgrims2020}
modeled the magnetic field by fitting the Planck 353 GHz dust-polarized
emission maps over the Galactic polar caps and concluded that the
magnetic field in each polar cap is almost aligned with the plane of the
sky. This result supports the validity of our approach, inserting the
observed velocity dispersion in Eq. (\ref{eq:B}) without a correction
for projection effects. We implicitly assume that the magnetic field is
aligned with the plane of the sky.  Our results do not necessarily
contradict those of \citet{Crutcher2010}, who consider Zeeman observations, where
the magnetic field is along the line of sight, while Eq. (\ref{eq:B}) is
valid for estimates of the field projected to the plane of the sky.
Zeeman measurements suffer from sensitivity limitations, while estimates
based on Eq. (\ref{eq:B}) are dominated by uncertainties in distance
and volume density.  The derived scaling with the magnetic field
strength is in our case $\delta B \propto n_{\mathrm{H}}^{0.59} $ , which is
consistent with the MHD simulations  of \citet{Seta2022} for the fluctuation dynamo in a
compressible multiphase medium. These authors
consider  a compression of the magnetic
field perpendicular to the magnetic field lines for the warm neutral medium (WNM), and therefore $B \propto
n_{\mathrm{H}}^{0.5}$ for cylindrical or filamentary geometry, and with
their simulations obtain $B \propto n_{\mathrm{H}}^{0.51}$.  A similar
power-law scaling was obtained by \citet{Ponnada2022} from their
simulations of filamentary cloud structures.

Studying the nature and the properties of the cold structures formed via
thermal instability in the magnetized atomic ISM, \citet{Gazol2018} and
\citet{Gazol2021} searched for clumps formed in forced MHD simulations
with an initial magnetic field ranging from 0 to 8.3~$\mu$G. These authors find
that a positive correlation between $B$ and $n_{\mathrm{H}}$ develops
for all initial magnetic field intensities. The density at which this
correlation becomes significant ($n_{\mathrm{H}} \la 30~{\rm cm^{-3}}$)
depends on the initial conditions but is not sensitive to the presence
of self-gravity. Figure 2 of \citet{Gazol2021} shows that the resulting
clumps have a wide range of magnetic field intensities, which is similar to the
large scatter that we observe here. The magnetic field in the simulations,
albeit weak, qualitatively affects the morphology by
producing filamentary structures.

\section{Estimating the mean magnetic field strength}
\label{rimLB_mean}

After considering estimates of the fluctuating magnetic field strength
$\delta B,$ we intend to discuss estimates of the mean magnetic field
strength $B_{\mathrm{POS}}$ in the plane of the sky. For this parameter, we
use four different estimates. 

\subsection{DCF-based methods}
\label{rimLB_CF}

\citet{Davis1951} and
\citet{Chandrasekhar1953} (DCF) introduced the basic approach to
determining $B_{\mathrm{POS}}$, which is usually formulated as
\begin{equation}
B_{\mathrm{DCF}} = \xi \sqrt{4 \pi \rho}  \frac{\Delta v_{\mathrm{LSR}} }
{\Delta \theta} .
\label{eq:B_DCF}
\end{equation}
Here, it is assumed that the velocity fluctuations are perpendicular to the
magnetic field, and therefore $\Delta v_{\mathrm{LSR}}$ is the 1D line-of-sight
dispersion in velocity (from turbulence, unaffected by thermal
broadening). $\Delta \theta$ is the dispersion of the orientation angles
in the plane of the sky, reflecting transverse fluctuations of the
magnetic field, and $ \xi $ is a correction factor, usually approximated as
$ \xi \sim 0.5$, that takes various physical and observational conditions
(density inhomogeneities, anisotropies on velocity perturbations,
observational resolution, and averaging effects) into account, (e.g.,
\citet{Zweibel1990}, \citet{Ostriker2001}, \citet{Heitsch2001},
\citet{Padoan2001} or \citet{Falceta2008}. The original DCF approach was
$ \xi = 1$. In the case where $\Delta \theta \la 25\deg$, a corrected factor $\xi
= 0.5 $ was determined by \citet{Ostriker2001} from MHD
simulations. However, various authors noticed that this factor may be
rather uncertain. Several attempts have been made to improve
Eq. (\ref{eq:B_DCF}) and we discuss three cases for comparison.

\citet{Heitsch2001} proposed the modification
\begin{equation}
B_{\mathrm{H}} = \sqrt{4 \pi \rho} \frac{\Delta v_{\mathrm{LSR}}}
{\Delta (\tan{\theta})} [1 + 3 \Delta (\tan{\theta})^2]^{1/4},
\label{eq:B_H}
\end{equation}
and verified this approach with numerical simulations.  These authors
concluded that their modified version yields magnetic field estimates in
molecular clouds accurate up to a factor of 2.5 even for the weakest
fields. 

\begin{figure*}[th] %%  25
  \centering
  \includegraphics[width=9cm]{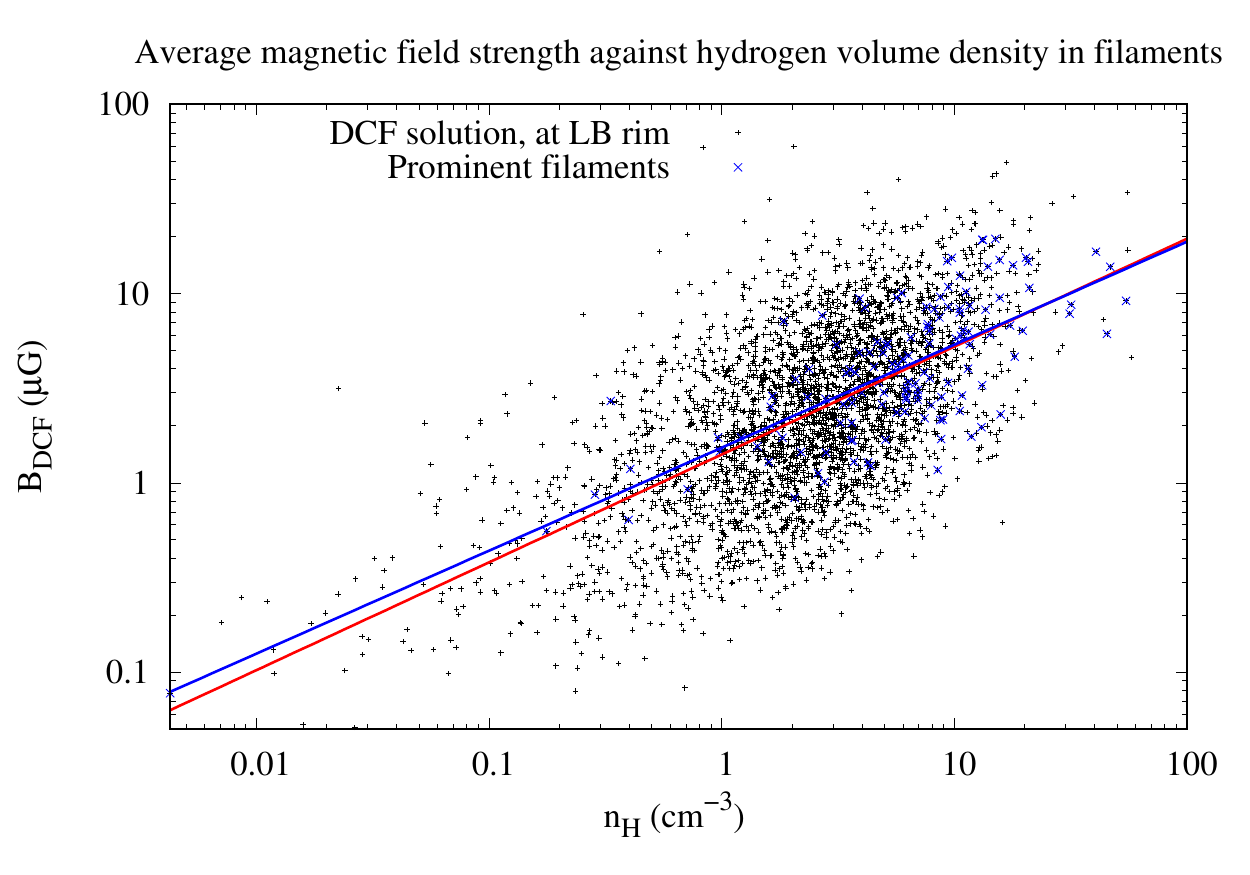}
  \includegraphics[width=9cm]{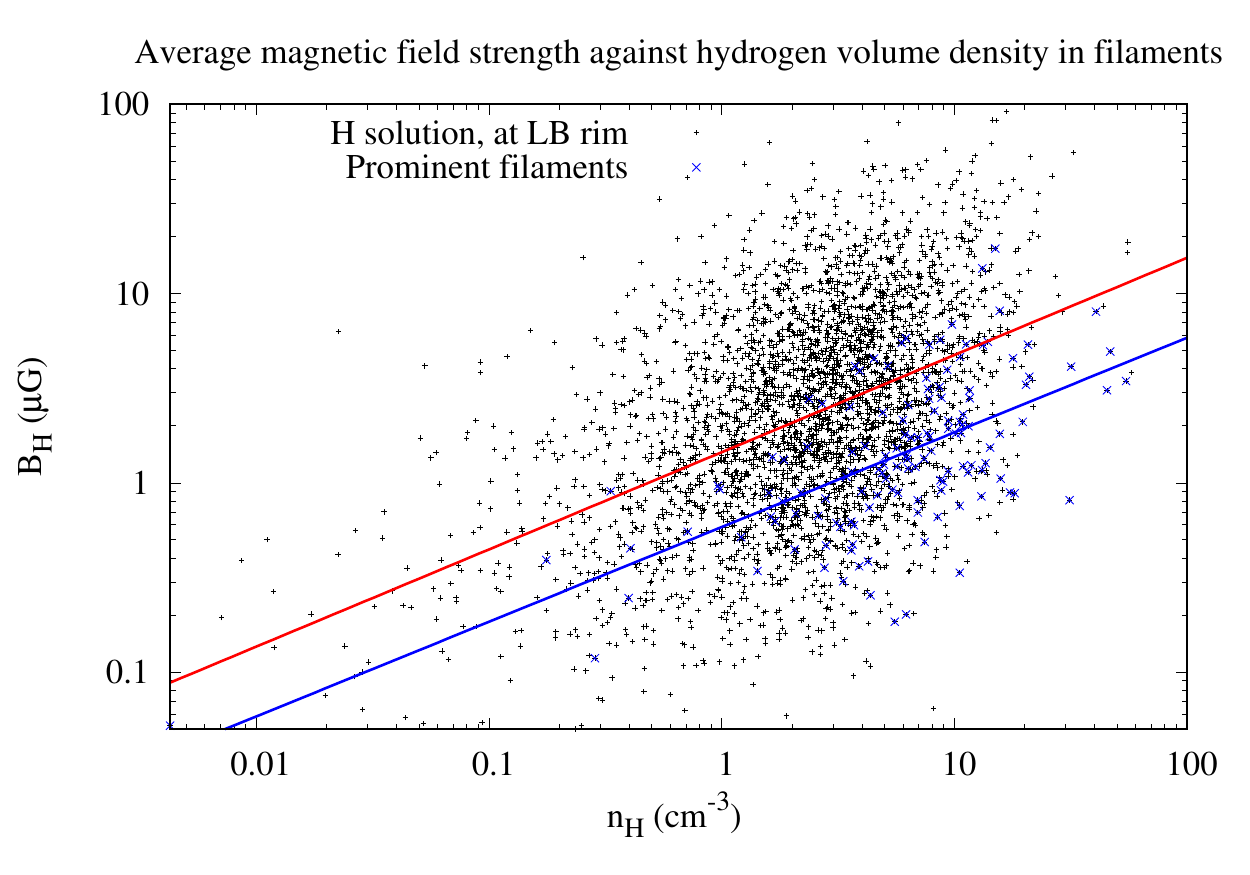}
  \includegraphics[width=9cm]{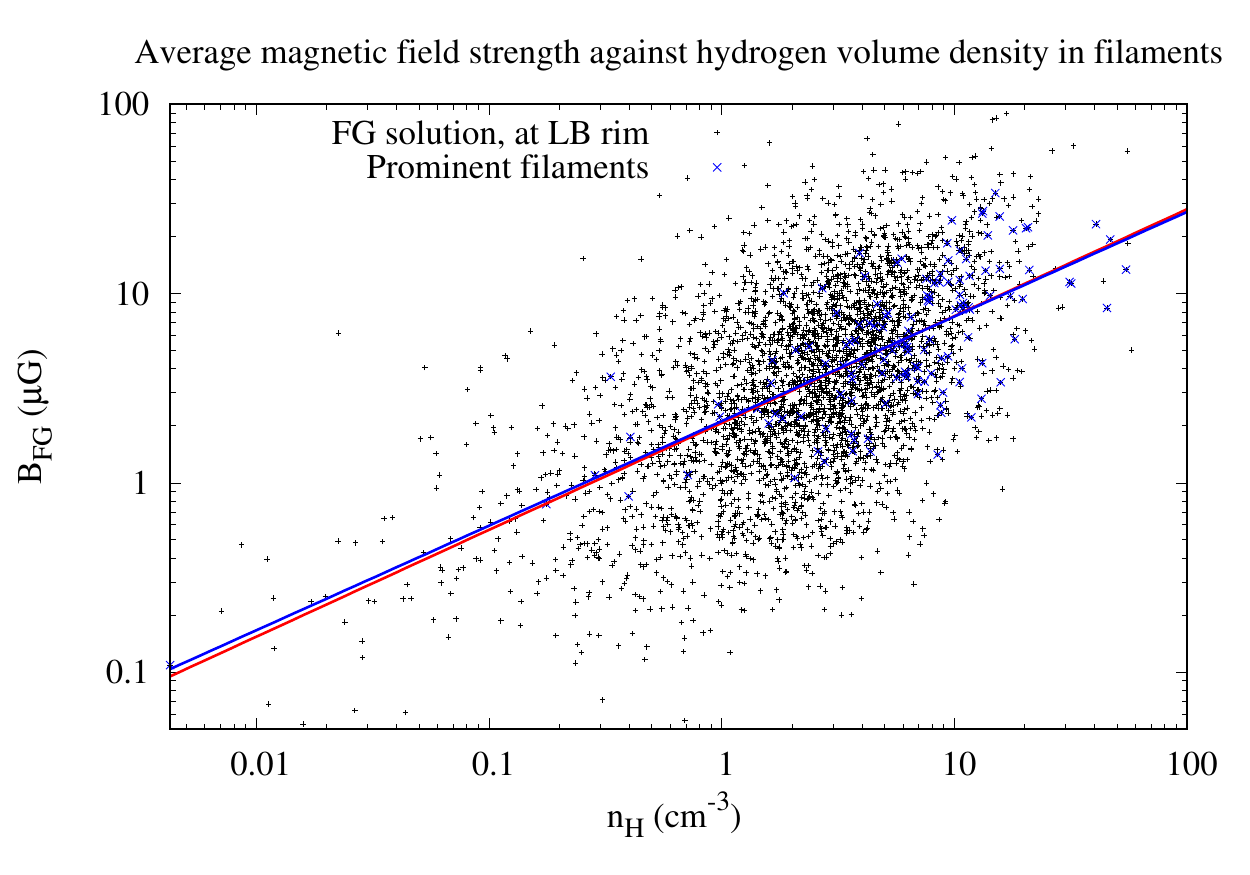}  
  \includegraphics[width=9cm]{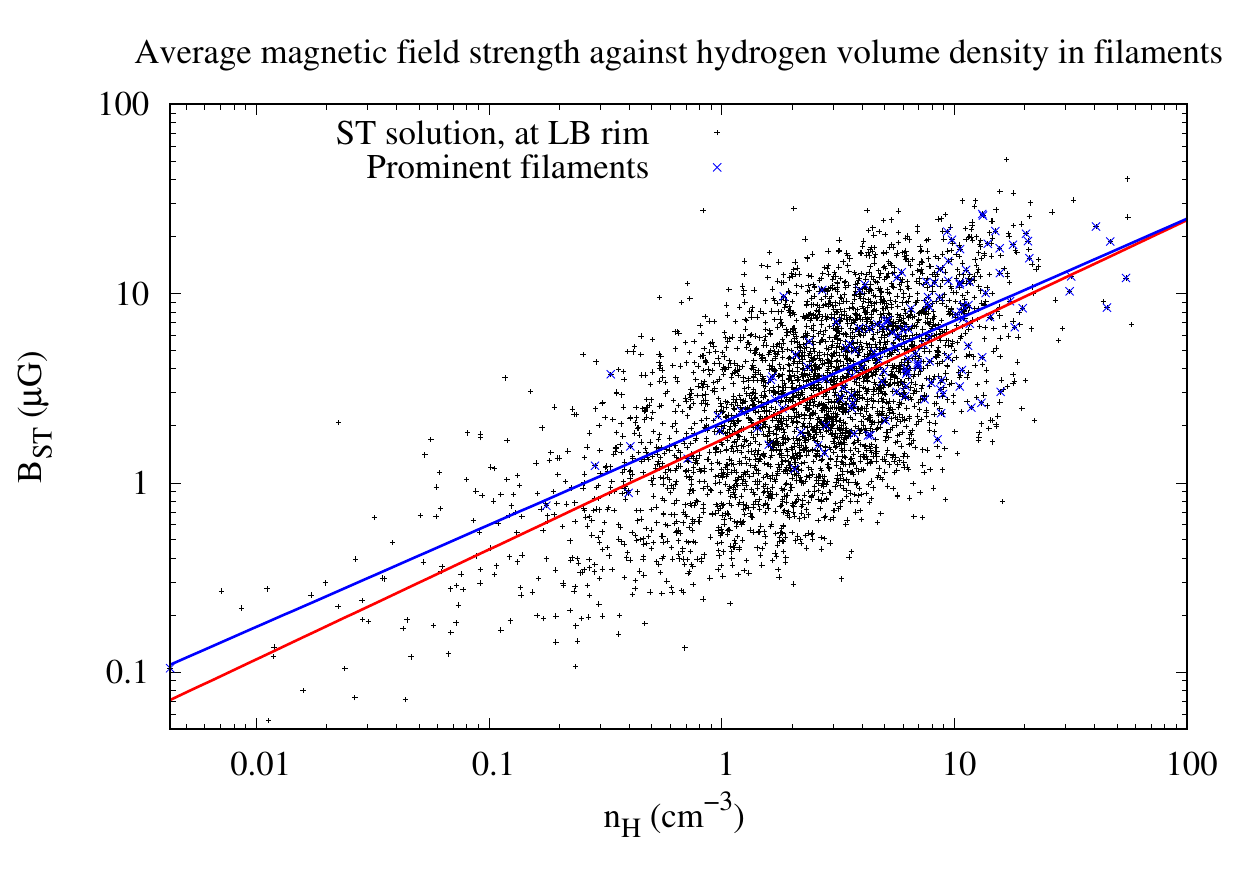}
  \caption{Average plane of the sky magnetic field strength in filaments
    $B_{\mathrm{POS}}$ as a function of the average hydrogen volume
    density for filaments located at the LB wall. We distinguish
    DCF approaches Eq. (\ref{eq:B_DCF}) (top left) from those of
    \citet[][H]{Heitsch2001}, Eq. (\ref{eq:B_H}) (top right),
    \citet[][FG]{Falceta2008}, Eq. (\ref{eq:B_FG}) (bottom left), and
    \citet[][ST]{Skalidis2021}, Eq. (\ref{eq:B_ST}) (bottom
    right). Fit parameters are given in Table \ref{tab:DCF}. }
\label{Fig_B_DFC_ST}
\end{figure*}

\begin{table*}
\centering
\caption{Derived parameters for distributions shown in
  Fig. \ref{Fig_B_DFC_ST}, mean magnetic field strengths, and
  parameters $a$ and $b$ for fitted regressions }
\begin{tabular}{lcccccc}
Approach & & All filaments & & & Prominent filaments  \\
\hline
 & $\langle B_{\mathrm{POS}} \rangle $  ($\mu$G) & $a$ & $b$ &
$\langle B_{\mathrm{POS}} \rangle$ ($\mu$G)  & $a$ & $b$      \\
\hline
Eq. (\ref{eq:B_DCF}) DCF  & 3.8  &  $ 0.35 \pm 0.02 $  & $ 0.57 \pm 0.01 $ & 5.3  & $
0.43 \pm 0.09 $ & $0.54 \pm 0.04$ \\
Eq. (\ref{eq:B_H}) H & 4.8  &  $ 0.37 \pm 0.02 $  & $ 0.51 \pm 0.02 $ & 2.1  & $ -0.5 \pm
0.1 $ & $0.50 \pm 0.06$ \\
Eq. (\ref{eq:B_FG}) FG& 6.1  &  $ 0.73 \pm 0.02 $  & $ 0.56 \pm 0.01 $ & 7.7  & $ 0.75 \pm
0.1 $ & $0.55 \pm 0.05$ \\
Eq. (\ref{eq:B_ST}) ST & 4.4  &  $ 0.53 \pm 0.02 $  & $ 0.58 \pm 0.01 $ & 7.1  & $ 0.73 \pm 0.09 $ & $0.54 \pm 0.04$ \\
\end{tabular}
\label{tab:DCF}
\end{table*}
\normalsize

Next, \citet{Falceta2008} proposed the modification
\begin{equation}
B_{\mathrm{FG}} = \sqrt{4 \pi \rho} \frac{\Delta v_{\mathrm{LSR}}} 
{\tan{\Delta (\theta})}.
\label{eq:B_FG}
\end{equation}
These authors noticed that,  in some cases, $B_{\mathrm{H}}$ tends to
systematically underestimate the magnetic field intensity.
They claim that their generalized equation for the DCF method 
allows them to determine the magnetic field strength from
polarization maps with errors $<20$\%.

The most recent reinvestigation of the DCF method was by
\citet{Skalidis2021}. These authors argue that the DCF method is based on the
assumption that isotropic turbulent motions initiate the propagation of
Alfv{\'e}n waves. \citet{Skalidis2021} also consider non-Alfv{\'e}nic
(compressible) modes that have been proven to be
important from observations but have not been considered in the DCF approach and
modifications. \citet{Skalidis2021} derive the relation
\begin{equation}
B_{\mathrm{ST}} = \sqrt{2 \pi \rho}  \frac{\Delta v_{\mathrm{LSR}} }
{\sqrt{\Delta \theta} } .
\label{eq:B_ST}
\end{equation}
Extensive numerical tests of this method were carried out by
\citet{Skalidis2021b}, who found that relaxing the
incompressibility assumption leads to far more reliable
magnetic field determinations for a broad range of parameters.

We use our data to determine the mean magnetic field strengths in the
plane of the sky and compare the four different approaches mentioned above.
Our approach is motivated by \citet{Pelgrims2020}, who concluded from
their best-fit model that, at high Galactic latitudes, the inclination of
the mean magnetic field has only a small angle $\gamma \sim 15\degr$ to
the plane of the sky. Projection effects caused by such a misalignment
should be negligible. We use the same distances to the LB rim as in
Sect. \ref{rimLB}. We determined the dispersion $\Delta \theta$ of the orientation
angles along individual filaments to be
\begin{equation}
\label{eq:defS}
\Delta \theta = 
\sqrt{\frac{1}{\sum_{i=1}^{N_i} N_j(i) } \sum_{i=1}^{N_i}\sum_{j=1}^{N_j(i)}\left[\theta(\boldsymbol{r_i}+\boldsymbol{\delta}_j)-\theta(\boldsymbol{r_i})\right]^2}
\, .
\end{equation}
This sum extends over all pixels $\boldsymbol{r_i}$ along the filament
with positions $ (\boldsymbol{r_i}+\boldsymbol{\delta}_j) $ within an
annulus centered on $\boldsymbol{r}$ and with inner and outer radii of
$\boldsymbol{\delta}/2$ and $3\boldsymbol{\delta}/2$, respectively.
According to the definition of the Hessian operator with adopted
Gaussian smoothing over five pixels in \citetalias{Kalberla2021}, we
select a lag of $\boldsymbol{\delta} = 18\arcmin$. 

Figure \ref{Fig_B_DFC_ST} shows the derived distributions of
$B_{\mathrm{POS}}$ as a function of $n_{\mathrm{S}}$ for the four
different estimates from Eqs. (\ref{eq:B_DCF}) to
(\ref{eq:B_ST}). Derived parameters for the ensemble averages $ \langle
B_{\mathrm{POS}} \rangle$ and from fitting $\ln(B_{\mathrm{POS}}) = a +
b \ln(n_{\mathrm{H}})$ are listed in Table \ref{tab:DCF}. Common to all
four approaches is a scaling relation between magnetic field
strength and volume density that is in conflict with the density-independent field strength determined by \citet{Crutcher2010}.  The ST
results are similar to the results from Sect. \ref{rimLB} and we adopt $B_{\mathrm{POS}} \propto n_{\mathrm{H}}^{0.58}$
in the following. The
most serious discrepancy in the derived scatter diagrams exists for
Eq. (\ref{eq:B_H}), the H solution. This may be explained by the fact that
\citet{Heitsch2001} put their focus on magnetized self-gravitating
molecular clouds. The filaments that we discuss here certainly do not
belong to this class. A significant discrepancy is also that these
authors report that filaments produced in their MHD simulations by
shocks do not show a preferred alignment with the magnetic field, which
is incompatible with our sample.
  
\subsection{Relations between mean and turbulent magnetic fields}
\label{rimLB_CF_relation}

\begin{figure}[th] %%  26
  \centering
  \includegraphics[width=9cm]{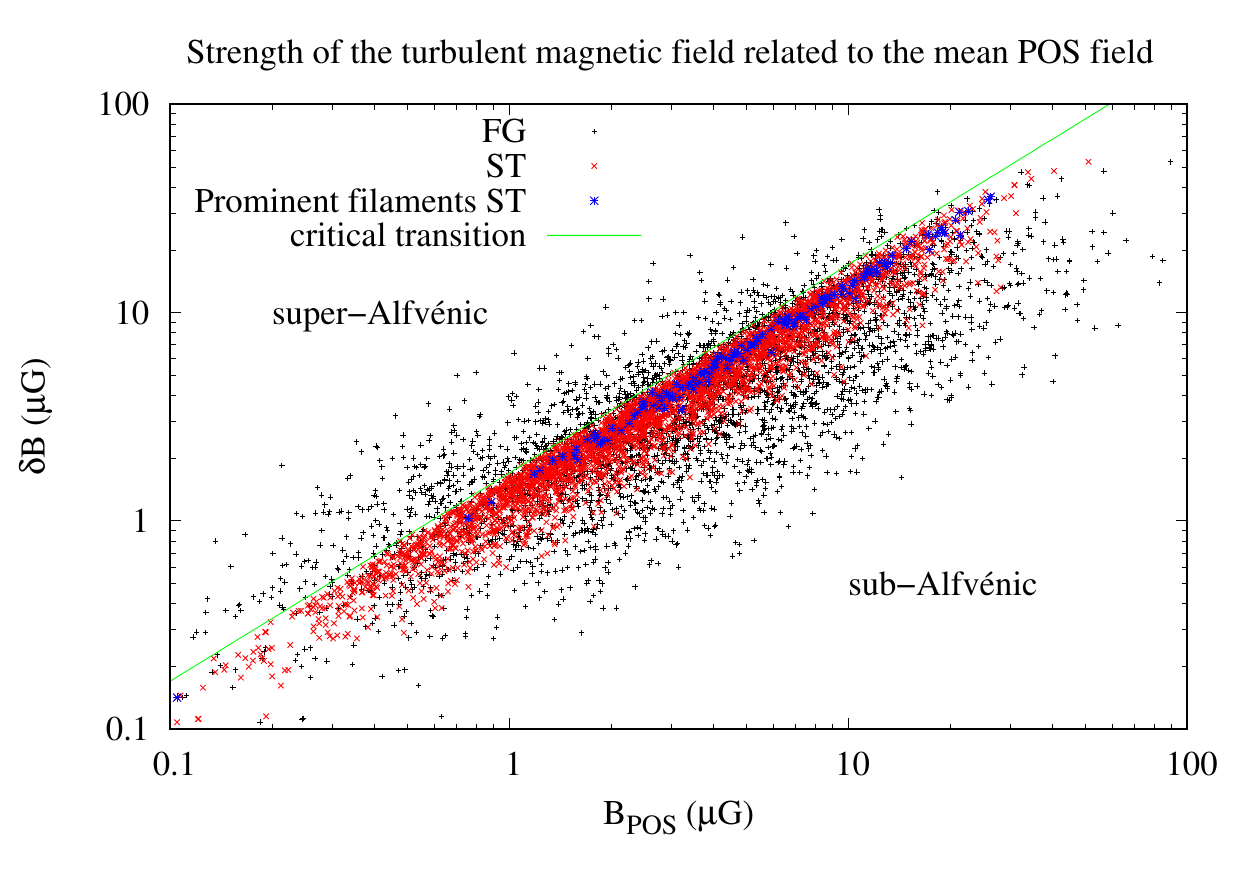}
  \caption{Turbulent magnetic field strength $\delta B$ as a function of the    mean magnetic field component $B_{\mathrm{POS}}$ from
    Eq. (\ref{eq:B_FG}, FG, black dots) and Eq. (\ref{eq:B_ST}, ST, red
    dots).  Overplotted in blue are prominent filaments with
    $n_{\mathrm{s}} > 500$ from Eq. (\ref{eq:B_ST}, ST). The green line
    indicates the critical transition for the magnetic field evolution
    according to Eq. (29) of \citet{Beattie2020}.  }
\label{Fig_mag_compare}
\end{figure}

The distributions of $B_{\mathrm{POS}}$ in Fig. \ref{Fig_B_DFC_ST} show
remarkable similarities to the distribution of the turbulent field from
Fig. \ref{Fig_B_LB}. We compare turbulent and mean field strengths in
Fig. \ref{Fig_mag_compare} for two cases, derived from Eqs.
(\ref{eq:B_FG}, FG) and (\ref{eq:B_ST}, ST). Figure \ref{Fig_mag_compare}   tells us that
turbulent and derived mean fields are correlated. We see the best
correlation in the ST case (red) and in particular for prominent
filaments (blue), with a correlation coefficient of 0.997. Derived
parameters for prominent filaments are most significant because statistical
errors are low in cases with a large number of surface elements
$n_{\mathrm{s}}$. The scatter of the point cloud in the case of the FG
solution is very large. Following the arguments of
\citet{Skalidis2021} and \citet{Skalidis2021b}, there are systematic
uncertainties in the DCF approach, as well as the H and FG derivatives, because compressible MHD wave modes have been ignored 
in
these cases. Our results
from Sect. \ref{dark} support the argument that compressibility is
important. Significant column-density fluctuations indicate phase
transitions; filaments indicate a highly compressible medium.

The approximate agreement between turbulent and mean magnetic field
strengths may imply that the prominent filaments in particular tend to
be critically affected by processes that regulate the energy balance
between kinetic energy and magnetic field strength. Dependencies between
the strength of turbulent and mean field were found in MHD simulations
by \citet{Federrath2016}. Three regions with different properties for
the growth of the turbulent magnetic field component were distinguished:
the dynamo regime, and the intermediate and strong guide-field regimes. An
approximate equilibrium between turbulent and mean magnetic fields was
found at the transition between weak and strong guide fields, separating
super- and sub-Alfv{\'e}nic regimes; see Fig 8 of
\citet{Federrath2016}. As shown by \citet{Beattie2020} in their Fig. 7,
the critical Alfv{\'e}nic Mach number where the turbulent magnetic field
gets damped by the mean field is in the range $ 1 \la M_{\mathrm{A}} \la
2$. Using  relation (29) of these latter authors to derive a more specific limit for the
transition between dominant turbulent kinetic energy and dominant
large-scale magnetic field energy, we obtain $|\delta B| /
|B_{\mathrm{POS}}| = 1.7$. This condition, identified by
\citet{Beattie2020} as the critical transition for the magnetic field
evolution, is indicated in Fig. \ref{Fig_mag_compare} with a green
line. This limit fits surprisingly well to the sharp boundary in
Fig. \ref{Fig_mag_compare} in the case of the ST solution (red). Velocity
fluctuations interact with the magnetic field, causing a fluctuating
magnetic field component. The interplay between strong magnetic fields
(sub-Alfv{\'e}nic) and supersonic velocity fluctuations
(super-Alfv{\'e}nic range) appears to be strongly affected at the critical
transition. The fluctuating magnetic field component is clearly
strongly limited. This limit, which can be explained with compressible
magnetohydrodynamic turbulence models, is only observable in the ST
solution, Eq. (\ref{eq:B_ST}), but does not exist in the case of
incompressible DCF models.

According to relation (29) of \citet{Beattie2020}, the critical
transition occurs in MHD simulations at an Alfv{\'e}nic Mach number of
$M_{\mathrm{Acrit}} = 1.44$; see also Fig. 7 of \citet{Beattie2022}. In
the following, we intend to also compare this model prediction with
observational data.

\subsection{Alfv{\'e}nic Mach numbers}
\label{AMach}

\begin{figure}[th] %%  27
  \centering
  \includegraphics[width=9cm]{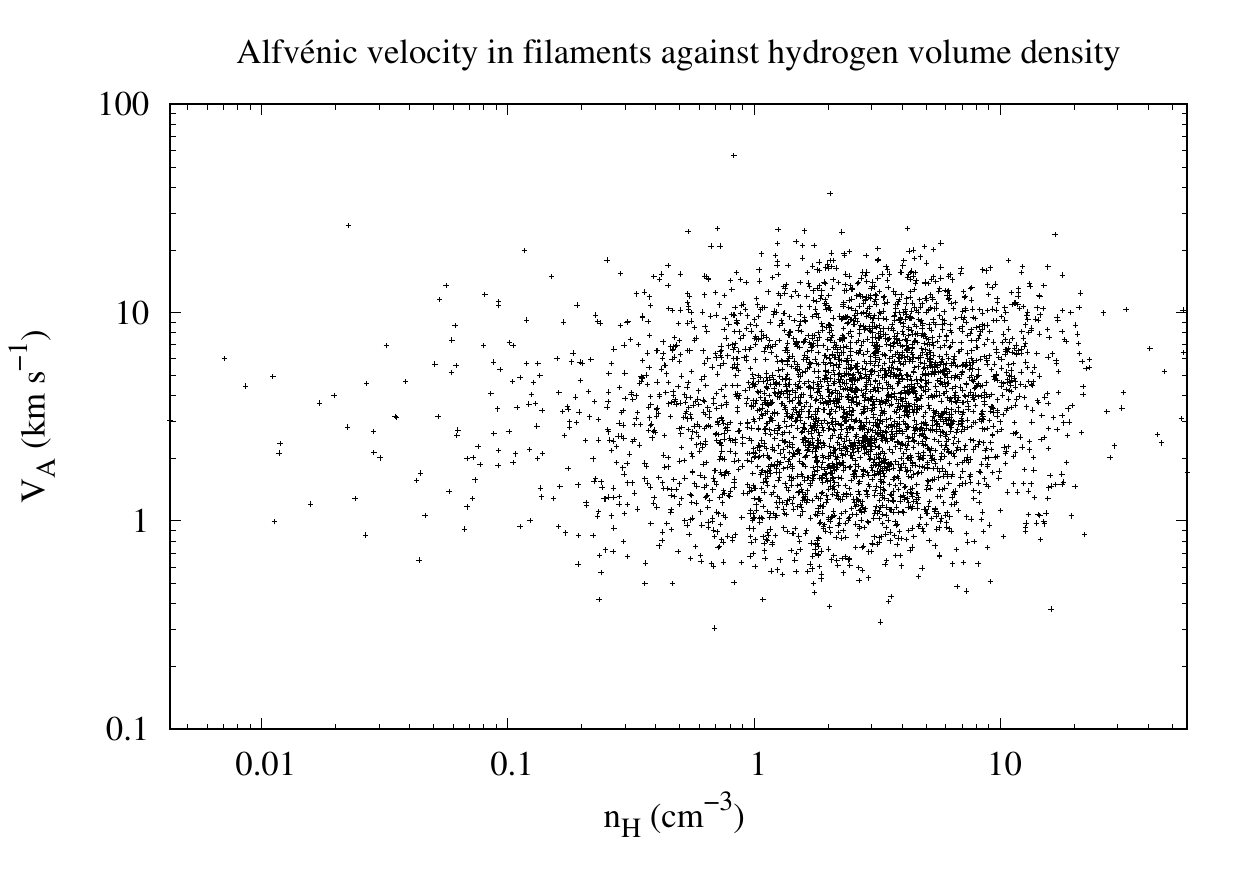}
  \caption{Distribution of the Alfv{\'e}n velocity $v_A$ according to
    \citet{Skalidis2021} as a function of the average hydrogen volume
    density $n_{\mathrm{H}}$ for filaments.    }
\label{Fig_VA}
\end{figure}

To test the self-consistency of our ST solution, we first  calculated the
Alfv{\'e}n velocities that have implicitly been used with
Eq. (\ref{eq:B_ST}); see Eq. (29) of \citet{Skalidis2021}:
\begin{equation}
  v_A = \frac{B_{\mathrm{ST}} } {\sqrt{4 \pi \rho} } = 
\frac{  \Delta v_{\mathrm{LSR}} } { \sqrt{2 \Delta \theta}}.
\label{eq:B_VA}
\end{equation}
The results are plotted in Fig. \ref{Fig_VA}. This scatter diagram
covers a large range in volume densities $n_{\mathrm{H}}$. This certainly reflects statistical uncertainties in the hydrogen
column density estimates by \citet{Kalberla2020} 
to some extent, but $v_A$ is else
unaffected by density effects. We also find no observable dependency of
$v_A$ with respect to the filament surfaces.  The mean Alfv{\'e}n
velocity is 4.7 \kms, which is slightly less than the mean turbulent velocity
dispersion of 5.24 \kms.

\begin{figure}[th] %%  28
  \centering
  \includegraphics[width=9cm]{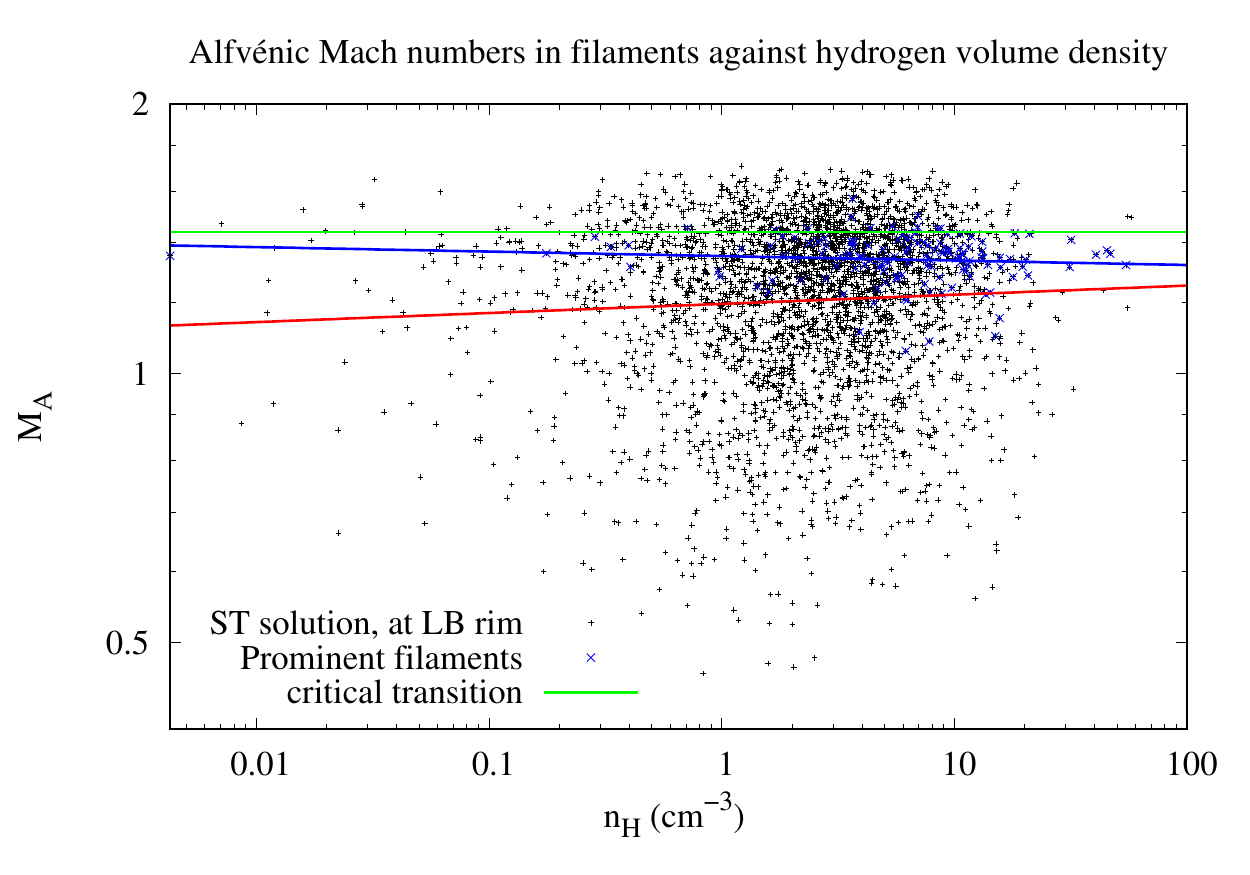}
  \caption{Alfv{\'e}nic Mach number in filaments as a function of the
    average hydrogen volume density for filaments located at the LB wall
    according to Eq. (\ref{eq:B_MA_ST}) based on \citet{Skalidis2021}.
    The fit $\ln(M_{\mathrm{AST}}) = 0.179 \pm 0.004 + (0.010 \pm 0.003)
    \ln(n_{\mathrm{H}})$ for all 6~568 filaments is indicated in
    red. Fitting only $n_{\mathrm{s}} > 500$ (right from the vertical
    blue line in Fig. \ref{Fig_B} top), we get $\ln(M_{\mathrm{AST}}) =
    0.30 \pm 0.01 + (-0.005 \pm 0.005) \ln(n_{\mathrm{H}}),$ as indicated
    by the blue line. The green line indicates the critical transition
    for the magnetic field evolution according to Eq. (29) of
    \citet{Beattie2020}.  }
\label{Fig_B_MA_ST}
\end{figure}

Next, we calculate the Alfv{\'e}nic Mach numbers.
For individual filaments, according to Eq. (7) of
\citet{Skalidis2021b},  we have
\begin{equation}
  M_{\mathrm{AST}} = \frac{ \Delta v_{\mathrm{LSR}} } { v_A } =
  \frac{\delta B} {B_{\mathrm{ST}}} = \sqrt{2 \Delta \theta}
\label{eq:B_MA_ST}
,\end{equation}
and Fig. \ref{Fig_B_MA_ST} shows the $M_{\mathrm{AST}}$ distribution.
There is no evidence that Alfv{\'e}nic Mach numbers depend significantly
on hydrogen volume densities.  Furthermore, the range covered by the
$M_{\mathrm{AST}}$ distribution is consistent with the range
$M_{\mathrm{A}} < 2,$ which was tested by \citet{Skalidis2021b} for the
validity and reliability of the ST approach.

 We derive a mean Alfv{\'e}nic Mach number of $\langle M_{\mathrm{AST}}
 \rangle = 1.23 $ for all filaments. The prominent filaments with
 $n_{\mathrm{s}} > 500$ have $\langle M_{\mathrm{AST}} \rangle = 1.34
 $. This is close to the critical Alfv{\'e}nic Mach number
 $M_{\mathrm{Acrit}} = 1.44$ from relation (29) of \citet{Beattie2020}.
 Consistent with Fig. \ref{Fig_mag_compare}, we observe an upper limit of $
 M_{\mathrm{AST}} < 1.7$.  Figure \ref{Fig_B_MA_ST} implies a well-developed trans-Alfv{\'e}nic turbulence  for most of
 the filaments. This is
 the regime of MHD turbulence where anisotropic structures are expected
 in the case of an incompressible medium \citep{Goldreich1995}. It is also
 the critical transition region where \citet{Federrath2016},
 \citet{Beattie2020}, and \citet{Beattie2022} found energy equipartition between the turbulent and
 large-scale magnetic field  in the case of
 compressible turbulence.

The observed sonic Mach numbers in filaments are typically $M_S \sim
3.7$ (\citealt{Heiles2005} and \citealt{Kalberla2016}). Our results for
prominent filaments shown in Fig. \ref{Fig_mag_compare} are consistent
with MHD simulations in a similar range, that is, $M_S \sim 3.5$ with
$M_{\mathrm{A}} \sim 1.7$ by \citet{Beattie2020}, see their Table 1.  In
the case of supersonic turbulent flows, shocks are expected with phase
transitions, which explains the observed exceptionally cold
\hi\ \citepalias[][Sect. 3.1]{Kalberla2021} and the large
$\overline{N_{\mathrm{Hcenter}}} / \overline{N_{\mathrm{Hoff}}}$ excess
that we discuss in Sect. \ref{dark}.  Filament motions excited by
turbulence are affected by the magnetic field, but in the observed
trans-Alfv{\'e}nic range, turbulent velocity fluctuations along the
filaments appear to remain confined such that the observed velocity
coherence is usually preserved.

\subsection{Results from the ST solution }
\label{All-sky}

Summarizing our results from the previous subsections, we derive a mean
magnetic field strength {for the ensemble average }that is consistent
with previous determinations (e.g., \citealt{Heiles2005} or
\citealt{Ferriere2015}).  The conventional DCF approach and the
modifications proposed by \citet{Heitsch2001} and \citet{Falceta2008}
show a larger intrinsic scatter than the ST solution
\citep{Skalidis2021b}.

We adopt the ST solution. Our ensemble averages of $\langle B_{\mathrm{ST}} \rangle
= 4.4~ \mu$G and $\langle B_{\mathrm{ST}} \rangle = 7.1~ \mu$G for
prominent filaments are consistent with the field strength of $B_0 = 6 \pm
1.8~\mu$G derived by \citet{Heiles2005}. For turbulent as well as for
mean magnetic field strengths, we find an average scaling relation of $B
\propto n_{\mathrm{H}}^{0.58}$. We confirm the models by
\citet{Federrath2016}, \citet{Beattie2020}, and \citet{Beattie2022} with
the prediction of a critical transition for the magnetic-field evolution
that imposes upper limits on the strengths of mean-field normalized
magnetic field fluctuations.
  
\begin{figure}[th] %%  29
   \centering
   \includegraphics[width=9cm]{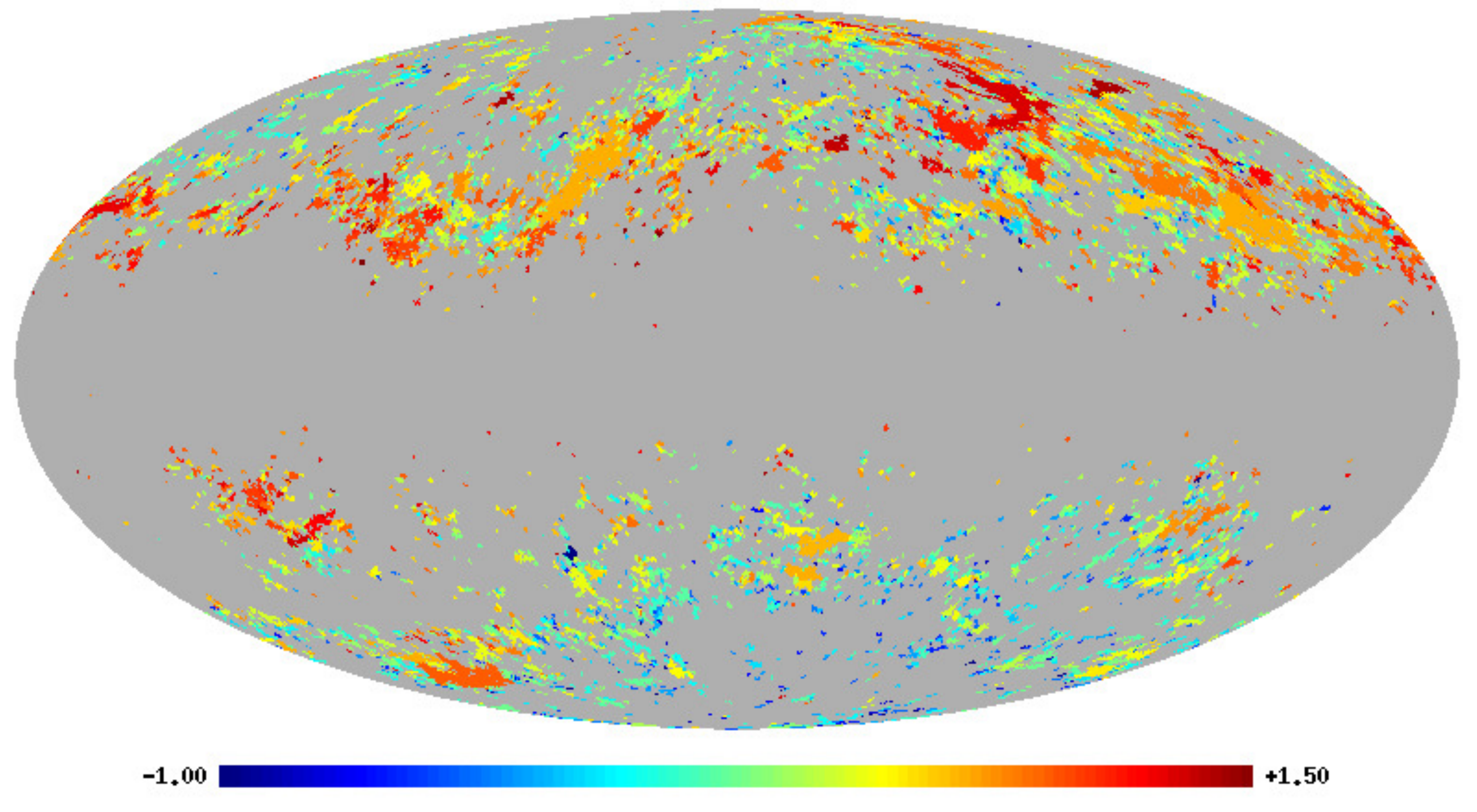}
   \caption{Spatial distribution of the derived average
     magnetic-field strengths of  filaments $\langle B_{\mathrm{ST}} \rangle$ in
     Mollweide projection. The color coding is logarithmic for a range
     $0.1 < \langle B_{\mathrm{ST}} \rangle < 32 \mu$G. }
   \label{Fig_Sky_BST}
\end{figure}

Figure \ref{Fig_Sky_BST} displays $\langle B_{\mathrm{ST}} \rangle$
averaged over a pixel size of 27\farcm5 and gridded to an nside=128
HEALPix array. Large-scale structures are visible that resemble to some
extent the large-scale magnetic-field distribution indicated by rotation
measures shown in Fig. 3 of \citet{Oppermann2012}, Fig. 7 of
\citet{Sobey2019}, or Fig. 1 of \citet{Dickey2022}.  The Faraday
rotation of linearly polarized radiation at radio wavelengths is
affected by the thermal electron density in the warm ionized medium and
the line-of-sight component of the magnetic field
\citep[e.g.,][Eq. 1]{Oppermann2012}.  The FIR/\hi\ filaments are embedded
in this diffuse medium and it is plausible that turbulent and mean
magnetic fields associated with filaments can be affected by the large-scale Galactic magnetic field.  The average large-scale magnetic field
strength in the Galactic halo is estimated to be $\sim 4 \mu$G by
\citet[][Table 3]{Sobey2019}, which is consistent with our estimate.

\subsection{Application to Loops I and IV}
\label{Loops}

\begin{figure}[th] %%  30
  \centering
  \includegraphics[width=9cm]{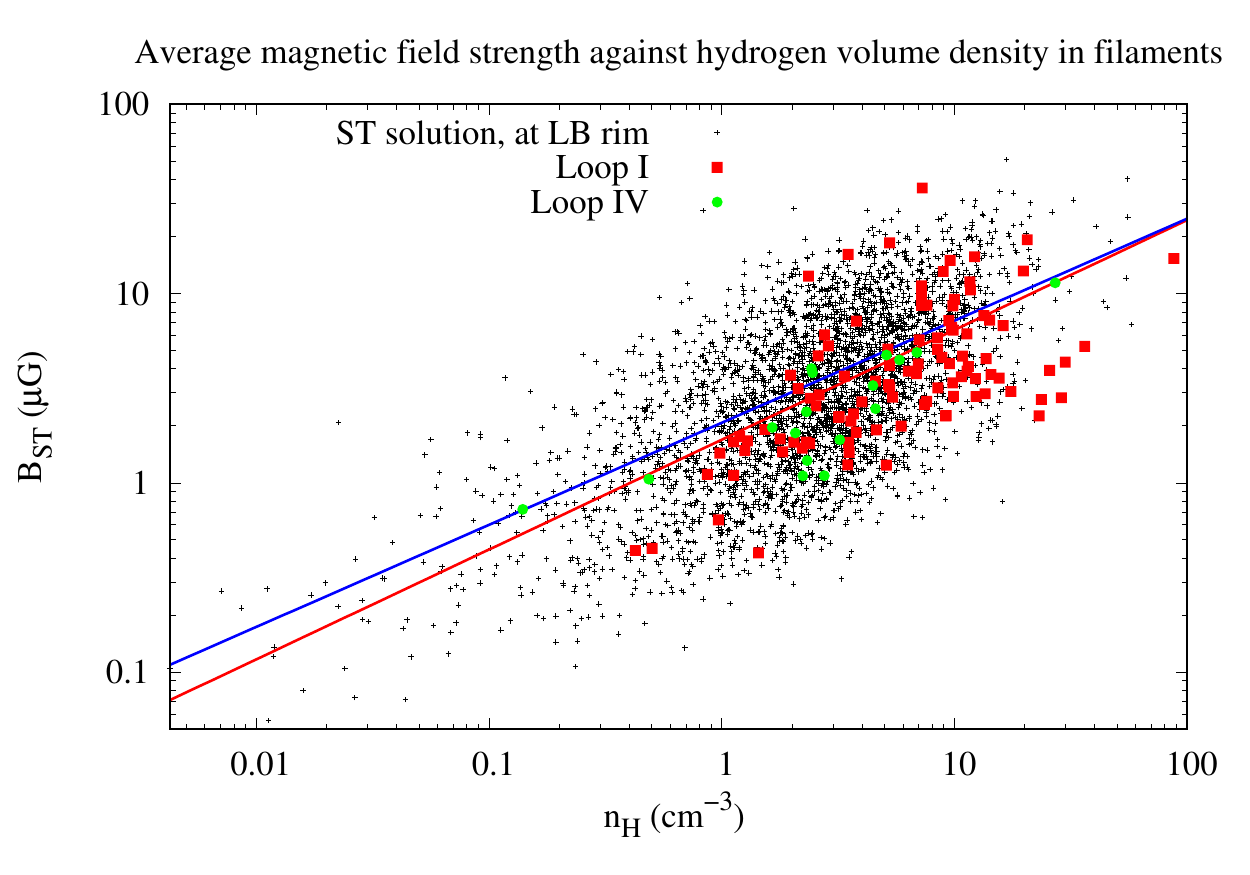}
  \caption{Average plane-of-the-sky magnetic field strength in
    filaments, replicating the bottom right panel of Fig. \ref{Fig_B_DFC_ST}  but
    annotating now $B_{\mathrm{ST}}$ derived for loops I (red) and IV
    (green) according to distances from Table \ref{tab:regions}. }
\label{Fig_B_LB_ST_loop}
\end{figure}

The radio sky at high Galactic latitudes is dominated by several bright
spurs and loops in radio continuum that are interpreted as synchrotron
radiation caused by supernova remnants \citep{Berkhuijsen1971}.  In a
recent analysis, \citet{Panopoulou2021} consider three tracers of the
magnetic field: stellar polarization, polarized synchrotron radio
emission, and polarized thermal dust emission. These authors use the alignment
between these three tracers to constrain estimates of the  distance to the
synchrotron spurs. In a similar way, the alignment of filamentary FIR and
\hi\ structures was analyzed in \citetalias{Kalberla2021}; both
publications followed the same basic assumption that an agreement in
angular orientation must be indicative of a physical correlation.
Distances obtained by \citet{Panopoulou2021} for fields with diameters
of 16\degr are summarized in Table \ref{tab:regions} together with our
determination of the average field strengths $\langle \delta B \rangle$ and
$\langle B_{\mathrm{ST}} \rangle$ according to Eqs. (\ref{eq:B}) and
(\ref{eq:B_ST}) for filaments at such distances. In addition, we display
$B_{\mathrm{ST}}$ for loops I and IV in Fig. \ref{Fig_B_LB_ST_loop}.
Except for a tendency toward higher volume densities, these results support
our previous results and conclusions as to the \hi\ counterparts
of the synchrotron loops.  These regions contain prominent
FIR/\hi\ filamentary structures (see Fig. 18 of
\citetalias{Kalberla2021}) and have particularly well-aligned orientation
angles between synchrotron emission and thermal dust. The \hi\ filaments
in loop I are dominated by a single spectral feature that accounts for
100\% of the hydrogen column density \citep{Das2020}. Systematic errors
are unexpected in this case.

\begin{table}[h]
\centering
\caption{Locations and distances for the loop\,I and IV regions according to
 \citet{Panopoulou2021} with derived magnetic field strengths. For the
 field Ic, we used a distance of 105 pc. Uncertainties are the
 expected errors in the case of a random distribution without systematic effects. }
\begin{tabular}{lcccc}
\hline
Region & Location & Distance  & $\langle \delta B \rangle $  &    $\langle B_{\mathrm{ST}} \rangle$        \\
 &$(l,b)$                    &(pc)       & ($\mu$G)  & ($\mu$G)       \\
\hline
$\mathrm{I}_{\mathrm{B}}$   &$33^{\circ}\!.95, 38^{\circ}\!.03$ & $112
\pm 17$\,pc & $ 8.5 \pm .1$ & $7.3 \pm .1$ \\
$\mathrm{I}_{\mathrm{C}}$   &$26^{\circ}\!.68, 58^{\circ}\!.73$ &$\leq
105$\,pc     & $5.3 \pm .1$ & $4.8 \pm .2$   \\
$\mathrm{I}_{\mathrm{D}}$   &$358^{\circ}\!.17, 74^{\circ}\!.06$ &$135
\pm 20$\,pc  & $4.1 \pm .1$ & $3.6 \pm .2$   \\
\hline
$\mathrm{IV}_{\mathrm{A}}$   & $328^{\circ}\!.74, 33^{\circ}\!.59$
&$180 \pm 15$\,pc & $4.0 \pm .1$ & $3.1 \pm .2$    \\
\hline
\end{tabular}
\label{tab:regions}
\end{table}
\normalsize

The distance to loop I is a matter of debate. \citet{Sofue2015}
determined a distance of at least 1 kpc and \citet{Mou2022} estimated a
magnetic field strength of $4 ~\mu$G for this case. We determine
accordingly $\langle \delta B \rangle = 2.2~ \mu$G and $\langle
B_{\mathrm{ST}} \rangle = 2.7~\mu$G for loop Ib. This does not agree
with the results listed in Table \ref{tab:regions}, nor with the results
in the previous subsection.

\subsection{Uncertainties and biases}
\label{Uncertainties}

The estimates of the magnetic field strengths $\delta B$ and
$B_{\mathrm{ST}}$ in the previous subsections are based on the
assumption of energy equipartition between magnetic field and kinetic
energy \citep{Alfven1949}. Such an approach may be invalid for
individual objects but should be justified on large scales or
alternatively for averages over a large number of objects
\citep{Beck2003}. In our case, we have 6~568 filaments with significant
individual uncertainties. In the case of a random error distribution,
uncertainties for the ensemble average are reduced by a factor of 80.

For a filamentary field geometry, the lateral component of the
velocity is effective \citep{Chandrasekhar1953}. The investigations by
\citet{Pelgrims2020} indicate that at high Galactic latitudes, the
inclination of the mean magnetic field has only a small angle $\gamma
\sim 15\degr$ to the plane of the sky. Uncertainties from projection
effects are at the level of 4\%. Statistical uncertainties from individual
volume density estimates are large but we expect that systematical
biases for averages in case of prominent filaments are limited to the
percentage level. Systematic errors from distance uncertainties can be
estimated by comparing the results from Sects. \ref{withinLB} and
\ref{rimLB}  and are below 6\%.

The question remains as to whether particular DCF-based methods might lead to
systematic errors. Table \ref{tab:DCF} indicates that the derived
$B_{\mathrm{POS}}$ can differ by a factor of up to
four. \citet{Crutcher2012} estimates that errors from the DCF approach
can amount to a factor of two on average.  Our investigations indicate
that systematic differences are introduced in particular by
compressibility issues.  The ST solution of Eq. (\ref{eq:B_ST}) is
incompatible with the other DCF-based methods of Eqs. (\ref{eq:B_DCF}) to
(\ref{eq:B_FG}). Furthermore, only $B_{\mathrm{POS}} = B_{\mathrm{ST}}$
from the ST solution shows an upper limit to $|\delta B| /
|B_{\mathrm{POS}}| $ for the expected equilibrium between turbulent and
mean magnetic fields as predicted by \citet{Federrath2016},
\citet{Beattie2020}, and \citet{Beattie2022}.  Our results strongly
support compressible quasi-static fluctuation models
(Fig. \ref{Fig_mag_compare}) but we cannot come to any conclusions about the level
of systematic discrepancies between these models and alternative models
that assume an incompressible medium  based on the present study. Tests by \citet{Skalidis2021b} demonstrated that the ST method of  Eq. (\ref{eq:B_ST}) has an accuracy of better than
50\%  for the
range $0.1 \la M_{\mathrm{A}} \la 2.0 $ . However, these authors admit that models with $ M_{\mathrm{A}} \ga
0.7 $ tend to underestimate the magnetic field strength
$B_{\mathrm{ST}}$. This may affect our results, shown in
Fig. \ref{Fig_mag_compare}, in the range close to $|\delta B| /
|B_{\mathrm{ST}}| = 1.7$ somewhat, but we did not apply any correction.

The typical strength of the random component of the local Galactic
magnetic field is $ 5 \pm 2~\mu$G; the estimate for the local regular
field amounts to $ 4 \pm 1~\mu$G, but \citet{Beck2003} caution that
systematical errors of up to a factor of two are possible.  Here, we notice
that also these estimates are consistent with a critical balance between
fluctuating and regular magnetic fields.  Altogether, our total averages
from Sects. \ref{rimLB} and the ST solution from \ref{rimLB_CF} are
compatible with these field strengths, and also with the value of $B_0 = 6 \pm
1.8~\mu$G derived by \citet{Heiles2005}.

We derived the field strengths for four fields with \hi\ counterparts of
the synchrotron loops I and IV. These regions contain prominent
FIR/\hi\ filamentary structures and have particularly well-aligned orientation
angles between synchrotron emission and thermal dust. The \hi\ filaments
in loop I are dominated by a single spectral feature that accounts for
100\% of the hydrogen column density \citep{Das2020}. Significant
systematic errors from observations and data reduction are expected to
be low in this case; the derived field estimates should at least be
internally consistent.

\section{Discussion}

\label{Discussion}

Our analysis is based on a Hessian analysis in \citetalias{Kalberla2021}
and here we discuss our results.
FIR filaments that are considered to be dusty magnetized structures are
associated with \hi\ filaments in narrow velocity intervals with a well-defined coherent velocity field along the filaments. Orientation angles
of the FIR at 857 GHz are closely related to \hi\ orientation angles
that define the filament velocities.  The distribution of
filament curvatures was found to match that expected from models for magnetic-field
curvatures induced by a Galactic small-scale turbulent dynamo
\citep{Schekochihin2004}. We conclude that these filaments must
have been shaped by interactions between the turbulent velocity field
and induced magnetic fields. We are refining the analysis from
\citetalias{Kalberla2021} and trace individual FIR filaments along the
plane of the sky. We also find well-defined and
coherent velocities for individual filaments,  with a typical dispersion of $\Delta
v_{\mathrm{LSR}} = 5.24 $ \kms. This value is comparable to
characteristic turbulent motions $\Delta v_{\mathrm{LSR}} = 5.9 $
\kms\ within the warm neutral medium \citep{Saury2014}.

The common FIR/\hi\ structures define a network of filaments and we
determined aspect ratios and filamentarities for 6~568 structures. We
find that aspect ratios and filamentarities are closely related to each
other. The observed distribution follows the ${\cal A}$-to-${\cal F}$
relation (\ref{eq:F_A}) expected for thin filaments. Filamentary
networks are self-replicating, and the merger or disruption of structures means a repositioning of 
individual fragments on the ${\cal A}$-to-${\cal F}$
track, but the general relation (\ref{eq:F_A}) shown in the top panel of
Fig. \ref{Fig_A_F12}  remains satisfied. Projection effects alter
the observed aspect ratios but do not lead to deviations from this
relation. We conclude that the observed continuous ${\cal A}$-to-${\cal
  F}$ distribution cannot be significantly affected by observational
artifacts or projection effects. Aspect ratios are measured as ratios
between length and widths and are therefore independent of distance. However, from
our independent width measurements, we find some indications
that filaments with ${\cal A} \ga 100 $ tend to be associated with
larger widths. These filaments may be at shorter distances but
Fig. \ref{Fig_Widths_lam} shows an increase in $W$ with increasing
eigenvalue $-\lambda_- $, implying stronger FIR/\hi\ ridges.  Results in \citetalias{Kalberla2021} suggest that stronger ridges are associated
with stronger magnetic fields and lower curvatures, as predicted by 
\citet{Schekochihin2004}. In Sect. \ref{field}, we find that 
the magnetic field is stronger in filaments with larger surfaces.
Comparing filament widths with distances to the LB wall, we find
  no indications for a distance dependency. 

Overall, we conclude that the observed FIR/\hi\ filaments populate a
well-defined continuous distribution in ${\cal A}$ and ${\cal F}$ with
aspect ratios of up to a few hundred. In our analysis, we meet only two
limitations. ${\cal A}$ and ${\cal F}$ are ill defined for surface
counts $ n_{\mathrm{S}} < 10$. Also, large structures with $
n_{\mathrm{S}} > 20~000 $ pixels cannot be analyzed because of
confusion. This does not imply that filamentary structures with large
aspect ratios cannot exist.  The famous dust filament shown in
Fig. \ref{Fig_fil_3} with ${\cal A} \ga 787 $ is a prominent example, but
was excluded from our analysis because of the exceedingly large total surface
of the filament.

In \citetalias{Kalberla2021}, we present evidence that FIR filaments are
associated with exceptionally cold \hi. Such \hi\ clouds are expected to
host molecular gas and in turn this \h2\ component is also expected to
be filamentary. We use the CO-dark \h2\ distribution derived
by \citet{Kalberla2020} by adopting the hypothesis of a constant
$E(B-V)/N_{\mathrm{H}}$ ratio and determine how the molecular gas
content depends on aspect ratio. We find a significant increase in the molecular
gas content with ${\cal A}$. Comparing $N_{\mathrm{H}}$ column
densities within filaments with that outside of them, the excess is
$\overline{N_{\mathrm{Hfil}}} / \overline{N_{\mathrm{Hoff}}} = 1.45 $.
For positions in the filament centers, this increases to a factor of
1.56. Accounting for CO-dark \h2\ only, we obtain an average column
density excess of 4.32 along the filaments and 5.29 for the filament
centers. Accounting only for prominent filaments with $n_{\mathrm{s}} >
500,$ we get another enhancement by $\sim 20$\%. In comparison to
brightness temperature enhancements by a factor of 1.3 for filaments in
the {\it Herschel} Gould Belt Survey \citep{Koch2015}, column density
fluctuations for filaments in the diffuse ISM are very strong. We
conclude that the filaments discussed here are part of a homogeneous
population of structures with strong density enhancements in gas and
dust.

\section{Summary and conclusions}
\label{Summary}

Assuming that the energy of the magnetic field is in equipartition with
the kinetic energy of turbulence \citep[e.g.][]{Groves2003}, we derive
an average scaling relation of $B \propto n_{\mathrm{H}}^{0.58}$ for both the
fluctuating and mean magnetic field.  We determine an
average filament volume density of $n_{\mathrm{H}} \sim 3.8~{\rm cm^{-3}}$
with a turbulent field strength  of $\langle \delta B \rangle = 5.3 ~\mu$G
and a mean field of $\langle B_{\mathrm{ST}} \rangle = 4.4 ~\mu$G, which are
good agreement with the canonical field strength of $B_0 = 6 \pm
1.8~\mu$G derived by \citet{Heiles2005} from \hi\ data.  Other
observations that support $B_0 \sim 5~\mu$G were discussed by
\citet{Ferriere2015}.

We used a Hessian analysis to determine filamentary structures in the
FIR and repeated this as a multi-channel analysis in \hi\ to determine
velocities and derived properties of the filaments.  A complete
morphological description of diffuse FIR/\hi\ structures demands a full
3D analysis which is not available from observations.  However, from
simulations, it is possible to consider shape finders for pancakes,
ribbons, and filaments \citep[e.g.,][]{Sahni1998}.  MHD simulations of
the fluctuation dynamo with morphological analyses in 3D by
\citet{Wilkin2007} indicate that ---at least at the kinematic stage--- the
magnetic field is concentrated into filaments rather than sheets or
ribbons.  \citet{Seta2020} find that the magnetic-field correlation
length scale increases as the field saturates. The increase in the
length of magnetic structures with saturation is consistent with the
finding by \citet{Schekochihin2004} that a linear velocity field (as
observed by us) allows the field to reach large aspect ratios between
folds.  \citet{Wilkin2007} and \citet{Seta2020} use for their
morphological analysis isosurfaces at a fixed value of the magnetic
field strength. This implies a preference for structures of stronger
fields with less curvature \citep{Schekochihin2004}. As shown by
\citet{Seta2020}, the magnetic correlation scale in the saturated
state  moves to larger scales.  The length scales associated with both the
greatest stretching and compression grow as the magnetic field saturates
\citep{Seta2021}. These results are consistent with the large aspect ratios
that we observed. Complex filamentary shapes in 3D may be responsible for
the filamentary network, as observed here in projection.

Concerning the 3D structure of the ISM, the existence of a correlation
between local \hi\ volume density enhancements as real physical objects
and magnetic field structures was questioned by \citet{Lazarian2018} and
\citet{Yuen2021}. \ According to
\citet{Lazarian2000}, a gaseous incompressible fluid in turbulent motion,
even when uniformly distributed in space, produces the fluctuating emission structures when observed
in a velocity slice of finite thickness. \citet{Clark2019} were the
first to demonstrate that this hypothesis is not verified by data. Furthermore, we showed  in \citetalias{Kalberla2021} that velocity
coherence along the filaments cannot be generated by random processes.

We interpret our \hi\ structures as filaments in a compressible medium,
and therefore as real physical entities. Such structures, which are observable in
projection with strong column density enhancements, cannot be generated
by random events that accumulate unrelated \hi\ along the line of sight.
We confirm the findings from compressible magnetohydrodynamic turbulence models by
\citet{Federrath2016}, \citet{Beattie2020}, and \citet{Beattie2022}; in
particular their prediction of a critical transition for the magnetic
field evolution with upper limits for the strength of mean-field
normalized magnetic field fluctuations.

%=========================================================================
%=========================================================================

\begin{acknowledgements}
We appreciate valuable comments and suggestions from the referee, which
helped us in improving the quality of the manuscript.
  P.K. acknowledges discussions with Susan Clark,
  Jürgen Kerp, Vincent Pelgrims, and Rafail Skalidis. In particular the
  discussions with Amit Seta were very helpful. HI4PI
  is based on observations with the 100-m telescope of the MPIfR
  (Max-Planck- Institut für Radioastronomie) at Effelsberg and the
  Parkes Radio Telescope, which is part of the Australia Telescope and
  is funded by the Commonwealth of Australia for operation as a National
  Facility managed by CSIRO. This research has made use of NASA's
  Astrophysics Data System.  Some of the results in this paper have been
  derived using the HEALPix package.
   \end{acknowledgements}

\end{document}